\documentstyle[12pt,epsf,psfig,a4]{report}

\def\part#1#2{\frac{\partial #1}{\partial #2}}
\def\xb{\bar\xi(a,x)}
\def\gaprox{\mbox{$\,$ 
\raisebox{0.5ex}{$<$}\hspace{-1.7ex}{\raisebox{-0.5ex}{$\sim$ }}$\,$} }

\begin{document}
\baselineskip 24pt
\textheight 8.6 true in
\textwidth 6.2 true in
\pagestyle{myheadings}
\markboth{}{}
\markright{\thepage}



\begin{titlepage}
\thispagestyle{empty}
\vfill
\begin{center}
\vskip 200pt
{\bf {\Huge Phenomena in Gravitational Clustering}}
\vskip 72pt
{\Large A thesis submitted to the}\\
{\bf{\LARGE University of Pune}}\\
{\Large for the degree of }\\
{\bf {\LARGE Doctor of Philosophy}}\\
{\Large (in Physics)}\\
{\Large by}\\
{\LARGE {\bf Sunu Engineer}}
\vfill
{\Large Inter-University Centre for Astronomy and Astrophysics}\\
{\Large Post Bag 4, Ganeshkhind, Pune 411 007, India}\\
\vfill
{\Large September 2003}\\
\end{center}
\end{titlepage}
\vfill\eject

\newpage
\pagenumbering{roman}
\tableofcontents
\listoffigures

\thispagestyle{empty} 
\chapter*{Declaration}
\addcontentsline{toc}{chapter}{Declaration}
{\large 
CERTIFIED that the work incorporated in the thesis
\begin{center}
{\bf Phenomena in Gravitational Clustering}\\
\end{center}
submitted by {\bf Sunu Engineer} was carried out by the candidate 
under my supervision. Such material as has been obtained from other 
sources has been duly acknowledged in the thesis.
\vskip 60 true pt
\noindent
Place: IUCAA, Pune{\hskip 120 true pt}{\bf T. Padmanabhan}\\
\noindent
Date: September 2003{\hskip 120 true pt}{(Thesis supervisor)}}
\newpage

\newpage
\vspace{1in}
\chapter*{Acknowledgments}
\addcontentsline{toc}{chapter}{Acknowledgements}
\begin{center}
{\scriptsize A teacher affects eternity; he can never tell where his influence stops -- Henry Adams}
\end{center}
\par
To paddy,
\begin{center}
{\it Akhandamandalakaram, vyaptham yena characharam }\\
{\it Tatpadam darshitam yena, tasmai shree gurave namah}
\end{center}
\par
Thanks to all my colleagues, who helped in this adventure of life, love and gravitation. 
\par
and to the people who started me off on this adventure and my friends 
who kept me on this songline until it was done.

\def\la{\mathrel{\mathchoice {\vcenter{\offinterlineskip
\halign{\hfil
$\displaystyle##$\hfil\cr<\cr\sim\cr}}}
{\vcenter{\offinterlineskip\halign{\hfil$\textstyle##$\hfil\cr
<\cr\sim\cr}}}
{\vcenter{\offinterlineskip\halign{\hfil$\scriptstyle##$\hfil
\cr<\cr\sim\cr}}}
{\vcenter{\offinterlineskip\halign{\hfil$\scriptscriptstyle##$\hfil
\cr<\cr\sim\cr}}}}}

\def\ga{\mathrel{\mathchoice {\vcenter{\offinterlineskip\halign{\hfil
$\displaystyle##$\hfil\cr>\cr\sim\cr}}}
{\vcenter{\offinterlineskip\halign{\hfil$\textstyle##$\hfil\cr>\cr
\sim\cr}}}
{\vcenter{\offinterlineskip\halign{\hfil$\scriptstyle##$\hfil\cr>
\cr\sim\cr}}}
{\vcenter{\offinterlineskip\halign{\hfil$\scriptscriptstyle##$\hfil
\cr>\cr\sim\cr}}}}}









\chapter*{Abstract}
\addcontentsline{toc}{chapter}{Abstract}
{\scriptsize We can lick gravity but sometimes the paperwork is overwhelming -- Werner von Braun }
\par
The problem of large scale structure formation in the universe is one of the
core problems of cosmology today. This thesis discusses some of the issues involved
in explaining how the observed large scale structure in the universe came to
be.

This thesis has two distinct parts. The first part (chapters 1--5) discusses the issues of structure
formation from the view point of standard model of structure formation. Chapter 6 discusses alternate cosmologies and structure formation scenarios
in them.

In the standard approach we will explore the problem of structure formation
through gravitational clustering in the universe.
The sizes of numbers involved renders a microscopic viewpoint cumbersome. 
We have to adopt a statistical procedure which essentially
averages over many particles and their individual behavior to give us some entities
which we shall call ``particles'' whose behavior we can describe more easily.
Alternatively  we can replace the discrete structure of particles by a continuum approximation,
thinking of the system of particles as a continuous fluid and apply the
theoretical models of fluid mechanics to describe cosmological systems. In either of the approaches we run into a similar problem.
When we frame the equations that describe the system, we find that the equation
has terms which are highly nonlinear, leading to an analytically intractable system of equations. 
An analytical solution to the relevant equations can be obtained only if one assumes that
the system is linear, {\it i.e}\/ one drops all the terms that are nonlinear in the
parameter of interest. But the degree of nonlinearity  associated with observed structures 
in the universe like galaxies is of the order of $10^3$. 
Thus we need to deal with the nonlinear terms in the equations to establish correspondence with observational
results.
The nonlinear terms elude a simple description, in that a general
solution is not yet found. But since we have to study the behavior of this equation
at highly nonlinear regions we try to model the behavior in a variety of ways. 

\begin{enumerate}
\item Various approximation schemes 
such as Zeldovich approximation, frozen potential approximation, frozen flow
approximation and  adhesion approximation  attempt to take us a little beyond linear theory.
\item Numerical simulation techniques  attempt to integrate the equations
either in terms of a particle based approach or a field based approach (or
a combination of the two). N Body simulations form the mainstay of this approach.
\item  An alternate technique treats evolution in time and space as
a mapping problem and tries to find an appropriate map that takes us
from one point in the history of the universe to another in a global sense.
Scaling laws and other ansatzes fall under this category.
\end{enumerate}
In this thesis we will be concerned with all the the three approaches but with
a preferential leaning towards the second and third approaches. Herein
we explore the utility of the third method
in various aspects of the study of structure formation and the insights it provides into dynamics 
of gravity in shaping the observed structures.

The first chapter gives a concise overview of the background. 

Chapter \ref{chap:nonlinear} of this thesis uses the third approach to generate nonlinear quantities
from their linear theory counterparts and applies it to the question of ``universal''
profiles in gravitational clustering. We addressed the problem by asking the
following question: Is it possible to populate the nonlinear universe with structures 
such that the two point correlation function
evolves as per linear theory in all regimes? We find that it is not possible
to have strict linear evolution but it is possible to find functional forms
such that approximate linear evolution to any desired order is possible. 
The earlier investigations had indicated that the evolution of density
contrast can be separated into three regimes namely linear, quasilinear and
nonlinear. We have found a functional form which evolves approximately linearly
in quasilinear and nonlinear ends of the two point correlation function. 
It is also conjectured in this chapter that it should be possible to find basis
functions which may be based on such approximately invariant forms such that
the nonlinear density fields can be decoupled to a large extent if expanded in terms of these
functions. We suggest
that the functions that we have derived which evolve approximately linearly
in all regimes might be a good candidate for the role of ``units of nonlinear
universe''.
Another aspect of the analysis tries  to discover universal aspects
of the structures formed via gravitational clustering such as density profiles.

Chapter \ref{chap:twod} critically examines the theoretical framework underlying 
two dimensional N body simulations. 
If one has to get requisite amount of dynamical range in force and mass one
requires large grids and large number of particles which is not possible given
the computing resources available. One attempts to get around this problem by
trying to do the simulations in two dimensions which brings the computational
requirements down by a considerable amount. So if it is possible to extract
fundamental principles which may be generalized to the case of three dimensional
gravity from such simulations then they are a good way of exploring the fundamental
features of gravity. 

There are three ways in which we can define  a two dimensional gravitational clustering scenario: (i) A set
of point particles interacting by \( 1/r^{2} \) force but with a special set
of initial conditions such that they are confined to a plane and have no velocity
components orthogonal to the plane (ii) A set of infinite parallel ``needles'' in which each particle interacts with \( 1/r^{2} \) force but the ``needles'' interact
with a \( 1/r \) force (iii) A description derived from Einstein's equations
in two dimensions. 

Approach (i) is highly contrived and we will not discuss it. The second
 approach  based on infinite ``needles'' suffers from the problem of manifest
anisotropy because the universe is considered to be expanding in three dimensions
while the clustering takes place only in two dimensions. In order to explore
the third approach we developed the formal theory of gravity in \( D+1 \) dimensions
and considered \( D=2 \) as a special case. The formal analysis of \( D+1 \)
dimensional gravity led us to the general expressions for scale factor and background
density in the \( D+1 \) dimensional universe.

Taking the usual Newtonian limit of the metric and writing down the equation
describing the growth of density perturbations in the universe via a fluid approach
made it possible to obtain the \( D+1 \) dimensional analogue of the equation
describing growth of density contrast. A corresponding formula for spherical
collapse model is also derived in this work. We  then specialize to the case
of \( D=2 \) and make the following observations. The linearized form of the
density contrast equations only yield a constant or decaying solution.This is
consistent with the result that perturbed gravitational potential does not couple
to density contrast \( \delta  \). The spherical collapse model solution yields
a similar result in the sense that it is not possible to have a gravitational
clustering model that grows in time. It is possible to obtain clustering by an {\it   ad hoc} approach
by making some assumptions but they also lead to inconsistent results in that
they give rise to singular solutions for the scale factor of the universe. Thus
we conclude that the infinite ``needle'' based approach is the only viable
way of simulating two dimensional gravity.

In Chapter \ref{chap:twodsim} some of the issues involved in two dimensional gravitational simulations
are discussed.  Nonlinear scaling relations, which have been identified in three dimensional simulations, 
define a mapping from initial linear theory values of two point correlation function to 
the final nonlinear values at a different length scale. This allows us to immediately 
compute the nonlinear parameter  at a  specific length scale knowing the value of 
the linear one at some other length scale. We wish to check the theoretical prediction  that similar nonlinear scaling relations 
hold in two dimensions. Another aspect of universal behavior
of gravitational clustering that is conjectured is called ``stable clustering''.
This is the conjecture that at late times structures have their gravitational
infall balanced by background expansion leading to a fixed profile.By applying
this conjecture to the theoretical model for the NSR (nonlinear scaling relations)
it is possible to derive the theoretical dependence of the NSR in two dimensions
if it exists. Our conclusions regarding two dimensional gravitational clustering
 based on the simulations are as follows. (i) The prediction is verified and a form of nonlinear scaling relation exists for the two point
correlation function in two dimensions as well. This NSR is independent of the
spectrum for the spectra (power laws) considered in this study. (ii) In the
quasilinear regime the theoretical model based on infall onto peaks predicts
a dependence that is confirmed by the numerical experiments. (iii) In the highly
nonlinear regime the results diverge from that predicted by  the ``stable clustering'' hypothesis. The ``stable clustering'' 
hypothesis demands that the ratio between infall velocity and expansion
velocity go to unity but we find that it is driven towards \( 3/4 \). Thus
we find that stable clustering is not a valid hypothesis in two dimensions.

Chapter \ref{chap:spherical}  addresses the question: What happens at the late stages of clustering?. 
The hypothesis of stable clustering asserts that at
some point the system `virialises'. We examine this process
of virialisation and stabilization in more detail by analyzing what happens to
a single spherically symmetric object as it goes through the cycle of expansion
and collapse. By including a term that describes the asymmetries that are generated
and enhanced during collapse we examine the final state of the system. We begin
by writing a modified equation for the spherical collapse of a system which
includes a term which takes into account the asymmetries that are generated during
the process of collapse and  we derive a functional form for this ``asymmetry''
term by an ansatz that this term depends only on density contrast \( \delta  \).
Since we have a relation connecting density contrast with the pairwise velocity
function \( h \), it is now possible to  close the system of equations. We require the form
of the \( h \) function before we can integrate this system of equations. By
using the fact that the system must reach a constant value for \( h \) we obtain
a functional form for the ``asymmetry'' term which allows us to integrate
the equation and analyze the collapse of a single object. This leads us to conclude
that the system reaches a constant value of 0.65 times the maximum radius which
is not widely off the value of 0.5 that is usually stipulated. Thus we demonstrate
that the growth of asymmetries can be used to stabilize the collapse.

In Chapter \ref{chap:QSSC} the question of approaching the problem of structure formation
from other alternate cosmological scenarios is discussed. To address this question
in the context of Quasi Steady state cosmology is the theme of the
work discussed here. An algorithm was devised which mimicked the basic physical
content of QSSC (Quasi Steady State Cosmology) model by the following geometrical method.
We generate N random points and require that in the next cycle the volume becomes
eight times the initial volume (due to the expansion)  and the particles generate
new mass particles in their vicinity. Then we select the central volume equal
to the initial volume and repeat this scale and shrink process. To apply this algorithm in the context of QSSC model we use the fact that only a fraction of the particles given by $f=3Q/P$ where $Q$ and $P$ are parameters of the QSSC model create new masses in the next generation. Consequently the  scaling used is $\exp(f)$. We have succeeded
in showing that this model reproduces the observed two point correlation function
with a slope of \( -1.8 \). Visual analysis shows clear indication of the clustering
growing and the initial smooth distribution separating into voids and clumps.
This approach is characterized by the fact that clustering in this model takes place without the help of gravity,
{\it i.e} gravity plays no role in inducing clustering.  

The final chapter of the thesis presents the conclusions and integrated discussion based on the previous chapters.
\par
\newpage
This thesis is based on the following publications.
\begin{enumerate}
\item T.~Padmanabhan, {\bf S.~Engineer}, {\it Nonlinear gravitational clustering: dreams of a paradigm}, Ap.J {\bf 493},(1998).
\item J.S.Bagla, {\bf S.~Engineer},T. Padmanabhan, {\it Scaling Relations for gravitational clustering in two dimensions}, Ap.J {\bf 498},(1998).
\item {\bf S.~Engineer}, K.Srinivasan, T. Padmanabhan, {\it A formal analysis of (2+1)-dimensional gravity}, Ap.J {\bf 512},(1999).
\item Ali Nayeri, {\bf S.~Engineer},Jayant Narlikar and F. Hoyle, {\it Structure formation in Quasi Steady State Cosmology: A toy model}, Ap.J {\bf 525},(1999).
\item {\bf S.~Engineer}, Nissim Kanekar and T. Padmanabhan, {\it Nonlinear density evolution from an improved spherical collapse model}, MNRAS {\bf 314},(2000).
\end{enumerate}

\pagenumbering{arabic}
\pagestyle{myheadings}
\markboth{}{}
\chapter{Introduction}
\label{chap:intro}

\baselineskip=12pt
{\scriptsize Duct tape is like the force. It has a light side, a dark side
\newline and holds the universe together. -- Carl Zwanzig}
\par
\begin{figure}[h]
\leavevmode\centering
\psfig{file=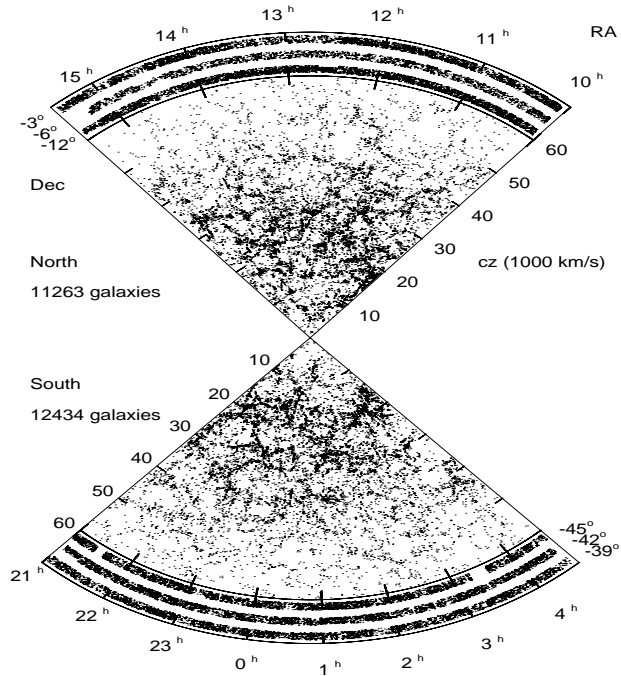,width=300pt,height=300pt,angle=0}
\caption[{\sf The Las Campanas Redshift Survey}]{\sf The Las Campanas Redshift survey (http://www.lcrs.com)}
\label{fig:chapter1:lcrs}
\end{figure}

\baselineskip=24pt
Observations of the universe in multiple wavelengths have revealed large  aggregates of matter on all length scales. Figure \ref{fig:chapter1:lcrs} shows the distribution of galaxies from Las Campanas Redshift survey which clearly shows clustering in the distribution of galaxies.
The formation and existence of large scale structures such as galaxy clusters and superclusters is an important problem in cosmology. In spite of the inhomogeneities on these scales it is also observed that on very large length scales ($ > 300 Mpc$) the universe appears homogeneous. This permits the universe to be modelled as being made up of statistically similar volumes of linear dimensions of about $300 Mpc$ or more. Standard cosmological models based on General Theory of Relativity are derived from this  assumption of statistical homogeneity and isotropy of the universe. The problem of large scale structure formation in these models thus involves explaining how the observed inhomogeneities and anisotropies were initially generated in this uniform background and how they grew into the observed structures. Once a mechanism for generating the initial perturbations of the smooth distribution of matter is postulated the transition from uniformity on large scales, to the highly non uniform structures at small scales, can be explained by  gravitational clustering. Observations by space experiments (COBE,WMAP etc.) have revealed the presence of small inhomogeneities in the uniform density fields of matter at early times. It is  assumed that these seeds evolved via gravitational clustering into the observed large scale structures and experienced a multitude of non gravitational phenomena at small scales (such as star formation) to evolve into the present day universe of clusters and galaxies.
\par
This chapter introduces the basic framework of this model in brief. It is organized as follows. 
The first section \ref{sec:chapter1:fried} summarizes the main features of the smooth background cosmological models. The following section \ref{sec:chapter1:dynamics} discusses the equations governing the dynamics of matter in this expanding background cosmology in discrete and continuum  approximations. Section \ref{sec:chapter1:linear} discusses  the  solutions of this nonlinear system of equations in linearised regime. Nonlinear approximation schemes such as Zeldovich approximation and  spherical top hat approximation are discussed in section \ref{sec:chapter1:nonlin}. Section \ref{sec:chapter1:nbody} deals with N body computer simulations as a way of modeling and studying some features of nonlinear structure formation.Various statistical measures used to quantify and analyze the formation of structures are discussed in section \ref{sec:chapter1:stats}. Section \ref{sec:chapter1:NSR} deals with an alternate approach to time evolution of some statistical quantities, thus obviating the need for expensive and time consuming  full scale N body simulations.
\par
\section{Background Cosmology}
\label{sec:chapter1:fried}
A model appropriate for a universe dominated by gravity on large scales, 
is best formulated in the context of General theory of Relativity. This model, which will be referred to as the `background cosmology', is described by the metric of the spacetime.The evolution of the metric is governed by the energy density distribution through Einstein's equation given by
\begin{equation}
\label{eqn:chapter1:einstein}
G^i_k=R^i_k-\frac{1}{2} \delta_k^i R =8 \pi G T_k^i
\end{equation} 
where $G^i_k$ is the Einstein tensor, $G$ the gravitational constant and $T_k^i$ the energy momentum tensor. 
The `cosmological principle' which assumes a statistically  homogeneous and isotropic universe is further used to  constrain the form of the metric of the universe to (\cite{structurebook},\cite{Probbook},\cite{Peeb80},\cite{peebles93} )
\begin{equation}
ds^2=dt^2-a^2(t)\left[ \frac{dr^2}{1-kr^2}+r^2(d\theta^2+\sin^2\theta d\phi^2) \right]
\label{eqn:chapter1:metric}
\end{equation}
(we use the convention $c=1$ consistently throughout unless explicitly indicated). This metric is completely characterized by a time dependent function  {\it scale factor} $a(t)$ and {\it curvature} $k$. 
The constant $k$ determines the geometry of the universe --- (the curvature of the spatial hypersurfaces of the Friedmann universe) --- taking the values $k>0$ in a closed universe, $k=0$ in a flat universe and $k<0$ in an open universe. The `fundamental observers' to whom the universe appears isotropic and homogeneous remain at constant $(r,\theta,\phi)$ with their physical separation increasing in proportion to scale factor $a(t)$ which determines the overall scale of the spatial metric. An observational  consequence of expansion is that the light emitted by a source at time $t$ is observed now, at $t=0$ with a cosmological `redshift' $z = a_0/a(t) -1$ where $a_0=a(t=0)$. Another quantity of primary significance in observational cosomology is the `Hubble Constant' $H$ defined by $H=\dot{a}/a$, which measures the rate at which the universe is expanding at a given point in time. The value of `Hubble constant' today is denoted by $H_0 = 100h km s^{-1} Mpc^{-1}$ with $0.52 < h < 0.62$ \cite{tam}.
\par
The `cosmological principle'  also constrains the source term in Einstein's equations, namely the energy momentum distribution, so as to make it consistent with the assumed homogeneity of the metric, leading to a diagonal form $T^{\alpha}_{\beta}={\rm dia}[\rho(t),-p(t),-p(t),-p(t)]$. This  form is motivated by the assumption of the source to be an ideal fluid with pressure $p$ and density $\rho$, whose stress tensor has the above form in the rest frame of the fluid. Under these assumptions, an extra equation of state connecting $p$ and $\rho$ allows the source to be completely determined. 
 
When this source term is plugged into equation \ref{eqn:chapter1:einstein}, the Friedmann equations
\begin{eqnarray}
\label{eqn:chapter1:friedmann}
\frac{\dot{a}^2+k}{a^2}=\frac{8\pi G}{3}\rho
\end{eqnarray}
\begin{equation}
{{2\ddot{a}} \over {a}} + {{\dot{a}^2+k} \over {a^2}} = -8 \pi G p\\
\end{equation}
for the scale factor $a(t)$ which determines the metric given the curvature $k$, are obtained.
\par
From equation \ref{eqn:chapter1:friedmann} we can define the value of curvature $k$ by
\begin{equation}
\label{eqn:chapter1:curvature}
\frac{k}{a_0^2}=\frac{8\pi G}{3}\rho_0-H_0^2=H_0^2(\Omega-1)
\end{equation}
where $a_0$ is the scale factor at present and $\Omega=\rho_0/\rho_c$ where $\rho_0$ is the present value of density and $\rho_c=3H_0^2 / 8 \pi G$ is called the critical density. Equation \ref{eqn:chapter1:curvature} shows that the universe will be open, closed or flat depending on the values of $\Omega$ being  less than, more than or equal to one respectively.
\par
 We assume that the total $\rho$ of the universe may be separated into contributions from the various constituents of the the cosmic fluid. Other than radiation and baryonic matter there is compelling evidence for the presence of a dominant component that does not interact with light and is consequently not directly observable, called the `dark matter'. There is also good observational evidence for the presence of a vaccum energy density $\rho_v$ \cite{perlmutter}. For each component of energy density $\rho$ with an equation of state given by $p=p(\rho)$, the density varies with the scale factor according to energy conservation
\begin{equation}
\label{eqn:chapter1:energycon}
d(\rho a^3)=-pd(a^3)
\end{equation}

For a generic equation of state given by $p=w\rho$, equation \ref{eqn:chapter1:energycon} gives $\rho\propto a^{-3(1+w)}$. The equations of state for radiation component of the fluid is 
$p=\frac{1}{3} \rho$ and for the non relativistic pressureless dust, $p=0$ \cite{structurebook}. 
Defining $\Omega_x=\rho_x/\rho_c$  where the subscript $x$ is $r$ for radiation, $nr$ for nonrelativistic matter which has contributions from both baryonic matter $\rho_B$ and `dark matter' $\rho_{DM}$, $v$ for the vaccum  energy density, we can write the complete time dependence of the scale factor $a(t)$ as
\begin{eqnarray}   
\label{eqn:chapter1:scale}
\rho_{tot}(a)&=&\rho_{crit}\left[\Omega_r\left(\frac{a_0}{a}\right)^4 +(\Omega_B+\Omega_{DM})\left(\frac{a_0}{a}\right)^3+\Omega_{v}\right] \\
\frac{\dot{a}^2}{a^2}&=&H_0^2\left[ \Omega_r\left(\frac{a_0}{a}\right)^4+\Omega_{nr}\left(\frac{a_0}{a}\right)^3+(\Omega-1)\left(\frac{a_0}{a}\right)^2+\Omega_{v} \right]
\end{eqnarray}
\par
 The radiation density varies as $a^{-4}$ whereas the matter density goes as $a^{-3}$. This has the interesting consequence that although radiation density is much smaller than matter density today, in the past the radiation field dominated. The point in time where the densities were equal is defined as the `matter-radiation equality'. 
The dynamics of the universe from this epoch is dominated by matter, and we speak of `radiation dominated epoch' as opposed to `matter dominated epoch'. The redshift of equality $z_{eq}$ and the value of Hubble constant at equality $H_{eq}$ are defined by, (neglecting curvature and radiation components)
\begin{eqnarray}
\frac{a_0}{a_{eq}}&=&\frac{\Omega_{nr}}{\Omega_r}=(1+z_{eq})\\
H_{eq}^2&=&2H_0^2\Omega_{nr}(1+z_{eq})^4 
\end{eqnarray}
Since  the curvature term and the vaccum energy terms do not contribute significantly in early times \cite{structurebook},  analytical solutions to \ref{eqn:chapter1:scale} valid for $t \gg t_{eq}$ and $t<t_{eq}$ may be obtained, where $t_{eq}$ is the time of equality. 
\begin{eqnarray}
\frac{a}{a_{eq}}&\approx&\left(\frac{3}{2\sqrt{2}}\right)^{2/3}(H_{eq}t)^{2/3} \qquad\qquad (for\;t \gg t_{eq})\\
\frac{a}{a_{eq}}&\approx&\left(\frac{3}{\sqrt{2}}\right)^{1/2}(H_{eq}t)^{1/2}\qquad\qquad (for\; t \ll t_{eq})
\end{eqnarray}

\begin{figure}[h]
\leavevmode\centering
\psfig{file=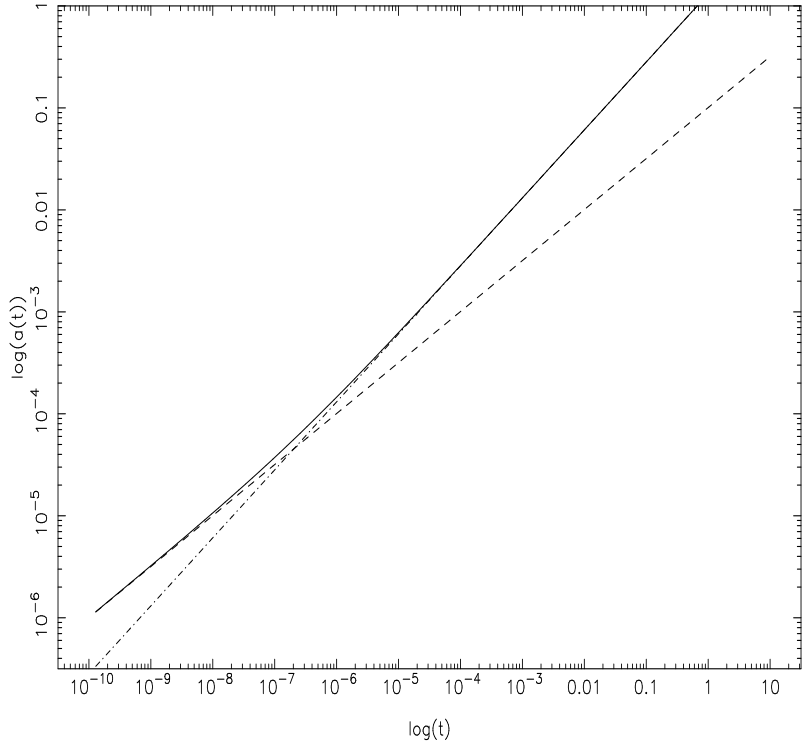,height=300pt,width=300pt,angle=-90}
\caption{\sf The scale factor (solid line) for a $\Omega=1$ universe with the solution for radiation dominated epoch  (dashed line) and matter dominated epoch (dot dashed line) superposed on it.}
\label{fig:chapter1:aoft}
\end{figure}
Figure \ref{fig:chapter1:aoft} shows the complete evolution for the scale factor $a(t)$ (obtained by integrating Eq.(1.8)with $\Omega_{v}$ and $(\Omega - 1)$ terms neglected) with the solutions in the matter dominated and radiation dominated regimes superposed.
\par
In the standard model,the initially hot universe whose evolution is dominated by radiation energy transitions to a matter dominated structure at $z \sim 10^4$. In this hot universe, electrons and photons couple to each other via scattering processes such as Thompson scattering. At $z \sim 1100 $ the temperature cools enough for neutral hydrogen to form and the photons and matter decouple. The photons then travel freely, cooling adiabatically, until they are observed at the present time as the $2.73 K$ cosmic microwave background (CMB) radiation. The observations of CMB ( \cite{cobe},\cite{bennet}) show that the universe at the time of formation of neutral hydrogen (last scattering surface)
was extremely uniform with fractional fluctuations in energy density ( and consequently in gravitational potential) of approximately $10^{-5}$. These small fluctuations, are the seeds of the large scale structures seen today. They are  amplified  through gravitational interaction, eventually leading to the formation of galaxies and stars.
\section{Dynamics of matter}
\label{sec:chapter1:dynamics}
To analyse the evolution of matter in this expanding background  a dynamical model  which describes how matter aggregates under the influence of force of gravity in an  expanding  universe is required.  
\par
If the system has to be described in it's full generality one must use a general relativistic description. But one can use a `Newtonian' approximation valid for certain regimes by finding the effective `Newtonian limit' of the  Friedmann metric \cite{Probbook}.
Applying the transformation to new variables  $R$ and $T$ defined by
\begin{eqnarray}
R&=&ra(t)\\ 
T&=&t-t_0+\frac{1}{2} a\dot{a}r^2+{\cal O}(r^4)
\end{eqnarray}
on eq \ref{eqn:chapter1:metric} we get 
\begin{eqnarray}
ds^2&\approx&(1-\frac{\ddot{a}}{a}R^2)dT^2
-(1+\frac{k}{a^2}R^2+\frac{\dot{a}^2}{a^2}R^2)dR^2- \nonumber\\
&&R^2(d\theta^2+\sin^2\theta d\phi^2)
\end{eqnarray}
Near $R=0$ one can assume a locally inertial coordinate system if one restricts to quadratic order in $R/d_H$ where $d_H=cH^{-1}$ is the `Hubble Radius'. In the weak field limit, we know that 
$g_{00}=(1+2\phi_N)$ where $\phi_N$ is the Newtonian potential. 
Hence the equivalent Newtonian potential of the FRW metric is
\begin{equation}
\phi_{FRW}(R,t)=-\frac{1}{2}\frac{\ddot{a}}{a}R^2 
\end{equation}
To proceed further one has to investigate how particles move in a self consistent manner in universe described by a perturbed Friedmann metric, with the perturbations introduced by the potential $\phi$ of the particles.  Since in the limit of weak gravity,the perturbed metric can be written as a linear superposition of the effective gravitational potential of the unperturbed metric and the perturbation $\phi$ due to matter, the effective Newtonian potential of the perturbed metric is 
\begin{equation}
\phi_N=\phi_{FRW}+\phi
\end{equation}
\par
In the transformed coordinate system we can write the equation of motion for a particle  in the Newtonian limit as
\begin{equation}
\frac{d^2 R_i}{dT^2}=\frac{d^2}{dT^2}(a{\bf x}_i)=-\nabla_R \phi_N
\end{equation}
where ${\bf x}_i$ is the coordinate of the particle $i$.
This equations when expanded, and the contribution from the effective Newtonian potential of the FRW metric cancelled, leads to 
\begin{equation}
a\frac{d^2}{dT^2}{\bf x}_{i}+2\dot{a}\dot{\bf x}_{i}=-\frac{1}{a} \nabla_{x} \phi 
\end{equation}
where $\phi$ is the perturbed potential which satisfies Poisson's equation
\begin{equation}
\label{eqn:chapter1:poisson}
\nabla_x^2 \phi=4\pi G a^2 \rho_{bm} \delta
\end{equation}
$\rho_{bm}$ is the smoothed density of the background and $\delta=(\rho/\rho_{bm})-1$ is the density contrast which defines the perturbations on the smooth background.

To the same order accuracy, we can replace $d^2/dT^2$ by $d^2/dt^2$  in the equation of motion leading to 
\begin{equation}
\ddot{\bf x}_{i}+2 \frac{\dot{a}}{a} \dot{{\bf x}_i}= -\frac{1}{a^2} \nabla_x \phi
\end{equation}
as the equation of motion governing the trajectory of a particle moving in an expanding universe in a potential created by the perturbation $\delta$.

 The density field $\rho({\bf x},t)$ of a set of point particles may be defined by the following expression 
\begin{eqnarray}
\label{eqn:chapter1:deltak}
\rho({\bf x})&=&\frac{m}{a^3(t)} \sum_i \delta_{Dirac}({\bf x}-{\bf x_i}(t)) \\
\end{eqnarray}
Taking derivatives, Fourier transforming to get the density modes in Fourier space $\delta_{\bf k}(t)$ and using the equation of motion for ${\bf x}$  we get \cite{Probbook}
\begin{equation}
\ddot\delta_{\bf k}+2 \frac{\dot a}{a} \dot \delta_{\bf k}=4\pi G \rho_{bm}\;\delta_{\bf k} + A_{\bf k} - B_{\bf k} \\
\end{equation}
where the $A_{\bf k}$ and $B_{\bf k}$ terms are given by
\begin{eqnarray}
\label{eqn:chapter1:deltakevolution}
A_{\bf k}&=&2\pi G \rho_{bm} \sum_{{\bf k'}\neq 0,{\bf k}} \delta_{{\bf k'}}\delta_{{\bf k}-{\bf k'}} \left[ \frac{{\bf k}\cdot{\bf k'}}
{k'^2} + \frac{{\bf k}\cdot ({\bf k}-{\bf k'})}{\vert {\bf k}-{\bf k'}\vert^2 }\right] \\
B_{\bf k}&=&\frac{m}{M}\sum_j \left({\bf k}\cdot\dot{\bf x}_j\right)^2 \exp\left[i{\bf k}\cdot{\bf x}_j(t)\right]
\end{eqnarray}
The two terms $A_{\bf k}$ and $B_{\bf k}$ are the nonlinear mode coupling terms which cause the evolution to have a nonlinear structure. The summation in term $B_{\bf k}$ is over all the particles in the system. Neglecting $A_{\bf k}$ and $B_{\bf k}$ when $\delta$ is small leads to a linear equation that has a solution that grows as well as a solution that decays in time.

 The equations that govern the mechanics of the fluid in the continuum are the {\em continuity equation}, the {\em Euler equation} and the {\em Poisson equation} which governs the evolution of the potential. 
\par
\begin{eqnarray}
{\frac{\partial \rho_m}{\partial t}}+\nabla_{r}\cdot\left(\rho_m{\bf U}\right)&=&0 \\
\frac{\partial {\bf U}}{\partial t}+\left({\bf U}\cdot\nabla\right){\bf U}=-\nabla \phi_{tot} &=& -\nabla \phi_{FRW}-\nabla\phi \\
\nabla^2_{r} \phi = 4 \pi G(\rho_m -\rho_{bm})&=&4\pi G\rho_{bm}\delta 
\end{eqnarray}

Using a  different time parameter $b(t)$ defined as  growing mode solution of the linearised equation for density contrast and the corresponding peculiar velocity $u=v/(a\dot{b})$ where $v=a\dot{\bf x}$ is the original peculiar velocity, results in a second order equation which governs the growth of density contrast $\delta$ \cite{Probbook}
\begin{equation}
\frac{d^2\delta}{db^2}+\frac{3A}{2b}\frac{d\delta}{db}-\frac{3A}{2b^2}\delta(1+\delta)=\frac{4}{3}\frac{1}{(1+\delta)}\left(\frac{d\delta}{db}\right)^2+(1+\delta)(\sigma^2-2\Omega^2)
\end{equation}
in the fluid approximation. $\sigma$ and $\Omega$ represent the shear and rotation of the peculiar velocity field of the fluid given by $\sigma^2=\sigma^{ab}\sigma_{ab}$ and $\Omega^2=\Omega_a\Omega^a$. 
and $A=(\rho_{bm}/\rho_{crit}) (\dot{a}b/a\dot{b})^2$.

Solutions to these formal mathematical equations  are rather difficult except in special cases, due to the presence of the highly nonlinear terms.Understanding the nonlinear terms is the key to understanding structure formation, since many of the observed structures are highly nonlinear entities with the parameters of interest such as density contrast of the order of a thousand or more.  
\section{Linear perturbation theory}
\label{sec:chapter1:linear}
Solutions to the equation governing the growth of $\delta$ may be obtained in the framework of linear perturbation theory \cite{structurebook}. In this approach the nonlinear terms $A_k$ and $B_k$ are assumed to be zero due to the small values of $\delta$ and velocities.
\par
The equation that governs the growth of $\delta_{\bf k}$ in the linear regime is 
\begin{equation}
\ddot\delta_{\bf k}+2 \frac{\dot a}{a} \dot \delta_{\bf k}=4\pi G \rho_{bm}\;\delta_{\bf k}
\end{equation}
\par
This equation as discussed before yields solutions which describe growth as well as decay of density contrast. The growing mode solution shows that $\delta$ in matter dominated phase evolves proportional to scale factor $a$ in an $\Omega=1$ universe.The solutions for time dependence of potential and velocity field indicate that the gravitational potential remains constant in time  and the velocity field in linear regime is proportional to the gradient of potential given by
\begin{equation}
{\bf v}=-\frac{2 f}{3 H \Omega} \frac{\nabla\phi}{a}
\end{equation}
where $f\approx\Omega^{0.6}$.

\par
\par
Since the regimes of interest are highly nonlinear, we require insight into the behaviour of the nonlinear terms. 
One of the important physical effects that can be readily mapped into the nonlinear terms is obvious from equation (\ref{eqn:chapter1:deltakevolution}). This equation clearly shows that mathematically speaking the nonlinear terms couple various Fourier modes of density contrast field in that it causes evolution of one mode to be dependent on the other. This gives an  equivalent way of identifying the linear regime as the period of evolution of a mode, where it evolves independently of the others. The slow growth of nonlinearity causes this independence to be broken until the strong coupling at highly nonlinear scales links together many modes of the system.
\par
\par
\section{Nonlinear approximations}
\label{sec:chapter1:nonlin}
\subsection{Zeldovich approximation}
The approximation schemes attempt to either extend the linear theory results such as constancy of potential, into nonlinear regimes or model the universe by analytically tractable structures such as spheres. The most fruitful of the approximation schemes of the first kind is the Zeldovich approximation \cite{Zeldo}. This is a simplistic yet effective approximation which assumes that the particles move in inertial trajectories, with their initial velocities given by the initial potential. These trajectories cross, leading to caustic surfaces where high density aggregates of matter are produced. The shapes of the caustic surfaces of the initial potential/density field determine the shapes of the large scale structures, which are formed in this approach. Thus this approximation scheme is a simple one time map from Lagrangian space ${\bf q}$  to the Eulerian space ${\bf x}$, given by
\begin{equation}
{\bf x}(t)={\bf q}+b(t){\bf f}({\bf q})
\end{equation}
where $b(t)$ is the growing mode solution from linear perturbation theory and {\bf f} is the initial velocity field $\propto \nabla\phi$.
The density in Eulerian space 
\begin{equation}
\frac{\rho({\bf x})}{\rho_0}= \left[ \left( 1- b(t)\;\alpha\right) \left( 1 -b(t)\;\beta\right) \left( 1-b(t)\;\gamma\right) \right]^{-1}
\end{equation}
where ($-\alpha, -\beta, -\gamma$) are the eigenvalues of $\partial^2{\bf f}/\partial{\bf q}^2$. This expression clearly shows that the collapse will take place along the axis defined by the largest negative eigenvalue. The distribution function of the eigenvalues for a gaussian random field indicates that collapse along one dimension in that has a much higher probability of occurring since $\alpha\;\gg\;\beta,\gamma$. Thus the first structures that form will be sheet like structures that are  likely to be highly warped. 
\par
The following figures (Fig.\ref{znb1} -- \ref{znb4}) show a slice of a full N body simulation and a corresponding scenario generated by the Zeldovich approximation. It is visually evident that Zeldovich approximation is fairly accurate. The point of breakdown of this approximation comes  when the particles after having reached the caustics start moving away from it, {\it i.e} the shells start crossing and passing through each other leading to a thickening and dissolution of the structures that form (Fig \ref{znb4}). This approximation when supplemented by an artificial viscosity term (the `adhesion approximation')  is more faithful to the N Body simulation even at late times. There are many variants on the basic Zeldovich approximation such as the truncated Zeldovich approximation  which are all consistent with N body simulations to various degrees.Some of the other approximation schemes driven by physical considerations are frozen potential approximation, frozen flow approximation and so on which extend the linear theory results related to density, velocity and potential field growth  in various ways. Nonlinear growth of a single object is also of interest because the universe at late stages can be modelled as a set of highly collapsed objects whose overall configuration in the universe is determined by the large scales in the power spectrum of perturbations. Spherical collapse is an approach in this direction. This model has been successfully used to derive evolution of mass spectrum, non linear scaling relations and so on.
\begin{figure}[t]

\leavevmode\centering
\epsfxsize=300pt
\epsfysize=300pt
\epsfbox{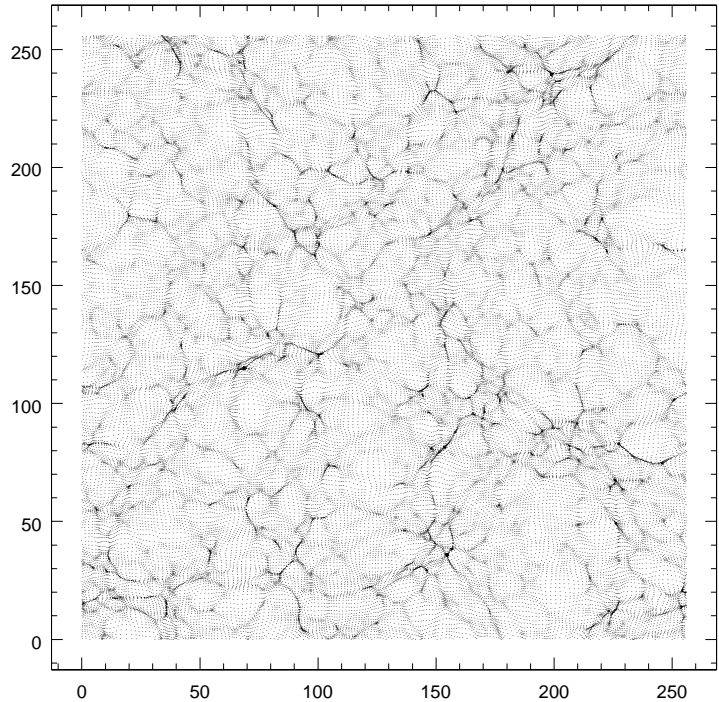}
\caption{\sf N-body simulation at a=0.25 (n=-1. Power law spectrum).}
\label{znb1}
\end{figure}
\begin{figure}[h]
\leavevmode\centering
\epsfxsize=300pt
\epsfysize=300pt
\epsfbox{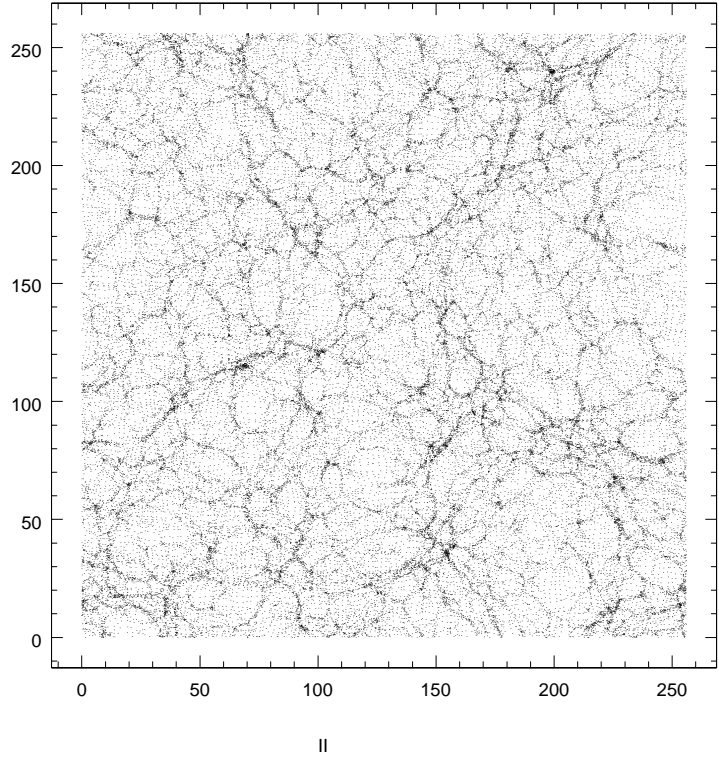}
\caption{\sf Zeldovich approximation at a=0.25 (n=-1. Power law spectrum). The correspondence between structures in the full Nbody simulation is evident.}
\label{znb2}
\end{figure}
\begin{figure}[t]

\leavevmode\centering
\epsfxsize=300pt
\epsfysize=300pt
\epsfbox{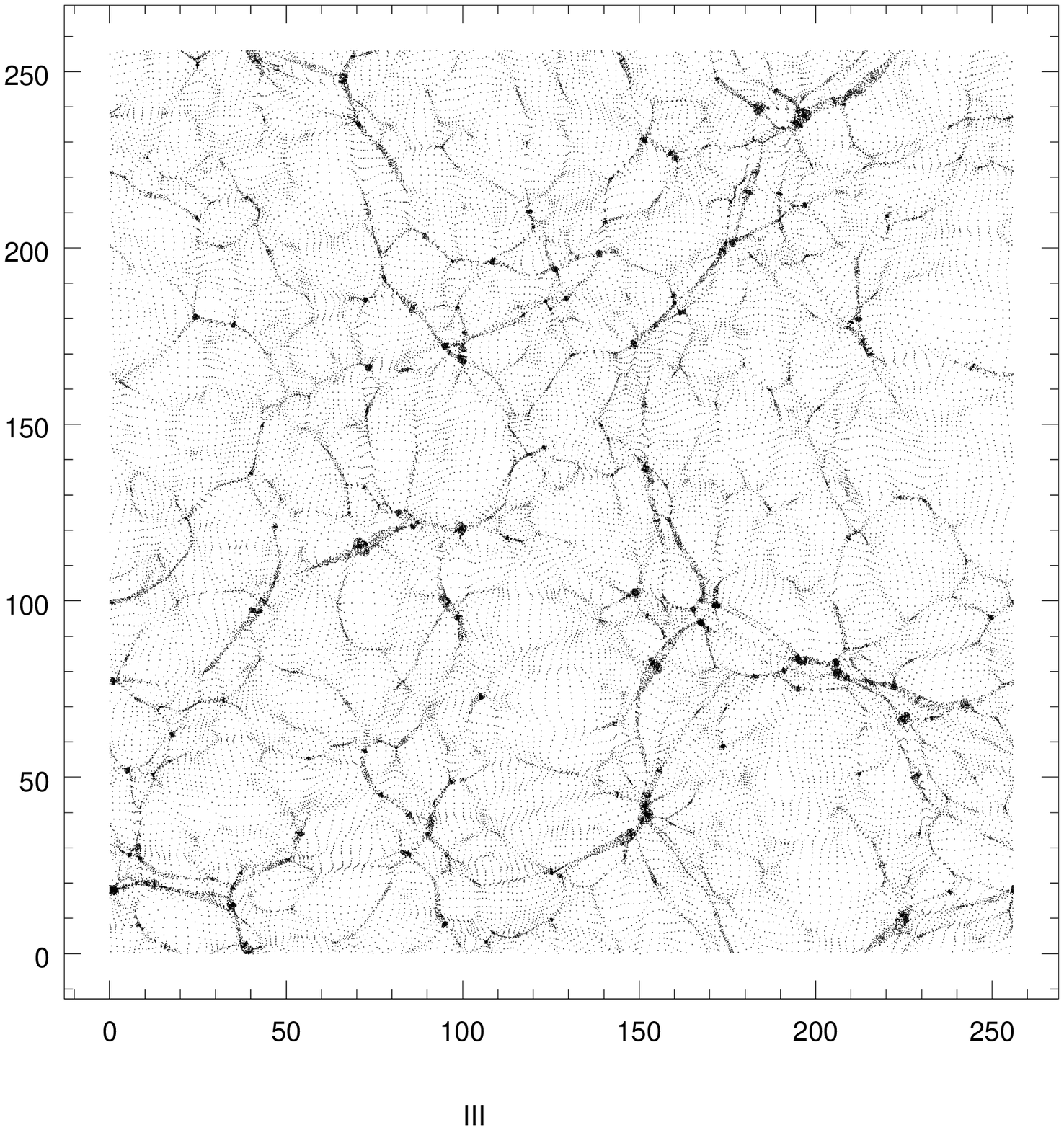}
\caption{\sf N-body simulation at a=0.5 (n=-1. Power law spectrum).}
\label{znb3}
\end{figure}
\begin{figure}[h]

\leavevmode\centering
\epsfxsize=300pt
\epsfysize=300pt
\epsfbox{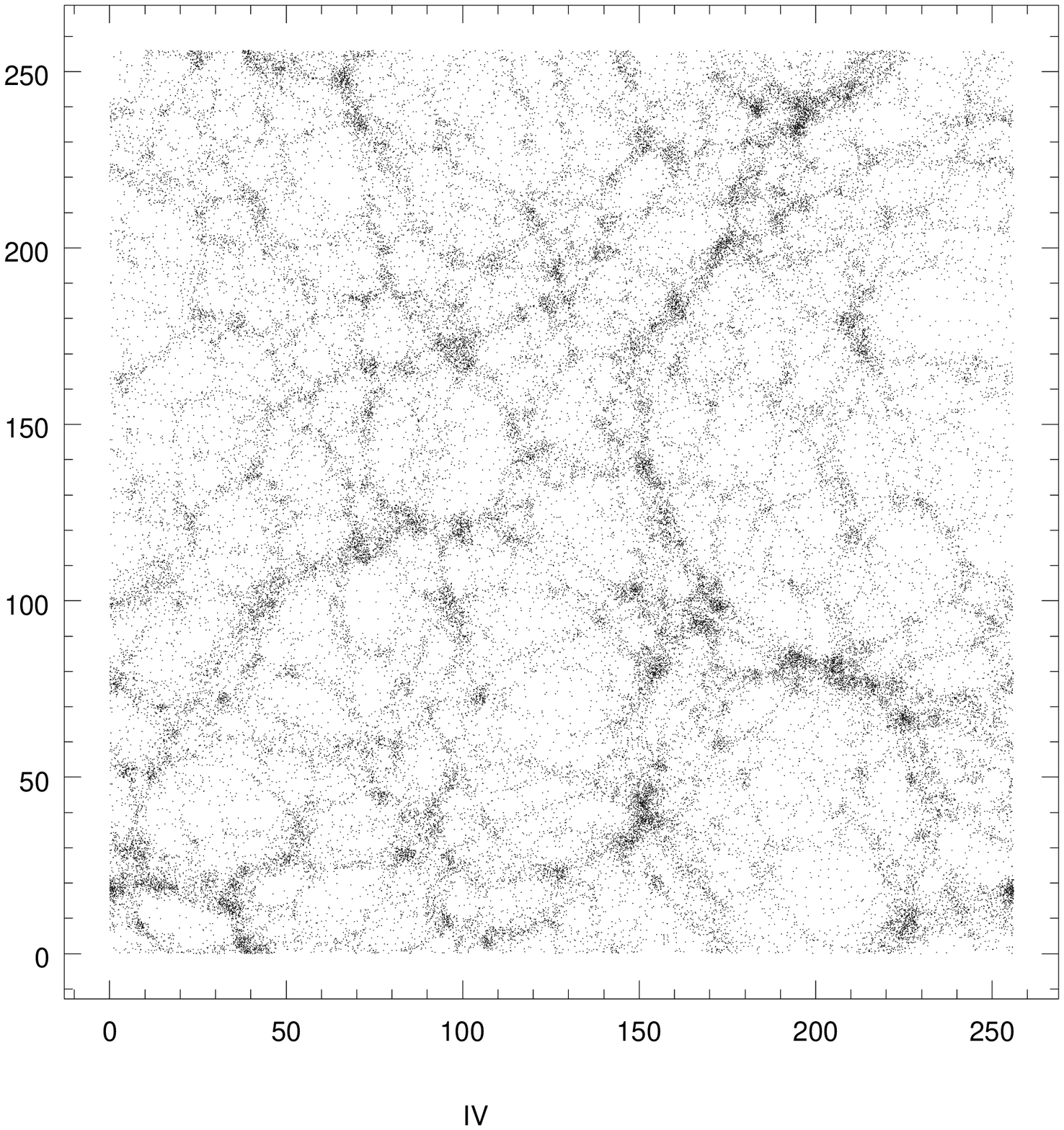}
\caption{\sf Zeldovich approximation at a=0.5 (n=-1. Power law spectrum). The thickening and dissolving of the structures formed is clear leading to breakdown of Zeldovich approximation.}
\label{znb4}
\end{figure}
\subsection{Spherical model}
A single virialised structure such as a galaxy or a cluster in the universe is modeled as a spherically symmetric density perturbation in this approximation. The assumption of spherical symmetry allows the system to be analytically solved  \cite{bertsph} and the whole history of evolution from initial conditions till the collapse and stabilization to be dealt with analytically. The system is modeled as consisting of concentric  spherical shells  of radius $R_{i}(t)$. It can be seen that this overdense/underdense sphere can be modeled as an embedded closed/open universe and the equation of motion for the shell can be identified with the equation of growth of scale factor. This analogy permits the solution for time evolution of a shell to be written in a parametric form as  
\begin{eqnarray}
R_{i}=A(1-\cos \theta)\\
t=B(\theta - \sin \theta)
\end{eqnarray}   
where $A^3= GMB^2$ ($M$ is the total mass inside $R_i$). Defining the density contrast in terms of the  average density within the sphere $\delta$,its time dependence may be approximated as 
\begin{equation}
\delta \simeq \frac{3}{20} \left(\frac{6t}{B}\right)^{2/3}
\end{equation}

\begin{figure}[h]

\leavevmode\centering
\epsfxsize=300pt
\epsfysize=300pt
\epsfbox{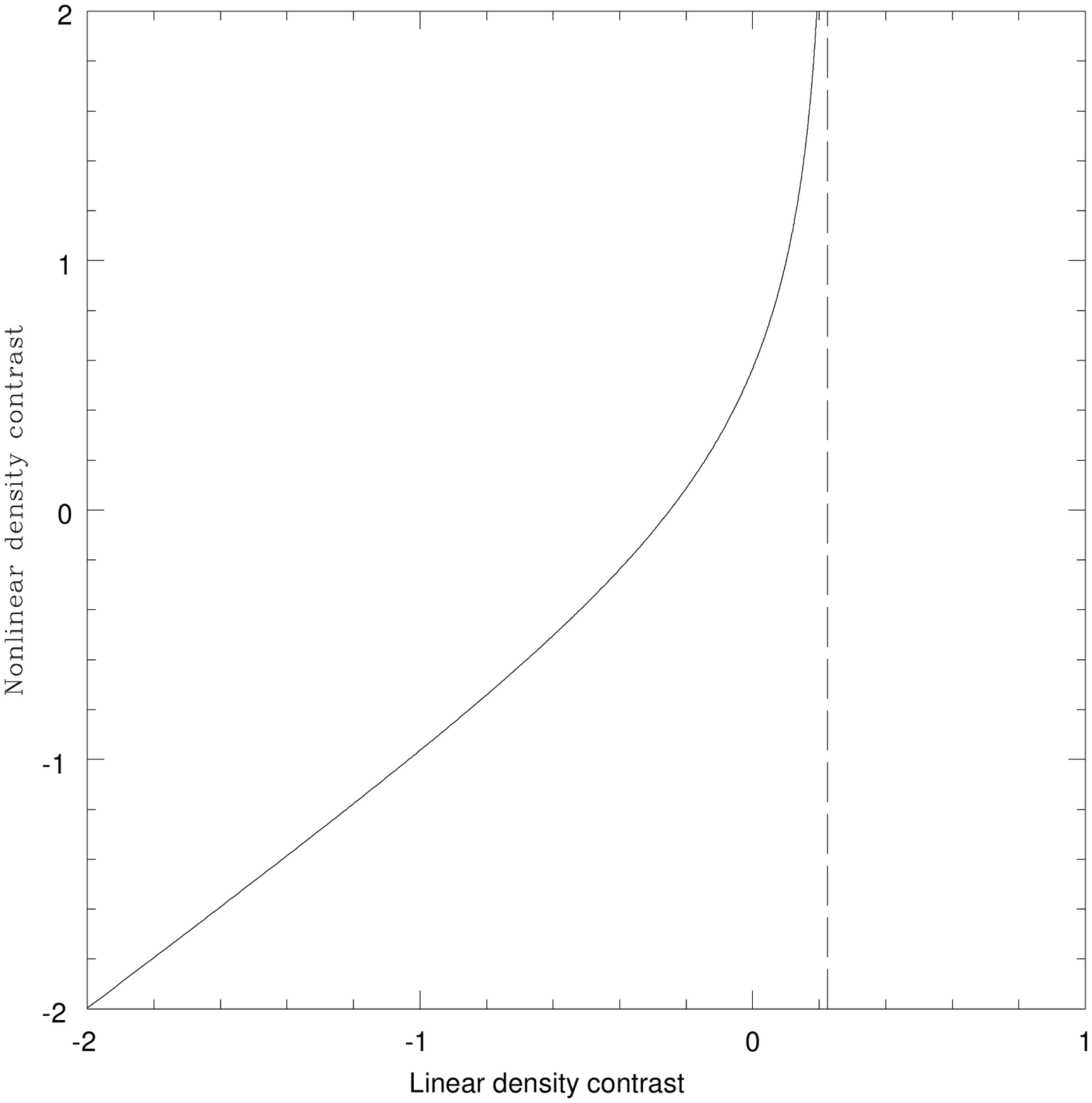}
\caption{\sf The nonlinear density contrast versus the linear density contrast for the spherical collapse model}
\label{spherical}
\end{figure}

Comparing with the linear theory results in an $\Omega =1$ universe it can be seen that the sphere will breakaway from the general expansion and reach a maximum radius at $\theta = \pi$. At this point the linear theory result for density contrast is $\delta_{lin}=1.06$ where as the nonlinear density contrast is $\delta_{sph}=5.6$. The final fate of the sphere will be an uninterrupted collapse towards a single point at $\theta = 2\pi$. The linear density contrast at this time is $\delta_{lin}=1.68$ where as the actual nonlinear density contrast essentially shoots up towards infinity (Figure \ref{spherical}). 
\par
To prevent the model from being driven towards a singularity one has to invoke a physical mechanism referred to as `virialisation'. This involves an {\it ad hoc} principle of introducing a stability criterion such that when  the potential energy is equal to twice the kinetic energy the collapse will stop and the system will reach a final density contrast of about $180$.
\par
This approximation has proved to very effective, despite its highly restrictive assumptions, in quantitatively describing measures such as mass functions and in theoretically modeling nonlinear scaling relations. However  spherical symmetry is a restrictive assumption and a  model which is more representative of the observed structures is to be sought. 
\section{N Body Simulations}
\label{sec:chapter1:nbody}
\par
An effective model of the phenomena so that the process can be better understood must have the following three components (1) a representation of the basic entities that are being modeled (2) a set of initial conditions (3) a dynamical system of equations to evolve the system. The various models differ in the ways in which these three requisites are met. Truly speaking, they also differ in the domain of the problem that they address but it can be traced down to the way in which they implement the above conditions.
\par
 (1) The basic entities must be chosen such that  the high degree of detail afforded by computer simulations is not  detrimental to  understanding these structures better. A complete model of the universe taking into account all the fundamental interactions between the particles is therefore ruled out. A more feasible and productive approach is to treat large mass aggregates as the ``particles'' of the system and describe the universe in terms of the dynamics of these massive ``particles'' which are governed by gravity. This approach forms the basis of N body simulations. An alternate approach involves averaging over the local velocities of the particles in a small volume and treating the volume itself as a fluid element and describe the system in terms of  dynamics of a fluid. Analytic approaches usually favor this picture. Observationally all the matter in the universe consists of  that which  is inferred from the various signals that we receive via the electromagnetic spectrum. Some observations such as rotation curves of galaxies as well as velocities of galaxies in clusters  indicate that there is a large, dominant component of matter in the universe, that has no interaction with light. This matter is referred to as `dark matter'. Many observational problems such as the observed flat rotation curves of galaxies and so on can be explained by a judicious invocation of this invisible ``matter'' whose constituents and interactions other than gravitational, is yet to be discovered. Since it is indicated that more than 90\% of the observed universe consists of dark matter,the models incorporate  the dynamics of the dark matter prior to that of visible matter. Thus the computer simulation model has three core components.  The Friedmann metric, the dark matter and  the visible matter. The dark matter forms the major component as far as gravitation is concerned.
\par
 The various forms of dark matter that have been used are (1) CDM [Cold dark matter]  (2) HDM [Hot dark matter] and a mixture of the two. The hot dark matter and the cold dark matter differ in the way power is distributed in various modes. The most notable physical difference is due to `free streaming' which causes small scale power to be suppressed in the case of hot dark matter. This qualitatively and quantitatively changes the way hierarchies of structures form in the two models. The next set of entities which are baryons (or the components of visible matter) are dealt with separately since their gravitational contribution is much smaller than that of dark matter but they have interactions which are non gravitational which govern their dynamics. Thus the basic entities and their interactions are as follows. The dark matter gravitationally clusters in an universe with an expanding background. The baryonic particles collect in the deep dark matter potentials with their structure and state dominated by additional non gravitational processes such as heating, cooling and so on.
\par
(2) The initial conditions: The universe that is observed is considered to be a single realization of an underlying gaussian random field which is predicted by various theories of early universe and has been observationally supported by the COBE data. The gaussian random field is wholly described by  the power spectrum defined as follows. 
\begin{equation}
P({\bf k},{\bf k'})=|\delta_{\bf k}|^{2}
\end{equation}
where $\delta_{\bf k})$ is $k$th Fourier mode of random field $\delta$.
\par
The quantities that are used to describe the structure of the universe are also statistical in nature. This implies that exact structure of the universe as observed today cannot be modeled by these approaches, but rather allows us to describe a statistical model of the universe.  
\par
(3) The dynamics; The dynamical equations that describe the growth of the system when it is treated as a set of interacting particles consist of the   equation of motion for the trajectory of  a single particle in an expanding background and an equation which allows us to evaluate the force on a particle due to all other particles. 
N Body simulations essentially define a number of particles and generate a phase space configuration for them based on the chosen initial conditions which are usually specified by the initial cosmological model, {\it ie} the parameters that describe the background universe such as density of matter, density of vacuum energy, Hubble's constant etc and the power spectrum of the initial perturbation field. The initial conditions also specify the epoch at which these conditions are given. This epoch is usually chosen to be the epoch of recombination at a redshift of approximately 1000.
\par
The initial conditions can be evolved  analytically to a point where the nonlinear effects start to increase in importance and from there the system is allowed to evolve according to the full system of equations until the desired point in time. This approach is detailed in the following review \cite{nbodyrev}. 
\par
N Body approach is complete but not feasible to a high degree of accuracy owing to the computational limitations that exist. The limitations are usually caused due to resource (memory and disk storage)  and computational (CPU power, time) requirements being inadequately met by existing technology. The large number of theoretical models and the continually increasing accuracy of observational data require that the system be modeled more and more accurately. In addition the outputs of simulations which consist  of the phase space of all the particles in the system are too detailed and a large amount of further computation has to be performed before statistical measures are derived from them which may be used to effect the comparisons.
\par
These limitations led to attempts to model the system under study by techniques which either bypass the simulation strategy (analytic and semi analytic techniques) or use some physically motivated ansatz to arrive at a final configuration from an initial  one, essentially by bypassing some of the steps in the simulation. The second category of approaches include the various numerical approximation schemes such as Zeldovich approximation, frozen potential approximation and so on. The other approach leads to  perturbative techniques for exploring the quasi and nonlinear regimes as well as some nonperturbative techniques that attempt to look at the problem of evolution in time as a one shot time dependent mapping which gives some of the statistical measures computable from the final configuration in terms of the measures computed over the initial configuration at one go.
Thus the three approaches, the complete N body approach, the approximation schemes and the mapping approach, all attempt to explain the formation of structure in a complimentary sense.
\par
The algorithms for N body simulations have been discussed in the references. The different categories of simulations are essentially based on the force computation method. This thesis will be primarily using a PM (particle mesh) code developed by Bagla and Padmanabhan \cite{nbodyrev}. The N body experiments and some of their results are discussed in this section. There are two kinds of PM N body simulations discussed here: the two dimensional simulations and the three dimensional  simulations. The 2D simulations allow us to explore a much higher dynamic range  in the given computational resources and consequently many experiments are conducted in two dimensions. The 3D simulations correspond to full scale models for the universe at hand. The essential nature of gravitational clustering in terms of how the ``particles'' of the system respond is easily  visualized in two dimensions. The following results reveal that initially the particles are distributed in a random manner according to chosen power spectrum. Then the clustering proceeds forming filamentary structures whose intersections serve as the points to which matter flows along the filaments to form clusters. In three dimensions this phenomenon is essentially repeated with sheets, filaments and clusters. 
\par
\begin{figure}[ht]

\leavevmode\centering
\psfig{file=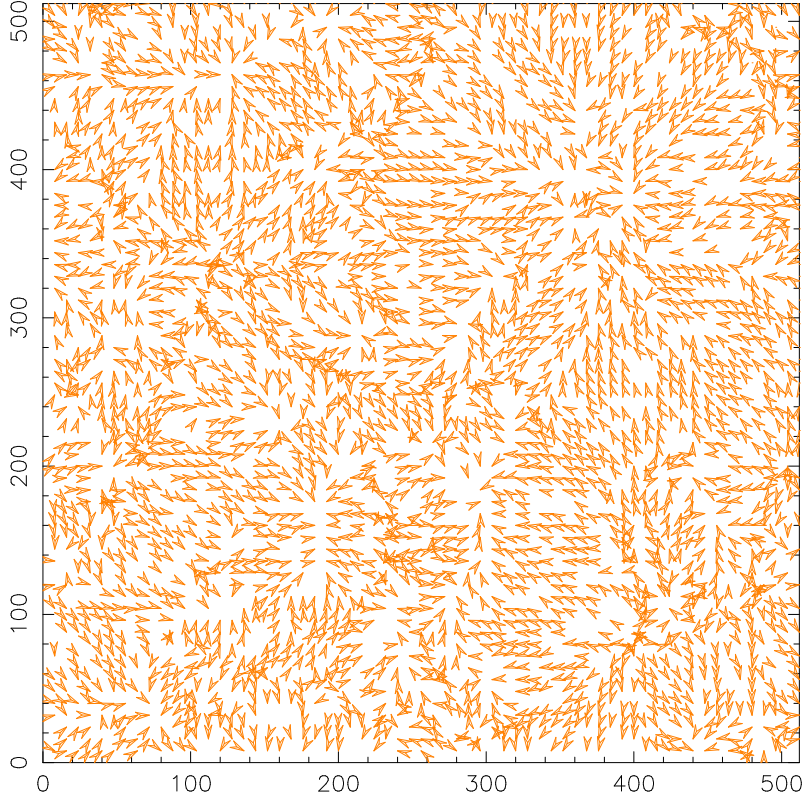,height=300pt,width=300pt,angle=-90}
\caption{\sf The flow field at a=0.5}
\label{d1}
\end{figure}
\begin{figure}[ht]

\leavevmode\centering
\psfig{file=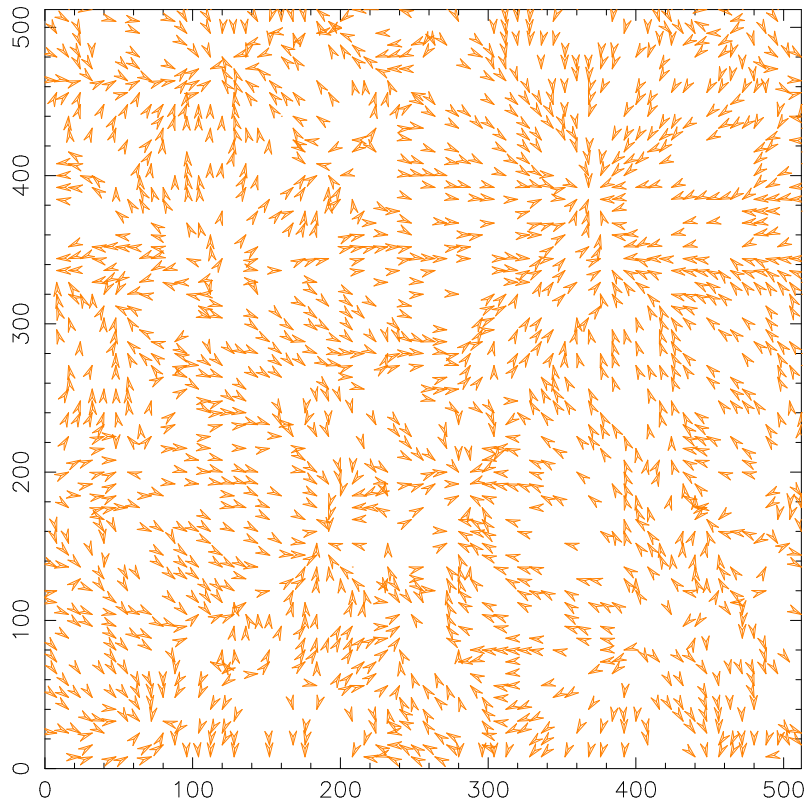,height=300pt,width=300pt,angle=-90}
\caption{\sf The flow field at a=1.5. It can be seen that the velocity vectors flow along the caustics (filaments) towards the cluster centers}
\label{d2}
\end{figure}
\begin{figure}[ht]

\leavevmode\centering
\psfig{file=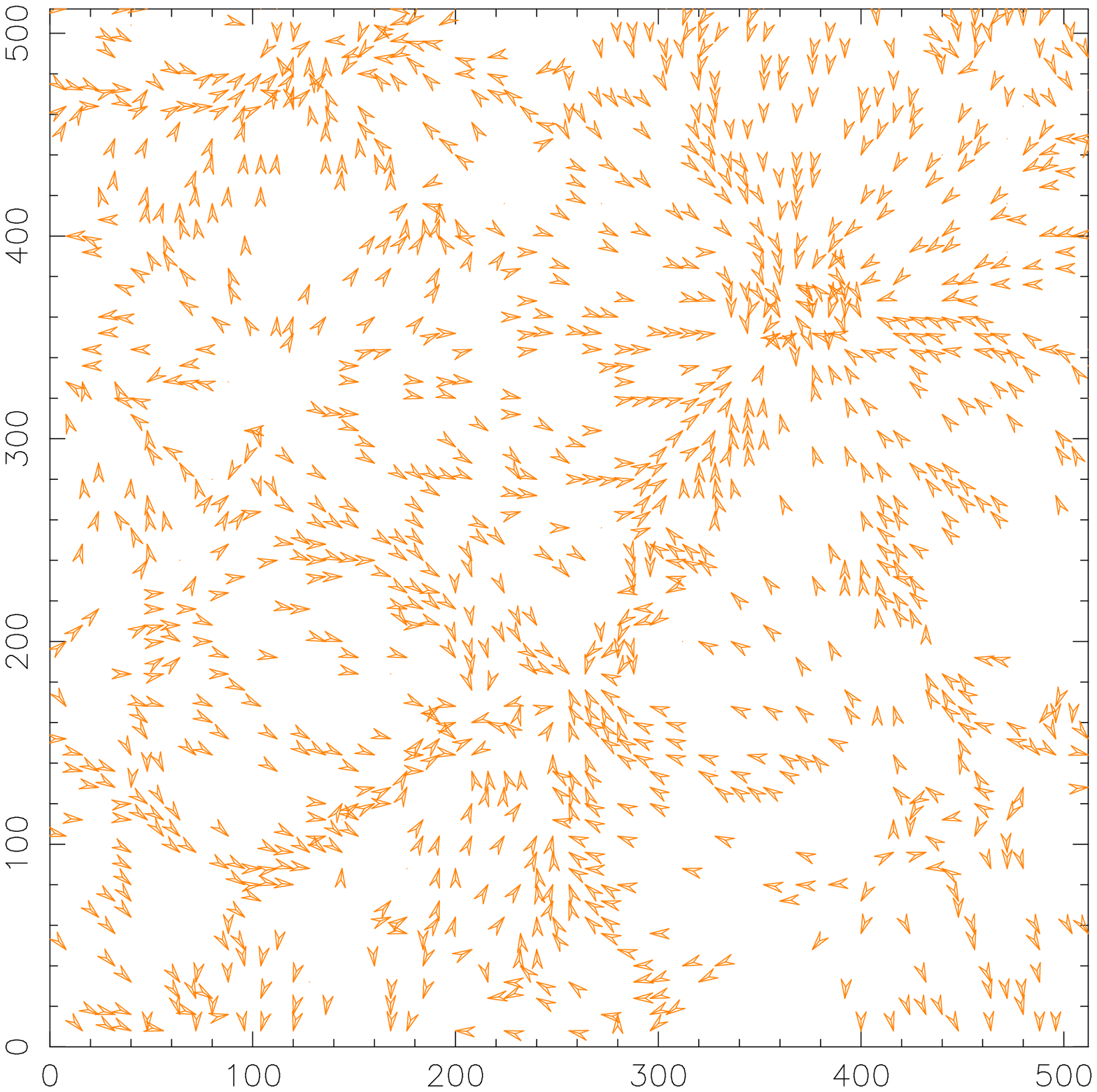,height=300pt,width=300pt,angle=-90}
\caption{\sf The flow field at very late times. The flow is almost completely concentrated at the centers. The particles participating in these flows transfer power from large scales to small scales}
\label{d3}
\end{figure}

\par
Figures(1.8 -- 1.10) which  show the results visually for a set of 2D simulations clearly reveal the initial peaks and the formation of long filamentary structures and the later collapse into the peaks themselves. To the first approximation  one can clearly understand the fact that the initial peaks in the system serve as the nucleation centres for structure formation and these nucleation centres themselves do move due to the power existing at length scales corresponding to their mean separation. The question of first structures that form, whether they are filamentary or sheet like in 3D and it's analogue in 2D has partial answer in terms of the Zeldovich approximation. This approximation scheme which essentially places the particles in inertial motion with their initial velocities given by the initial conditions indicates that the initial structures are sheet like (string like in 2D) which is roughly concordant with what the simulations indicate. Observationally too we see the presence of very large structures (100 Mpc) which may be sheet like or filamentary. A simple analysis based on the Zeldovich approximation which is a valid approximation until shell crossing takes place gives the statistical answer, that for a gaussian random field it is more probable that sheets would form rather than filaments in 3D and correspondingly in 2D.  The pictures below clearly reveal the initial formation of filamentary structures (the arrows indicate the direction of velocity) and the damping of the velocity  transverse to the filament. This causes the matter to flow along the filaments into the junctions where filaments meet which are essentially where the roughly spherically   symmetric clusters form. This picture is expected to hold good in three dimensions also.
\par
\section{Statistical indicators}
\label{sec:chapter1:stats}
In developing the framework discussed in the earlier sections, we have made some simplifying assumptions regarding homogeneity, Newtonian limit of the metric and so on. Since these approximations are valid only at different length scales (homogeneity on large length scales and Newtonian approximation on small length scales) it is clear that our models for the universe as a whole is built from smaller models at different length scales. To extend the results derived from the smaller models to the universe as a whole as well as correlate them with observations, we need to use statistical approaches. Thus the models allow us to predict various statistical quantities which may be compared to observations. 
\par 
The statistical quantities that are used to analyse and compare the N body data with other simulations (as well as with observations) are computed from the positions and velocities. 
The statistical quantities are usually measures of deviation from the assumed smoothness and distributions of such quantities such  two point correlation function, power spectrum and higher moments of the distribution functions. Another approach followed when  models are being compared to observational data, is to constrain the parameters of the model by using statistical goodness of fit and other tests of significance.In this section we shall briefly discuss some of the quantities relevant to later discussions.
\par
The underlying gaussian field is completely characterised by the power spectrum $P(k)$ defined by 
$$
 P({\bf k}) = |\delta_{\bf k}|^2 = P(k)
$$
One can further define other related quantities such as two point correlation function in terms of the power spectrum 
\begin{equation}
\xi({\bf x})=\int \frac{d^3 {\bf k}}{(2 \pi)^3} P({\bf k}) e^{i{\bf k}\cdot {\bf x}} = \int_0^\infty \frac{dk}{k} \left( \frac{k^3 P(k)}{2 \pi^2}\right) \left(\frac{\sin{kx}}{kx}\right)
\end{equation}
\begin{equation}
\bar\xi(r)= \frac{3}{r^3}\int_0^r dx\;x^2\;\xi(x)
\end{equation}
The assumed statistical isotropy of the universe implies that the two point correlation function and power spectrum are functions of magnitude of $r$ and $k$ respectively and is independent of direction. 
Some of the other measures of departures from homogeniety are the variance of the density field after an appropriate smoothing using a window function such as 
\begin{eqnarray}
W_{sph}(k,R)&=&\frac{3}{k^3 R^3}\left[ \sin{kR}-kR \cos{kR} \right] \\
W_{gauss}(k,R)&=&\exp\left({-{1 \over 2} k^2 R^2}\right)
\end{eqnarray}
giving
\begin{equation}
\sigma_{sph}^2(R)=\int \frac{d^3 k}{(2 \pi)^3} P(k) W_{sph}(k,R)^2
\end{equation}
\begin{equation}
\sigma^2_{gauss}(R) = \int_0^\infty \frac{dk}{k} \frac{k^3 P(k)}{2 \pi^2} W_{gauss}(k,R)^2
\end{equation}
Figure 1.11 shows the quantities discussed above computed for a standard COBE normalized CDM power spectrum  given by
\begin{equation}
P(k)=\frac{A k^n}{(1+B k+C k^{1.5}+D k^{2})^2}
\end{equation}
where the parameters $A,B,C,D,n$ are equal to $(24/h)^4 Mpc^4$, $1.77/(\Omega h^2)  Mpc$, 
$9 (\Omega h^2)^{-1.5} Mpc^{3/2}$, $(\Omega h^2)^{-2} Mpc^2$  and $1$ respectively ($h=0.65$)\cite{structurebook}.
\par
A more complex quantity is the Press Schecter mass function (Fig.1.12) which gives the number of collapsed objects of mass M as a function of redshift
\begin{equation}
N(M) dM=-\frac{\bar \rho}{M} \left(\frac{2}{\pi}\right)^{1/2} \frac{\delta_c}{\sigma^2} \left(\frac{\partial \sigma}{\partial M}\right) \exp\left(-\frac{\delta_c^2}{2\sigma^2}\right) dM
\end{equation}

\begin{figure}[ht]

\leavevmode\centering
\psfig{file=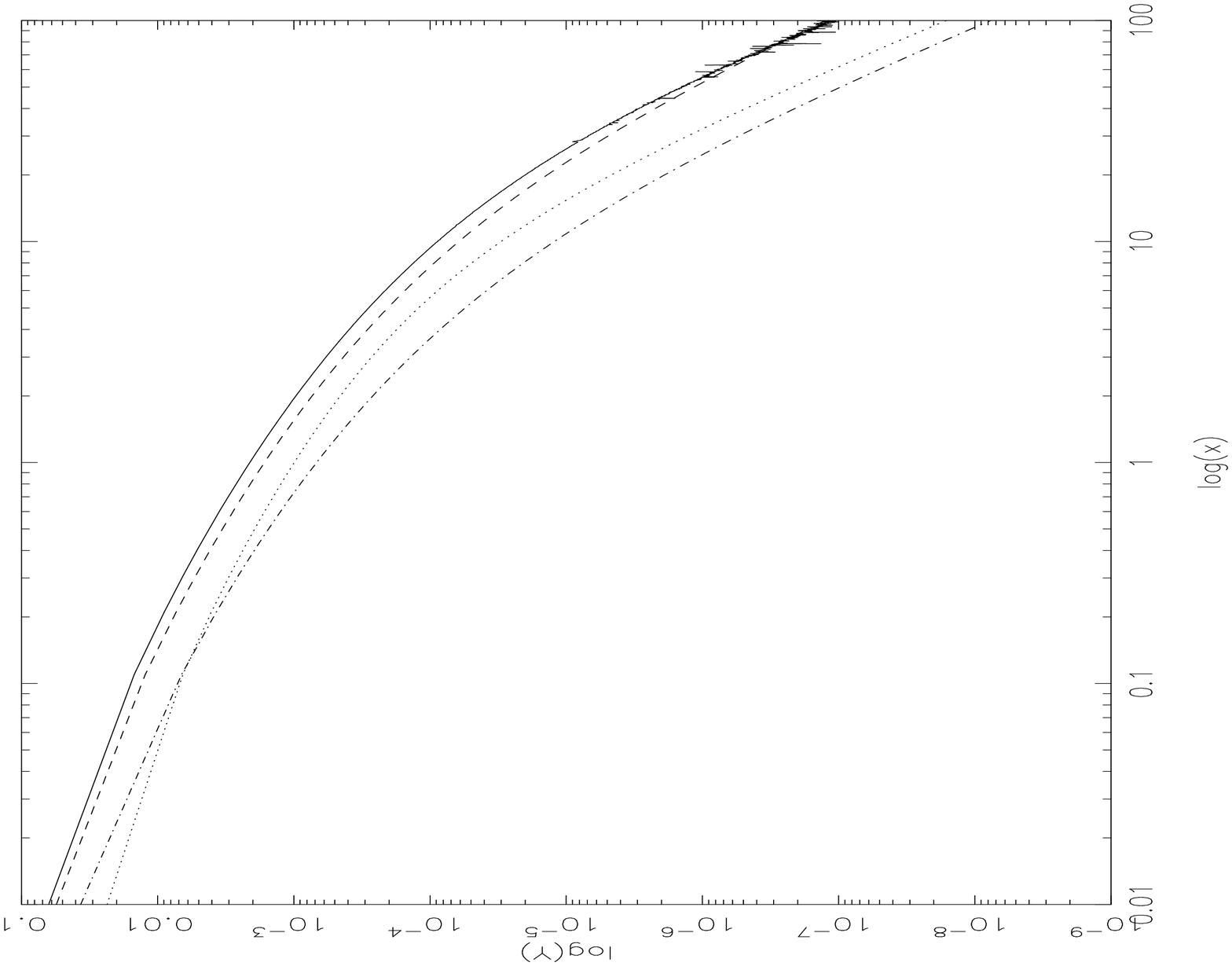,height=300pt,width=300pt,angle=-90}
\caption{\sf Different statistical indicators such as the two point correlation function and $\Delta^2(k)=k^3 P(k)/(2 \pi^2)$.($\Delta(k)$ -- dashed line, $\bar\xi$ -- solid line, $\sigma_{sph}$ -- dotted line, $\sigma_{gauss}$ -- dash dot line for a standard CDM power spectrum with $k=1/l$}
\label{stats}
\end{figure}

\begin{figure}[ht]
\leavevmode\centering

\epsfxsize=300pt
\epsfysize=300pt
\epsfbox{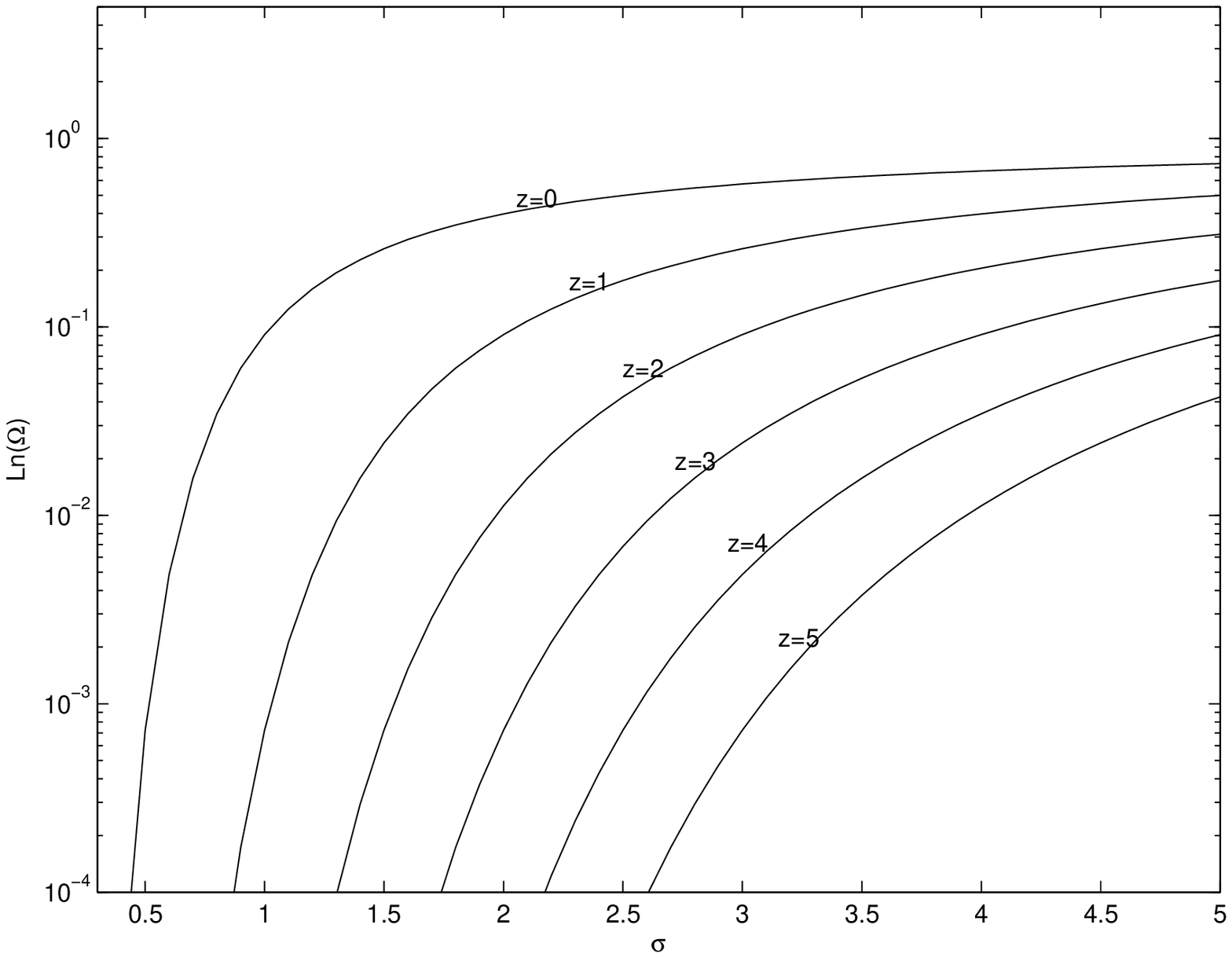}
\caption{\sf The mass function plotted as the log of fraction of mass in structures more massive than $M$ ($\ln(\Omega)$) as a function of $\sigma(M)$ for different redshifts z=0,1,2,3,4,5 from top to bottom.}
\end{figure}
\label{presss}
\par
As has been indicated, all these different measures which are low order moments of the underlying distributions are not independent of each other. 
Higher order moments contain more detailed information which is required to discriminate between  models for instance when the lower order correlations are unable to do so. There exist statistical measures such as minimum spanning tree, percolation statistics and so on which attempt to deal with the whole hierarchy of moments in a holistic manner. They can be used to characterise complementary aspects in the statistical description of large scale structure.
Various other quantities are also used to quantify  morphologies and topological properties of large scale structure such as minkowski functionals which are related to the moments of the distributions.
\section{Nonlinear scaling relations}
\label{sec:chapter1:NSR}
The N body approaches form an expensive way of arriving at the statistical properties of the late universe given the statistical quantities describing the early universe. An alternate approach which is a one time mapping approach to nonlinear evolution is referred to as `Nonlinear scaling relations'. It was demonstrated empirically from simulations by Hamilton {\it et al} \cite{Ham} that the averaged two point correlation function in the nonlinear regime may be obtained from the initial average two point correlation function via  a non local map  given by
\begin{equation}
\bar\xi_{NL}(x)=f_{NL}(\bar\xi_{L}(l))
\end{equation}
where $x$ and $l$ are related by $l^3 = (1+\bar\xi_{NL}(x))x^3$. The functional form for $f_{NL}$ proposed by Hamilton {\it et al} \cite{Ham} is as follows
\begin{equation}
f_{NL}(z)=\frac{z+0.358\;z^3+0.0236\;z^6}{1+0.0134\;z^3+0.0020\;z^{9/2}}
\end{equation}
It is possible to give a theoretical explanation for the existence of such a mapping \cite{RajPad94} through an argument based on the pair conservation equation which leads to the result that the universal mapping function for the two point correlation function may be obtained from  the behaviour of $h = -v/(\dot{a} r)$, ratio of average pair velocity to Hubble velocity $\dot{a} r$ under an assumed closure condition. This allows us to express the mapping function as an integral over the $h$ function. 
\par
A theoretical modeling of the mapping based on infall on to high peaks in the intermediate regimes \cite{TPMNRAS} allows the empirical relation to be well approximated by a discontinous  powerlaw approximation with three different values for the power law indices in the three regimes, linear, quasilinear and nonlinear. The power law index in the nonlinear regime is related to the value of $h$  in the highly nonlinear end which might not be unity as is implied by the stable clustering hypothesis. This approach permits the notion of a more generalised stable clustering to be introduced with a corresponding slope for the nonlinear end of the mapping.
The power law representation  in three dimensions is 
\begin{equation}
f_{NL}(z)=\left\{
\begin{array}{ll}
1 & (z \ll 1) \\
z^{3} & (1 \gaprox z \gaprox 200) \\
z^{3/2} & (200 \ll z ) \\
\end{array}
\right. 
\end{equation}
\par
Figure 1.13 shows the mapping function in three dimensions and power law approximations to the same. Figure 1.14 shows the evolution of average two point correlation function for a CDM power spectrum at a late epoch and the linearly evolved two point correlation function at the same time. The nonlinear evolution of the two point correlation function may be seen at small scales.

\begin{figure}[h]

\leavevmode\centering
\psfig{file=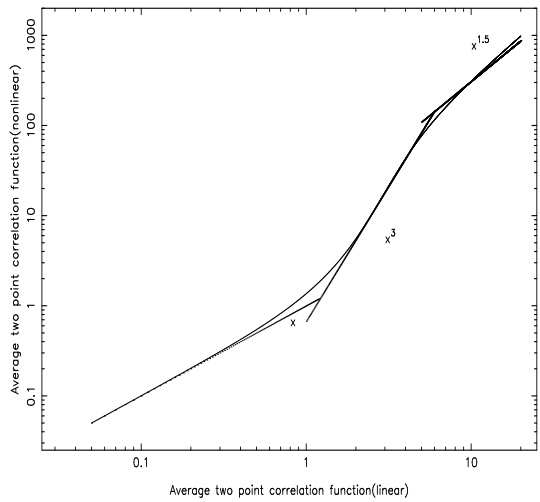,height=200pt,width=200pt,angle=-90}
\caption{\sf The nonlinear scaling relation between nonlinear and linearly scaled two point correlation function.}
\label{ham123}
\end{figure}
 
\begin{figure}[h]

\leavevmode\centering
\psfig{file=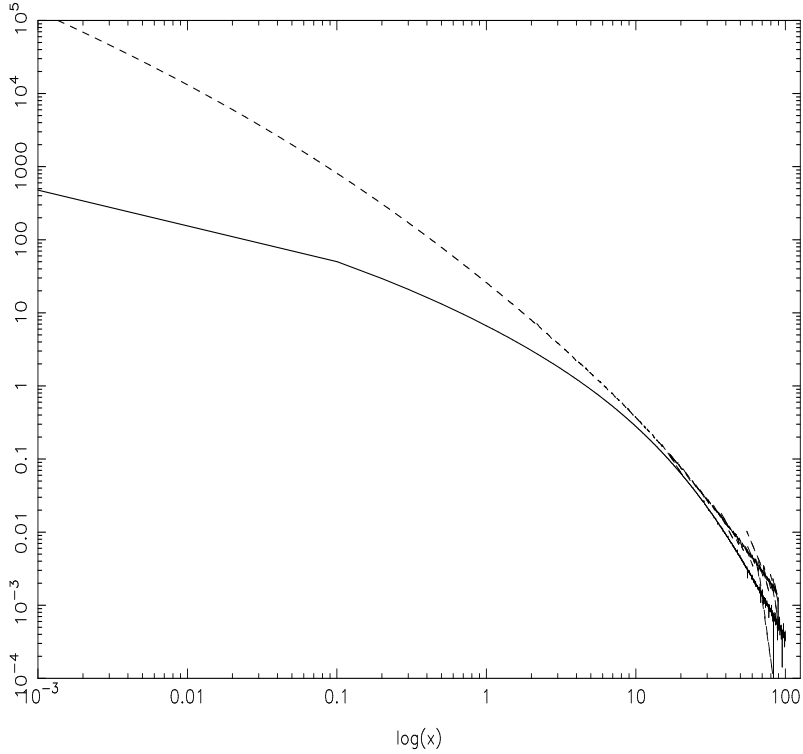,height=300pt,width=300pt,angle=-90}
\caption{\sf The linearly scaled two point correlation function (solid line) and the nonlinear two point correlation function of  CDM power spectrum at a late epoch (a=1)}
\label{CDMNSR}
\end{figure}
\par
In brief the approach taken towards studying structure formation in the universe is a) a model for the underlying smooth background universe b) a model for the seed perturbations c) a system of equations for describing the growth of these perturbations, their linear approximations and  analytical solutions d) statistical quantities to describe the various uncertainties and selection effects e) nonlinear approximations of various kinds to obtain physical insight and f) full scale N body simulations in an expanding background. 
\par
The following chapters  detail the approaches we have taken and progress made, using the theoretical model sketched herein. 

\chapter{ Nonlinear gravitational clustering: Dreams of a paradigm}
\label{chap:nonlinear}
{\scriptsize Humor is the only test of gravity and gravity of humor -- Aristotle}
\section{Introduction}

As is clear from the first chapter the evolution of  large number of particles under their mutual
gravitational influence is a well-defined mathematical problem. But, if such a
system occupies a finite region of phase space at an initial instant, and evolves
via Newtonian gravity, then it does not reach any sensible `equilibrium' 
state. The core region of the system will keep on shrinking and will be
eventually be dominated by a few `hard binaries'. Rest of the particles will
evaporate away to large distances, gaining kinetic energy from the shrinking
core [for a discussion of  such systems, see \cite{PadPhyRep} ].

The situation is drastically different in the presence of an expanding 
background universe characterised by an expansion factor $a(t)$. Firstly, the 
expansion tends to keep particles apart 
thereby exerting a civilising influence against Newtonian attraction. Secondly,
it is now possible to consider an infinite region of space filled with 
particles.
 The average density of particles will contribute to the expansion of the 
background universe and the deviations from the uniformity will lead to
clustering. Particles evaporating from a local overdense cluster cannot
escape to ``large distances'' but necessarily will encounter other deep
potential wells. Naively, one would expect the local overdense regions to 
eventually 
form gravitationally bound objects, with a hotter distribution of particles
hovering uniformly all over. As the background expands, the 
velocity dispersion of the second component will keep decreasing and they will
be captured  by the deeper potential wells. Meanwhile, the clustered component 
will also evolve dynamically and participate in, e.g mergers. If the background 
expansion and  the initial
conditions have no length scale, then it is likely that the clustering  will 
continue
in a hierarchical manner {\it ad infinitum}.
\par

Most of the practising cosmologists will broadly agree with
the above  picture of gravitational clustering in an expanding universe. It is, 
however, not
easy to translate these concepts into a well-defined mathematical formalism
and provide a more quantitative description of the gravitational clustering.
One of the key questions regarding this system which needs to be addressed is 
the following: Can one make any general statements about the very late stage
evolution of the clustering ? For example, does the power spectrum at late times
`remember' the initial power spectrum or does it possess some universal
characteristics which are reasonably independent of initial conditions ?
[This question is closely related to the issue of whether gravitational 
clustering leads to density profiles which are universal.   
\cite{Navarro}].

\par
In this chapter we address some of the above issues and show that it is possible to 
provide (at least partial) answers to  these questions based on a simple 
paradigm.
The key assumption we shall make is the following: Let ratio 
 between
mean relative pair velocity $v(a,x)$ and the negative hubble velocity ($-\dot a 
x$) 
be denoted by $h(a,x)$ and let $\bar\xi(a,x)$ be the mean correlation
function averaged over a sphere of radius $x$. {\it We shall assume that
$h(a,x)$ depends on $a$ and $x$ only through   
$\bar\xi(a,x)$; that is, $h(a,x)=h[\bar\xi(a,x)]$.} With such  a minimal 
assumption, we will be able to obtain several conclusions regarding the
evolution of power spectrum in the  universe. Such an assumption was originally
introduced --- in a different form --- by Hamilton \cite{Ham}.
The present form, as well as its theoretical implications were discussed
in  \cite{RajPad94}, and a theoretical model for the
scaling was attempted by Padmanabhan
\cite{TPMNRAS}. 
It must be noted that 
simulations indicate a dependence of the relation $h(a,x)=h[\bar\xi(a,x)]$
on the intial spectrum and also on cosmological parameters 
(\cite{PeaDodd},\cite{PeaDodd96},\cite{PadOst}, \cite{MoJain}). Most of our discussion 
is independent of this fact or can be easily generalised to such cases.
When we need to use an explicit form for $h$ we shall use the original
ones suggested by Hamilton \cite{Ham} because of its simplicity. 

Since this chapter addresses several independent but related  questions,
we provide here a brief summary of how it is organised. Section 2.1 studies
some aspects of nonlinear evolution based on the assumption mentioned above.
We begin by summarising some previously known results in subsection 2.2.1 to
set up the notation and collect together in one place the formulas we need later.
Subsection 2.2.2 makes a brief comment about the critical indices in gravitational
dynamics
so as to motivate later discussion. In section 2.3, we discuss the relation between density profiles
of halos and correlation functions and derive the conditions under which one
may expect universal density profiles in gravitational clustering. In section
2.4 we show that gravitational clustering does {\it not} admit self similar
evolution except in a very special case. We also discuss the conditions for
approximate self-similarity to hold. Section 2.5 discusses the question as whether
one can expect to find power spectra which evolve preserving their shape, even in the nonlinear
regime. We first show, based on the results of section 2.4, that such {\it exact}
solutions cannot exist. We then discuss the conditions for the existence of
some {\it approximate} solutions. We obtain one prototype approximate solution
and discuss its properties. The solution also allows us to understand the
connection between statistical mechanics of gravitating systems in the small
scale and evolution of correlation functions on the large scale. Finally,
section 2.6 discusses the results. 
\vskip 0 pt plus -20pt
\section{General features of nonlinear evolution}
Consider the evolution of the system 
starting from a gaussian initial fluctuations with an initial power
spectrum, $P_{in}(k)$. The fourier transform of the power spectrum 
defines the correlation function $\xi(a,x)$ where $a\propto t^{2/3}$
is the expansion 
factor in a universe with $\Omega=1$. It is more convenient to work with the
average correlation function inside a sphere of radius $x$, 
defined by
\begin{equation}
\label{avxi}
\bar{\xi}(a,x)\equiv {3\over x^3}\int^{x}_{0}\xi(a,y)\: y^2 dy  
\end{equation}
This quantity is related to the power spectrum $P(a,k)$ by
\begin{equation}
\bar\xi(x,a)=\frac{3}{2 \pi^2 x^3} \int_{0}^{\infty} \frac{dk}{k} P(a,k)\left[ 
\sin(k\:x)-k\:x\:\cos(k\:x)\right]
\end{equation}
with the inverse relation
\begin{equation}
P(a,k)=\frac{4 \pi}{3 k}\int_{0}^{\infty} dx\: x\: \bar\xi(a,x)\left[ 
\sin(k\:x)-k\:x\:\cos(k\:x)\right]
\end{equation}
In the linear regime we have $\bar{\xi}_L(a,x)\propto
a^2\bar\xi_{in}(a_i,x)$.
\par
We now recall that the conservation of pairs of particles gives an exact
equation satisfied by the correlation function \cite{Peeb80}:
\begin{equation}
\label{pairconserv}
{\partial\xi \over\partial t}+{1\over ax^2}{\partial\over\partial
x}[x^2(1+\xi)v]=0
\end{equation}
where $v(a,x)$ denotes the mean relative velocity of pairs at
separation $x$ and 
epoch $a$. Using the mean correlation function $\bar\xi$ and a
dimensionless pair velocity $h(a,x) \equiv - (v/\dot{a}x)$, equation
(\ref{pairconserv}) can be written as
\begin{equation}
\label{redefpair}
({\partial\over\partial \ln a}-h{\partial\over\partial \ln x})\,\,\,
(1+\bar{\xi})=3h(1+\bar{\xi})
\end{equation}
This equation can be simplified by introducing the variables
\begin{equation}
A=\ln a,\qquad X=\ln x ,\qquad D(X,A) = \ln (1+\bar{\xi})
\end{equation}
in terms of which we have \cite{RajPad94}

\begin{equation}
\label{qkey}
\frac{\partial D}{\partial A}-h(A,X)\frac{\partial D}{\partial
X}= 3h(A,X)
\end{equation}
At this stage we shall introduce our key assumption, viz. that
$h$ depends on $(A,X)$ only through $\bar\xi$ (or, equivalently, $D$).
Given this single assumption, several results follow which we shall
now summarise.
\subsection{{\it Formal solution}}
Given that $h=h[\bar\xi(a,x)]$, one can easily integrate the equation
 (\ref{redefpair})
to find the general solution \cite{RajPad94}. The characteristics 
of this equation (\ref{redefpair}) satisfy the condition 
\begin{equation}
\label{qxandl}
x^3(1+\bar{\xi})=l^3
\end{equation}
where $l$ is another length scale. When the evolution is linear at all
the relevant scales, $\bar{\xi}\ll 1$ and $l\approx x$. As clustering
develops,
$\bar{\xi}$ increases and $x$ becomes considerably smaller than $l$.
The behaviour of clustering at some scale $x$ is then
determined by
the original {\it linear} power spectrum at the scale $l$ through the
``flow of information'' along the characteristics.
This suggests that {\it we can  express the true
correlation function $\bar\xi(a,x)$ in terms of the linear correlation
function $\bar\xi_L(a,l)$ evaluated at a different point}.
This is indeed true and the general solution can be expressed as
a nonlinear scaling relation (NSR, for short) between $\bar\xi_L(a,l)$ and $\bar\xi(a,x)$ with $l$ and $x$ related 
by equation(\ref{qxandl}). To express this solution we define
two functions ${\cal V}(z)$  and ${\cal U}(z)$ where ${\cal V}(z)$  is related 
to the function $h(z)$ by

\begin{equation}
{\cal V}(z)=\exp\left(\:\frac{2}{3} \int^{z} \frac{d z}{h(z)\,(1+z)} \right)
\end{equation} 
and
${\cal U}(z)$
is the inverse function of ${\cal V}(z)$. Then the solution to the
equation (\ref{redefpair}) can be written in either of two equivalent forms as:
\begin{equation}
\label{mapfun}
\bar\xi(a,x)={\cal U}\left[\bar\xi_L(a,l)\right];\qquad \bar\xi_L(a,l)={\cal 
V}\left[\bar \xi(a,x)\right]
\end{equation}
where $l^3=x^3 (1+\bar\xi)$
\cite{RajPad94}. Given the form of $h(\bar\xi)$
this allows one to relate the nonlinear correlation function to the linear
one.

From general theoretical considerations \cite{TPMNRAS} it can be shown 
that ${\cal V}(z)$ has the form:

\begin{equation}
\label{Vapprox}
{\cal V}(z)=\left\{
\begin{array}{ll}
1 & (z \ll 1) \\
z^{1/3} & (1 \gaprox z \gaprox 200) \\
z^{2/3} & (200 \ll z ) \\
\end{array}
\right. 
\end{equation}
In these three regions $h(z)\approx[(2z/3),2,1]$ respectively. We shall
call these regimes, linear, intermediate and nonlinear respectively. More exact fitting 
functions to ${\cal V}(z)$ and ${\cal U}(z)$ have been suggested in literature. 
(\cite{Ham},\cite{MoJain},\cite{PeaDodd}). When needed in this chapter, we 
shall use the one  given in Hamilton \cite{Ham}:
\begin{equation}
\label{ham1}
{\cal V}(z)=z 
\left(\frac{1+0.0158\:z^2+0.000115\:z^3}{1+0.926\:z^2-0.0743\:z^3+0.0156\:z^4}
\right)^{1/3}
\end{equation}
\begin{equation}
\label{ham2}
{\cal U}(z)=\frac{z+0.358\:z^3+0.0236\:z^6}{1+0.0134\:z^3+0.0020\:z^{9/2}}
\end{equation}
Equations (\ref{mapfun}) and (\ref{ham1},\ref{ham2}) implicitly determine 
$\bar\xi(a,x)$ in terms of $\bar\xi_L(a,x)$.

\subsection{{\it Critical indices}}

These NSR already allow one to obtain some general conclusions regarding
the evolution. To do this most effectively, let us define a local 
index for rate of clustering by
\begin{equation}
n_a(a,x)\equiv \part{\ln \xb}{\ln a}
\end{equation}
which measures how fast $\xb$ is growing. When $\xb\ll 1$, then $n_a=2$
irrespective of the spatial variation of $\xb$ and the evolution preserves the shape of $\xb$. However, as clustering develops, the growth rate will
depend on the spatial variation of $\xb$. Defining the effective spatial
slope by
\begin{equation}
-[n_{eff}(a,x)+3]\equiv \part{\ln \xb}{\ln x}
\end{equation}
one can rewrite the equation (\ref{redefpair}) as
\begin{equation}
\label{naeqn}
n_a=h(\frac{3}{\xb} -n_{eff})
\end{equation}
At any given scale of nonlinearity, decided by $\xb$, there exists a critical
spatial slope $n_c$ such that $n_a>2 $ for $n_{eff}<n_c$ [implying rate of growth is faster
than predicted by linear theory] and 
$n_a<2 $ for $n_{eff}>n_c$ [with the rate of growth being slower
than predicted by linear theory]. The critical index is fixed by setting $n_a=2$ in  equation (\ref{naeqn}) at any instant. This feature will tend  to ``straighten out" correlation functions  towards the critical slope.
[We are assuming that $\xb$ has a slope that is decreasing with
scale, which is true for any physically interesting case]. From the fitting function it is easy to see that in the range $1 {\mbox{\gaprox}} 
\bar\xi {\mbox{\gaprox}} 200$, the critical index is $n_c\approx -1$
and for $200 \gaprox \bar\xi$, the critical index is $n_c\approx -2$ \cite{JsbTp}.
This clearly suggests that the local effect of evolution is to
drive the correlation function to have a shape with $(1/x)$ behaviour
at nonlinear regime and $(1/x^2)$ in the intermediate regime. Such a 
correlation function will have $n_a\approx 2$ and hence will grow at
a rate close to $a^2$.
 
\section{ Correlation functions, density profiles and stable clustering}

Now that we have a NSR giving  $\xb$ in terms of $\bar\xi_L(a,l)$  
we can ask the question:
How does $\xb$ behave at highly nonlinear scales or, equivalently, at any
given scale at large $a$ ? 
\par
To begin with, it is easy to see that we must have $v=-\dot a x$ or  $h=1$ for 
sufficiently large $\bar\xi(a,x)$ {\it if we assume} that the
evolution gets frozen in proper coordinates at highly nonlinear scales. 
Integrating equation (\ref{redefpair}) with $h=1$, we get $\bar\xi(a,x)=a^3 F(ax)$;
we shall call this phenomenon ``stable clustering''. There are two points
which need to be emphasised about stable clustering:
\par
(1) At present, there exists some evidence 
from simulations \newline
\cite{PadOst} that 
stable clustering does {\it not} occur in a $\Omega=1$ model. In a {\it formal} sense, numerical simulations cannot disprove [or
even prove, strictly speaking] the occurrence of stable clustering, because of the finite dynamic
range of any simulation. 

(2). Theoretically speaking, the ``naturalness'' of stable clustering is
often overstated. The usual argument is based on the assumption that
at very small scales --- corresponding to high nonlinearities --- the structures
are ``expected to be" frozen at the proper coordinates. However, this argument does not
take into account the fact that mergers are not negligible at {\it any scale} in
an $\Omega=1$ universe. In fact,  stable clustering
is more likely to be valid in models with $\Omega<1$ --- a claim which seems to 
be again supported by simulations \cite{PadOst}.

{\it If} stable clustering {\it is} valid, then the late time  behaviour of $\xb$ 
{\it cannot}
be independent of initial conditions. In other words the two requirements:
(i) validity of stable clustering at highly nonlinear scales and
(ii) the independence of late time behaviour from initial conditions, 
are mutually
exclusive. This is most easily seen for initial power spectra which
are scale-free. If $P_{in}(k)\propto k^n$ so that $\bar\xi_L(a,x)\propto a^2 
x^{-(n+3)}$, then it is
easy to show that $\xb$ at small scales will vary as
\begin{equation}
\bar\xi(a,x) \propto a^{\frac{6}{n+5}} x^{-\frac{3(n+3)}{n+5}};\qquad (\bar\xi 
\gg 200)
\end{equation}
if stable clustering is true. Clearly, the power law index in the nonlinear 
regime ``remembers''
the initial index. The same result holds for more general initial conditions.

What does this result imply for the profiles of individual halos?
To answer this question, let us start with the simple assumption that the density field $\rho(a,{\bf x})$ at late stages  can 
be expressed as a superposition
of several halos, each with some density profile; that is, we take
\begin{equation}
\label{haloes}
\rho(a,{\bf x})=\sum_{i} f({\bf x}-{\bf  x}_i,a)
\end{equation}
where the $i$-th halo is centered at ${\bf x}_i$ and contributes
an amount $f({\bf x}-{\bf  x}_i,a)$  at the location ${\bf x}_i$  [We can easily generalise this equation to the situation in which there are halos with
different properties, like core radius, mass etc by summing over the number
density of objects with particular properties; we shall not bother to
do this. At the other extreme, the exact description merely corresponds to taking
the $f$'s to be Dirac delta functions]. The power spectrum for the 
density contrast, $\delta(a,{\bf x})=(\rho/\rho_b-1)$, corresponding to the 
$\rho(a,{\bf x})$ in (\ref{haloes})  can be expressed as
\begin{eqnarray}
\label{powcen}
P({\bf k},a) &\propto& \left( a^3 \left| f({\bf k},a)\right| \right)^2 \left| 
\sum_i \exp -i {\bf k}\cdot{\bf x}_i(a) \right|^2   \\
\label{powcen1}
& \propto & \left( a^3 \left| f({\bf k},a)\right| \right)^2 P_{\rm cent}({\bf 
k},a)
\end{eqnarray}
 where $P_{\rm cent}({\bf k},a)$
denotes the power spectrum of the distribution of centers of the halos.
\par

If  stable clustering is valid, then the density profiles of halos are
frozen in proper coordinates and we will have $f({\bf x} -{\bf x}_i,a)=
f(a\:({\bf x}-{\bf x}_i))$;
hence the fourier transform will have the form $f({\bf k},a)=f({\bf k}/a)$. On 
the other
hand, the power spectrum at scales which participate in stable clustering
must satisfy $P({\bf k},a)=P({\bf k}/a)$ [This is merely the requirement 
$\xb=a^3F(ax)$
re-expressed in fourier space]. From equation (\ref{powcen1}) it follows that we 
must have
$P_{\rm cent}({\bf k},a)={\rm constant} $ independent of ${\bf k}$ and $a$ at 
small length scales. This can arise in the special case of 
random distribution of centers or --- more importantly --- because  we are 
essentially probing the interior of a single halo at sufficiently small scales. 
[Note that we must {\it necessarily} have $P_{\rm cent} \approx {\rm constant}$, for length scales smaller than typical halo size, by definition]. 
We can relate the halo profile to the correlation function
using  (\ref{powcen1}).  In particular, if the halo profile is a power law with 
$f\propto r^{-\epsilon}$,  it
follows that the $\xb$ scales as $x^{-\gamma}$ ( \cite{silkmac},\cite{SethJain}) where
\begin{equation}
\label{gammep}
\gamma=2\epsilon-3
\end{equation}
\par
Now if the {\it correlation function} scales as $[-3(n+3)/(n+5)]$, then
 we see that
the halo density profiles should be related to the initial power law
index through the relation 
\begin{equation}
\epsilon=\frac{3(n+4)}{n+5}
\end{equation} 
So clearly,  
the halos of
highly virialised systems still ``remember'' the initial power 
spectrum.
\par

Alternatively, one can try to ``reason out'' the profiles of the individual
halos and use it to obtain the scaling relation for correlation functions.
One of the favourite arguments used by cosmologists to obtain such a ``reasonable'' halo profile is based on spherical, scale invariant,
collapse.  It turns out
that one can provide a series of arguments, based on spherical collapse, to
show that --- under certain circumstances --- the {\it density profiles} at the
nonlinear end scale as $[-3(n+3)/(n+5)]$. The simplest variant of this argument
runs as follows: If we start with an initial density
profile which is $r^{-\alpha}$, then scale invariant spherical collapse
will lead to a profile which goes as $r^{-\beta}$ with $\beta=3\alpha/
(1+\alpha)$ \cite{TPMNRAS}. Taking the intial slope
as $\alpha=(n+3)/2$ will immediately give $\beta=3(n+3)/(n+5)$. [Our definition of the stable clustering in the last section 
is based on the scaling of
the correlation function and gave the
slope of $[-3(n+3)/(n+5)]$ for the {\it correlation} function. The spherical
collapse gives the same slope for {\it halo profiles}.] In this case, when the halos have the slope of $\epsilon=3(n+3)/(n+5)$,
then the correlation function should have slope
\begin{equation}
\gamma=\frac{3(n+1)}{n+5}
\end{equation}
Once again, the final state ``remembers'' the initial index $n$.

Is this conclusion true? Unfortunately, simulations do not have sufficient
dynamic range to provide a clear answer but there are some claims\cite{Navarro} that
the halo profiles are ``universal'' and independent of initial conditions.
The theoretical arguments given above are also not very rigourous (in spite
of the popularity they seem to enjoy!). The argument for correlation function to scale as
$[-3(n+3)/(n+5)]$ is based on the assumption of $h=1$ asymptotically, which
may not be true. The argument, leading to density profiles scaling as
$[-3(n+3)/(n+5)]$, is based on scale invariant spherical collapse which
does not do justice to nonradial motions. Just to illustrate the situations
in which one may obtain final configurations which are independent of
initial index $n$, we shall discuss two possibilities:

(i) As a first example we will try to see when the slope of the correlation
function is universal and obtain the slope of halos in the nonlinear limit
using our relation (\ref{gammep}). Such an interesting situation can develop {\it if we assume that $h$ reaches a 
constant
value asymptotically which is not necessarily unity}. In that case, we can
integrate our equation (\ref{redefpair}) to get   $\xb=a^{3h}F[a^h x]$ where $h$ now
denotes the constant asymptotic value of of the function. For an initial
spectrum which is scale-free power law with index $n$, this result translates
to 
\begin{equation}
\bar\xi(a,x)\propto a^{\frac{2 \gamma}{n+3}} x^{-\gamma}
\end{equation} where $\gamma$ is given by 
\begin{equation}
\gamma=\frac{3 h (n+3)}{2+h(n+3)}
\end{equation}
We now notice that one can obtain
a $\gamma$  which is independent of initial power law index provided
$h$ satisfies the condition $h(n+3)=c$, a constant.  In this case, the nonlinear 
correlation
function will be given by
\begin{equation}
\bar\xi(a,x)\propto a^{\frac{6c}{(2+c)(n+3)}} x^{-\frac{3c}{2+c}}
\end{equation} 
The halo index will be independent of $n$
and will be given by
\begin{equation}
\epsilon=3\left( \frac{c+1}{c+2} \right) 
\end{equation}
Note that we are now demanding the asymptotic value of $h$ to {\it explicitly 
depend} on the initial conditions though the {\it spatial} dependence of $\xb$ 
does not.
In other words, the velocity distribution --- which is related to $h$ --- still 
``remembers'' the initial
conditions. This is indirectly reflected in the fact that the growth
of $\xb$ --- represented by $a^{6c/((2+c)(n+3))}$ --- does depend on the
index $n$.

\par
As an example of the power of such a --- seemingly simple --- analysis note the 
following: Since $c \geq 0 $, it follows that $\epsilon > (3/2)$; invariant 
profiles
with shallower indices (for e.g with $\epsilon=1$) are not consistent 
with the evolution described above.
\par

(ii) For our second example, we shall make an ansatz for the halo profile
and use it to determine the correlation function. 
We assume, based on small scale dynamics, that
the density profiles of individual halos 
should resemble that of isothermal spheres, with $\epsilon=2$, irrespective of 
initial conditions. Converting this halo profile to correlation function
in the {\it nonlinear} regime is straightforward and is based on equation
(\ref{gammep}):
If $\epsilon=2$, we must have $\gamma=2 \epsilon-3=1$ at 
small scales; that is $\bar\xi(a,x)\propto x^{-1}$ at the nonlinear regime.
Note that this corresponds to the critical index at the nonlinear
end, $n_{eff}=n_c=-2$ for which the growth rate is $a^2$ --- same as in linear
theory. 
[This is, however, possible for initial power law spectra, only if 
$\epsilon=1$, i.e $h(n+3)=1$ at very nonlinear scales.
Testing the conjecture that $h(n+3)$ is a constant is probably a little
easier than looking for invariant profiles in the simulations but the
results are still uncertain].

The corresponding analysis for the intermediate regime, with $1\gaprox\xb\gaprox 200$, is
more involved.
This is clearly
seen in equation (\ref{powcen1}) which shows that the power spectrum [and
hence the 
correlation
function] depends {\it both} on the fourier transform of the halo profiles as
well as the power spectrum of the distribution of halo centres. In general, 
both quantities will evolve with time and we cannot
ignore the effect of $P_{\rm cent}(k,a)$ and relate $P(k,a)$ to $f(k,a)$. 
The density profile around a {\it local maxima} will
scale approximately as $\rho\propto\xi$ while the density profile around
a {\it randomly} chosen point will scale as $\rho\propto\xi^{1/2}$. [The relation
$\gamma=2 \epsilon-3$   expresses the latter scaling of $\xi\propto\rho^2$]. 
There is, however,
reason to believe that the intermediate regime (with $1 \gaprox \bar\xi \gaprox 200$) is dominated by the
collapse of high peaks \cite{TPMNRAS} . In that case, we expect the
correlation function and the density profile to have the same slope
in the intermediate regime with $\xb\propto (1/x^2)$. Remarkably enough,
this corresponds to the critical index $n_{eff}=n_c=-1$ for the intermediate
regime for which the growth is proportional to $a^2$.

We thus see that if: (i) the individual halos are isothermal spheres
with $(1/x^2)$ profile and (ii) if $\xi\propto\rho$ in the intermediate regime
and $\xi\propto\rho^2$ in the nonlinear regime, we end up with a correlation
function {\it which grows as $a^2$ at all scales}. Such an evolution, of course,
preserves the shape and is a good candidate for the late stage evolution of
the clustering.

While the above arguments are suggestive, they are far from conclusive. It
is, however, clear from the above analysis and it is not easy to provide
{\it unique} theoretical reasoning regarding the shapes of the halos. 
The situation gets more complicated if we include the fact that all halos
will not all have the same mass, core radius etc and we have to modify our
equations by integrating over the abundance of halos with a given value of
mass, core radius etc. This brings in more ambiguities and depending on
the assumptions we make for each of these components [e.g, abundance for halos of a particular mass could be based on Press-Schecter or Peaks formalism],
and the final results have no real significance.
It is, therefore, better [and 
probably easier] to attack the question based on the evolution equation for
the correlation function rather than from ``physical'' arguments for density profiles as done next.

\section{ Self-similar evolution}

Since the above discussion motivates us to look for correlation functions
of the form $\xb=a^2L(x)$, we will start by asking a more general question:
Does equation (\ref{redefpair}) possess 
self-similar 
solutions of the form
\begin{equation}
\label{xiansatz}
\xb=a^{\beta}\:F(\frac{x}{a^{\alpha}})=a^{\beta} F(q)
\end{equation}
where $q\equiv x a^{-\alpha}$ ?. Defining $Q=\ln q=X -\alpha A$ and changing 
independent variables to from $(A,X)$ to $(A,Q)$  we can tranform our equation 
(\ref{redefpair}) to the form:
\begin{equation}
\left( \frac{\partial \bar \xi}{\partial A}\right)_{Q}-(h+\alpha) \left( 
\frac{\partial \bar \xi}{\partial Q}\right)_{A}=3 (1+\bar \xi)\:h(\bar \xi)
\end{equation}
Using the relations $({\partial \bar \xi}/{\partial A})_Q=\beta \bar \xi$,
 $({\partial \bar \xi}/{\partial Q})_{A}=(\bar \xi/F)(d F/d Q)$
we can rewrite this equation  as 
\begin{equation}
\label{KQeq}
\frac{\beta \bar \xi-3 (1+\bar \xi)h(\bar \xi)}{\left[\alpha+h(\bar 
\xi)\right]\bar \xi}=\frac{1}{F} \frac{d F}{d Q}\equiv K(Q)
\end{equation}
The right hand side of this equation depends only on $Q$ and hence will
vanish if differentiated with respect to $A$ at constant $Q$. Imposing this 
condition on
the left hand side and noticing that it is a function of $\xb$ we get
\begin{equation}
\left( \frac{\partial \bar \xi}{\partial A} \right)_Q \frac{d}{d \bar \xi}({\rm 
Left\:Hand\:Side})=0
\end{equation}
To satisfy this condition we either need (i) $(\partial \bar\xi/\partial 
A)_Q=\beta \bar\xi=0$ implying $ \beta=0 $ or (ii) the left hand side must be a 
constant.
Let us consider the two cases separately.

(i)
The simpler case corresponds to $ \beta=0 $ which implies that
$ \xb=F(Q) $. Setting $ \beta=0 $ in equation (\ref{KQeq}) we get

\begin{equation}
\left( \frac{d \bar \xi}{d Q} \right)=-\frac{3(1+\bar \xi)h(\bar 
\xi)}{\left[\alpha+h(\bar \xi)\right]}
\end{equation}
which can be integrated in a straightforward manner to give a relation
between $q=\exp Q$ and $\bar\xi$:
\begin{eqnarray*}
\label{qeq}
q&=&q_0 (1+\bar \xi)^{-1/3} \exp\left( -\frac{\alpha}{3} \int \frac{d \bar 
\xi}{(1+\bar \xi)h(\bar \xi)} \right)\\
&=&q_0 (1+\bar \xi)^{-1/3} {\cal V}(\bar\xi)^{-\alpha/2}
\end{eqnarray*}
Given the form of $h[\xb]$, this equation can be in principle
inverted to determine $\bar\xi$ as a function of $q=x a^{-\alpha}$.
\par
 To
understand when such a solution will exist, we should look at the limit
of  $\bar \xi\ll 1$. In this limit, when linear theory is valid,
we know that $h\approx (2/3) \bar \xi $ \cite{Peeb80}. Using this in 
equation
 (\ref{qeq})  we get the solution to be $\ln {\bar \xi}=-(2/\alpha) \ln q $ or
\begin{equation}
 {\bar \xi} \propto q^{-\frac{2}{\alpha}}\propto x^{-\frac{2}{\alpha}} a^2  
\propto a^2 x^{-(n+3)} 
\end{equation}
with the definition $\alpha\equiv 2/(n+3)$. This clearly shows that our solution 
is valid, {\it if and only if} the linear
correlation function  is a scale-free power law. In this case, of course, it is 
well known that solutions of the type $\bar\xi(a,x)=F(q)$ with $q=x 
a^{-\frac{2}{(n+3)}}$  exists. [Equation
 (\ref{qeq})  gives the explicit form of the function $F(q)$]. This result shows that this is the {\it only} possibility. It should be noted that, even though
we have no explicit length scale in the problem, the function $\bar\xi(q)$
--- determined by the above equation --- does exhibit different behaviour at
different scales of nonlinearity. Roughly speaking, the three regimes in
equation (\ref{Vapprox}) translates into nonlinear density contrasts in the ranges
$\delta<1,1<\delta<200 $ and $\delta >200$ and the function $\bar\xi(q)$
has different characteristics in these three regimes. This shows that
gravity can intrinsically select out a density contrast of $\delta\approx 200$
which, of course, is well-known from the study of spherical tophat collapse. 

(ii) Let us next consider the second possibility, {\it viz.} that the left hand
side of equation (\ref{KQeq}) is a constant. If the constant  is denoted by
$\mu$, then we get $F=F_0\:q^{\mu}$ and 
\begin{equation}
\beta\:\bar \xi-3\:(1+\bar \xi)\:h(\bar \xi)=\mu\:\alpha\:\bar \xi+\mu\:h\:\bar \xi
\end{equation}
which can be rearranged to give
\begin{equation}
\label{hform}
h=\frac{(\beta-\alpha \mu)\bar \xi}{3+(\mu+3)\bar \xi}
\end{equation}
This relation shows that solutions of the form 
$\bar\xi(a,x)=a^{\beta}\,F(x/a^{\alpha})$ with $\beta\neq 0$ is
possible only if $h[\xb]$ has a {\it very specific} form given by (\ref{hform}). 
In this form,
$h$ is a monotonically increasing function of $\xb$. There is, however,
firm theoretical and numerical evidence \cite{Ham},\cite{TPMNRAS}
to suggest that $h$ increases with
$\xb$ first, reaches a maximum and then decreases. In other words, the
$h$ for actual gravitational clustering is {\it not} in the form suggested by
equation (\ref{hform}). {\it We, therefore, conclude that solutions of the form 
in equation (\ref{xiansatz}) with $\beta\neq0$
cannot exist in gravitational clustering}.

By a similar analysis, we can prove a stronger result: There are no  solutions 
of the form $\bar \xi(a,x)=\bar \xi(x/F(a))$ except
when $F(a) \propto a^{\alpha}$. So self-similar
evolution in clustering is a very special situation.

This result, incidentally, has an important implication. It shows that power-law initial
conditions are very special in gravitational clustering and may not represent
generic behaviour. This is because, for power laws, we have a strong constraint
that the correlations etc can only depend on $q=x a^{-2/(n+3)}$. 
For more realistic --- non-power law --- initial conditions the shape can be
distorted in a generic way during evolution. 

All the discussion so far was related to finding {\it exact} scaling
solutions. It is however possible to find {\it approximate} scaling solutions
which are of practical interest. Note that we normally expect
constants like $\alpha,\, \beta,\, \mu$ etc  to be of order unity while $\xb$ can take arbitrarily
large values. If $\xb\gg 1$ then equation (\ref{hform}) shows that $h$ is 
approximately
a constant with $h=(\beta-\alpha\mu)/(\mu+3)$. In this case
\begin{equation}
\bar\xi(a,x)=a^{\beta} F(q) \propto a^{\beta} q^{\mu}\propto a^{(\beta-\alpha 
\mu)} x^{\mu} \propto a^{h(\mu+3)} x^{\mu}
\end{equation}
which has the form $\xb=a^{3h}F(a^hx)$ which was obtained earlier by
directly integrating equation (\ref{redefpair}) with constant $h$. We shall say 
more about such approximate solutions in the next section.

\section{Units of the nonlinear universe}

Having reached the  conclusion that {\it exact} solutions of
the form $\xb=a^2 G(x)$ are not possible, we will ask the question:
Are there such {\it approximate } solutions? And if so, how do they
look like? We will see  that such profiles --- which we shall call
``pseudo-linear profiles''--- that  evolve very close to the the above form
indeed exist.
In order to obtain such a solution and check its 
validity,
it is better to use the results of section 2.2.1  and proceed as
follows:

We are trying to find an approximate solution of the form 
$ \bar\xi(a,x)=a^2 G(x)$ 
to equation (\ref{redefpair}). Since the linear correlation function 
$\bar\xi_L(a,x)$ does grow as $a^2$ at fixed $x$, continuity demands that 
$\bar\xi(a,x)=\bar\xi_L(a,x)$ for all $a$ and $x$. [This can be proved more 
formally as follows: Let $\bar\xi=a^2 G(x)$ and $\bar\xi_L=a^2 G_1(x)$ for some 
range $x_1<x<x_2$. Consider a sufficiently early epoch $a=a_i$ at which all the 
scales in the range $(x_1,x_2)$ are described by linear theory so that 
$\bar\xi(a_i,x)=\bar\xi_L(a_i,x)$. It follows that $G_1(x)=G(x)$ for all 
$x_1<x<x_2$. Hence $\bar\xi(a,x)=\bar\xi_L(a,x)$ for all $a$ in $x_1<x<x_2$. By 
choosing $a_i$ sufficiently small, we can cover any range $(x_1,x_2)$. So 
$\bar\xi=\bar\xi_L$ for any arbitrary range. {\it QED}]. Since we have a formal 
relation (\ref{mapfun}) between nonlinear and linear correlation functions, we 
should be able to determine the form of $G(x)$. 

To do this we shall invert the form of the linear correlation function
 $\bar \xi_L(a,l)=a^2 G(l)$
and   write 
$l=G^{-1}(a^{-2} \bar \xi_L)\equiv F(a^{-2} \bar \xi_L)$
where $F$ is the inverse function of $G$. We also know that the linear correlation 
function $\bar\xi_L(a,l)$ at scale $l$ can be
expressed as ${\cal V}[\bar\xi(a,x)]$ in terms of the true correlation 
function $\bar\xi(a,x)$ at scale $x$
where 
\begin{equation}
\label{lx}
l=x (1+\bar \xi(a,x))^{1/3}
\end{equation}
So we can write
\begin{equation}
\label{lequ1}
l=F\left[\frac{\bar \xi_L(a,l)}{a^2}\right]=F\left[\frac{{\cal 
V}[\bar\xi(x,a)]}{a^2} \right]
\end{equation}
But $x$ can be expressed as $x=F[\bar\xi_L(a,x)/a^2]$;
Substituting this in (\ref{lx}) we have  
\begin{equation}
l=F\left[\frac{\bar \xi_L(a,x)}{a^2}\right]\left[
1+\bar \xi\right]^{1/3}
\end{equation}
From our  assumption $\bar\xi_L(a,x)=\bar\xi(a,x)$ ;
therefore this relation can also be written as
\begin{equation}
\label{lequ2}
l=F\left[\frac{\bar \xi(a,x)}{a^2}\right] \left( 1+\bar\xi \right)^{1/3}
\end{equation}
Equating the expressions for $l$ in (\ref{lequ1}) and (\ref{lequ2}) we get an 
implicit 
functional 
equation for $F$:
\begin{equation}
F\left[ \frac{{\cal V}[\bar \xi]}{a^2} \right]=F\left[ \frac{\bar \xi}{a^2} 
\right] \left(1+\bar\xi\right)^{1/3}
\end{equation}
which can be rewritten as
\begin{equation}
\label{feq}
\frac{F\left[{\cal V}(\bar \xi)/{a^2}\right]}{F\left[{\bar \xi}/{a^2}\right]}=
(1+\bar \xi)^{1/3}
\end{equation}
This equation should be satisfied by the function $F$ if we need to maintain the 
relation $\bar\xi(a,x)=\bar\xi_L(a,x)$. 
\par
To see what this implies, note that the left hand side 
 should not vary with $a$ at fixed $\bar \xi$. This is possible only if 
$F$ is a power law: 
\begin{equation}F(\bar \xi)=A \bar\xi^{m}\end{equation}
which in turn constrains the form of ${\cal V}(\bar \xi)$ to be 
\begin{eqnarray}
\label{Vform}
{\cal  V}(\bar \xi)=\bar \xi\:(1+\bar \xi)^{1/3m}
\end{eqnarray}
Knowing the particular form for ${\cal V}$ we can compute the corresponding 
$h(\bar\xi)$ from the relation
\begin{equation}
\frac{d \ln {\cal V}}{d \bar\xi}=\frac{2}{3} \;\frac{1}{(1+\bar\xi)\,h(\bar\xi)}
\end{equation}
For the ${\cal V}(\bar\xi)$ considered in equation (\ref{Vform}) we get
\begin{equation}
\label{hiform}
h=\frac{2 \bar\xi}{3+(3+1/m) \bar\xi}
\end{equation} 
which is the same result obtained by putting $\beta=2\;,\alpha=0$ in equation
(\ref{xiansatz}). We thus recover  our old result --- as we should --- that {\it exact\ } solutions 
of the form $\bar\xi(a,x)=\bar\xi_L(a,x)=a^2\;G(x)$ are {\it not} possible because the 
correct ${\cal V}(\bar\xi)$ and $h(\bar\xi)$ do not have the forms in 
equations (\ref{Vform}) and (\ref{hiform}) respectively. But, as in the
last section, we can look for approximate solutions.
\par
We note from equation (\ref{Vform}) that for $\bar\xi \gg 1$, we have
\begin{equation}
\label{anotherV}
{\cal V}(\bar\xi)={\bar\xi}^{(1+1/3m)};\qquad F(\bar\xi)\propto\bar\xi^m;\qquad 
G(\bar\xi) \propto \bar\xi^{1/m}
\end{equation}
This can be rewritten as 
\begin{equation}
\label{yetanotherv}
{\cal V}(\bar\xi)=\bar\xi^{\nu};\qquad 
F(\bar\xi)\propto\bar\xi^{1/3(\nu-1)};\qquad 
G(\bar\xi)\propto\bar\xi^{3/(\nu-1)}
\end{equation}
In other words if ${\cal V}(\bar\xi)$ can be approximated as $\bar\xi^{\nu}$, we 
have an approximate solution of the form
\begin{equation}
\label{approxsol}
\bar\xi(a,x)=a^2\;G(x)=a^2\;x^{3(\nu-1)}
\end{equation}
Since the ${\cal V}$ in equation (\ref{ham1}) is well  
approximated by the power laws in (\ref{Vapprox}) so that
\begin{eqnarray}
\label{approximateV}
{\cal V}(\bar\xi) &\propto & {\bar\xi}^{1/3} \;\;(1 \mbox{\gaprox} \bar\xi 
\mbox{\gaprox} 200)\\
&\propto&\bar\xi^{2/3} \;\;(200 \mbox{\gaprox} \bar\xi)
\end{eqnarray}
we can take $\nu=1/3$ in the intermediate regime and $\nu=2/3$ in the nonlinear regime. 
It follows from (\ref{yetanotherv}) that the approximate solution should have 
the form 
\begin{eqnarray}
\label{approxF}
F(\bar\xi) & \propto & {\frac{1}{\sqrt{\bar\xi}}} \;\;(1 {\mbox{\gaprox}} 
\bar\xi {\mbox{\gaprox}} 200)\\
& \propto & \frac{1}{\bar\xi}\;\;\;\;\;\;(200 \gaprox \bar\xi)
\end{eqnarray}
This gives the approximate form of a pseudo-linear profile which will grow
as $a^2$ at all scales.

There is another way of looking at this solution which is probably more physical and throws light on the scalings of pseudo-linear profiles. We recall that, in the study of finite gravitating systems made of point particles and
interacting via Newtonian gravity, isothermal spheres play an important
role. They can be shown to be the local maxima of entropy  
\cite{PadPhyRep} and hence dynamical
evolution drives the system towards an $(1/x^2)$ profile. Since one expects
similar considerations to hold at small scales, during the late stages of evolution of the universe, we may hope that isothermal spheres with
$(1/x^2)$ profile may still play a role in the late stages of evolution of 
clustering in an expanding background. However, while converting the profile to correlation, we have to take note of the issues discussed in section 2.
In the intermediate regime, dominated by scale invariant radial collapse \cite{TPMNRAS}, the density will scale as the correlation function and
we will have $\bar\xi\propto (1/x^2)$. On the other hand, in the nonlinear
end, we have the relation $\gamma=2\epsilon -3$ [see equation (\ref{gammep}) ] which
gives $\bar\xi\propto (1/x)$ for $\epsilon=2$. Thus, if isothermal spheres
are the generic contributors, then we expect the correlation function to
vary as $(1/x)$ and nonlinear scales, steepening to $(1/x^2)$ at intermediate
scales. Further, since isothermal spheres are local maxima of entropy, a configuration like this should remain undistorted for a long duration. This
argument suggests that a $\bar\xi$ which goes as $(1/x)$ at small scales
and $(1/x^2)$ at intermediate scales is likely to be a candidate for pseudo-linear profile. It was found that this is indeed the case.

To go from the scalings in two limits given by equation
(\ref{approxF}) to an actual profile, we can use
some fitting function. By making the fitting function sufficiently complicated,
we can make the pseudo-linear profile more exact. We shall, however, choose
the simplest interpolation between the two limits and try the ansatz:
\begin{equation}
\label{ansatzF}
F(z)=\frac{A}{\sqrt{z}\;(\sqrt{z}+B)}
\end{equation}
where $A$ and $B$ are constants. Using the original definition 
$l=F[\bar\xi_L/a^2]$ and the condition that $\bar\xi=\bar\xi_L$, we get
\begin{equation}
\frac{A}{\sqrt{\bar\xi/a^2}\;(\sqrt{\bar\xi/a^2}+B)}=l
\end{equation}
This relation implicitly fixes our pseudo-linear profile. Solving for $\bar\xi$, we get
\begin{equation}
\label{xisolution}
\bar\xi(a,x)=\left(\frac{Ba}{2}\;\left(\sqrt{1+\frac{L}{x}} -1\right)\right)^2
\end{equation}
with $L=4A/B^2$. Since this profile is not a pure power law, this will satisfy the equation (\ref{feq})
 only approximately. We choose $B$ such that the relation 
\begin{equation}
\label{justaneq}
F\left( \frac{{\cal V}(\bar\xi)}{a^2} \right) = F\left(\frac{\bar\xi}{a^2}
\right) 
\left(1+\bar\xi\right)^{1/3}
\end{equation}
is satisfied to greatest accuracy at $a=1$.
\begin{figure}[htbp]
\epsfxsize=300pt
\epsfysize=300pt
\epsfbox{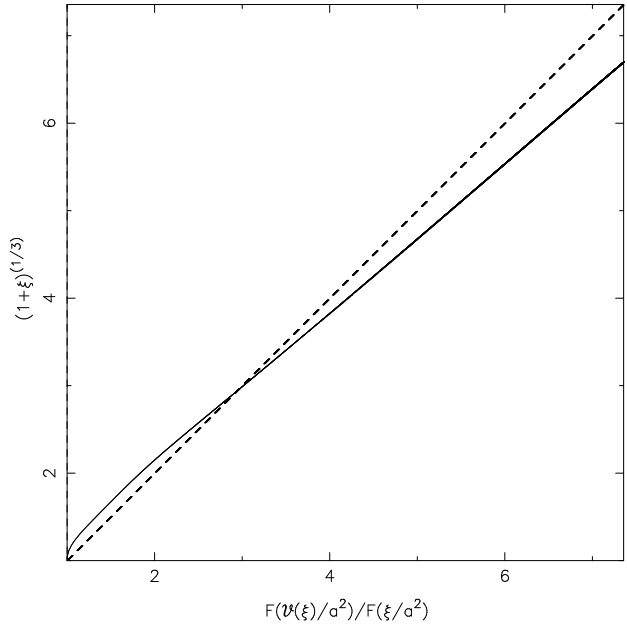}
\caption{\sf The approximate solution to the functional equation
determining the pseudo-linear profile.}
\end{figure}
This approximate profile works reasonably well. Figures 2.1 and 2.2 show this 
result. In figure 2.1 we have plotted the 
ratio 
$F( {\cal V}(\bar\xi)/{a^2})/F({\bar\xi}/{a^2})$ on the x-axis and
the function $(1+\bar\xi)^{1/3}$ on the y-axis. If the function in
(\ref{xisolution}) satifies equation (\ref{feq}) exactly, we should get a 45-degree line
in the figure which is shown by a dashed line. The fact that our curve is
pretty close to this line shows that the ansatz in (\ref{xisolution}) satisfies equation 
(\ref{feq}) fairly well. The optimum value of $B$ chosen for this figure is
$B=38.6$. When $a$ is varied from $1$ to $10^3$, the percentage of error
between the 45-degree line and our curve is less than about 20 percent in the
worst case.
 It is clear that our profile in (\ref{xisolution})  satisfies equation 
(\ref{justaneq}) quite well for a dynamic range of $10^6$ in $a^2$.

Figure 2.2 shows this result more directly. We evolve the pseudo linear profile 
form $a^2=1$ to $a^2\approx 1000$ using the NSR, and plot 
$[\bar\xi(a,x)/a^2]$ against $x$.
The dot-dashed, dashed and two solid curves (upper one for $a^2=100$ and lower one for $a^2=900$) are for $a^2=1,9,100$ and $900$
respectively. The overlap of the curves show that the profile does grow 
approximately as 
$a^2$. Also shown are lines of slope $-1$ (dotted)  and $-2$ (solid); clearly $\bar\xi\propto 
x^{-1}$ for small $x$ and $\bar\xi\propto x^{-2}$ in the intermediate regime.
\begin{figure}[h]
\epsfxsize=350pt
\epsfysize=300pt
\epsfbox{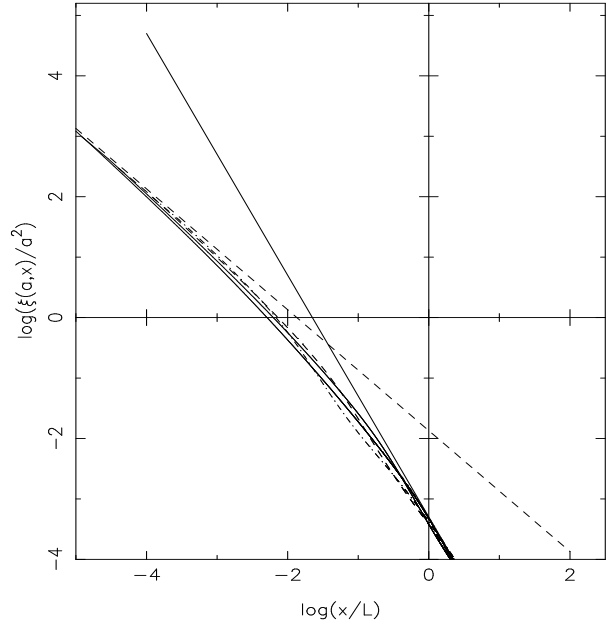}
\caption{\sf The dot-dashed,dashed and two solid curves (upper one for $a^2=100$ and lower one for $a^2=900$) are for $a^2=1,9,100$ and $900$.
The dotted straight line is of slope -1 and the solid one is of slope-2 showing
both the $1/x$ and $1/x^2$ regions of the profile.}
\end{figure}

We emphasis that we have chosen in equation (\ref{xisolution}) the simplest kind of ansatz combining the
two regimes and we have used only two parameters $A$ and $B$. It is quite
possible to come up with more elaborate fitting functions which will solve
our functional equation far more accurately but we have not bothered to do
so for two reasons: (i) Firstly, the fitting functions in equation (\ref{Vapprox}) 
for ${\cal V}(z)$ itself is approximate and is probably accurate only at
10-20 percent level. There has also been repeated claims in literature
that these functions have weaker dependence on $n$ which we have ignored for simplicity. (ii) Secondly, one must remember that only those
$\bar\xi$ which correspond to positive definite $P(k)$ are physically meaningful. This happens to be the case our choice [which can be verified by explicit numerical integration with a cutoff at large $x$] but this may not be true for arbitrarily complicated  fitting functions. Incidentally, another simple fitting function for the pseudo-linear profile is
\begin{equation}
\bar\xi(a,x)=a^2 \frac{A'}{(x/L')[(x/L')+ 1]}
\end{equation} 
with $A'=B^2$ and $L'=L/4$.

If a more accurate fitting is required, one can obtain it more directly from
equation (\ref{naeqn}). Setting $n_a=2$ in that equation predicts the instantaneous
spatial slope of $\xb$ to be
\begin{equation}
\part{\ln\xb}{\ln x}=\frac{2}{h[\xb]}-3(1+\frac{1}{\xb})
\end{equation}
which can be integrated to give
\begin{equation}
\ln\frac{x}{L}=\int_{\bar\xi[L]}^{\bar\xi[x]}\frac{h d\bar\xi}
{\bar\xi(2-3h)-3h}
\end{equation}
at $a=1$ with $L$ being an arbitratry integration constant. Numerical integration of this equation will give a profile which is
varies as $(1/x)$ at small scales and goes over to $(1/x^2)$ and then to
$(1/x^3),(1/x^4)....$ etc with an asymptotic logarithmic dependence. In the regime
$\xb>1$, this will give results reasonably close to our fitting function. 

It should be noted that equation (\ref{feq}) reduces to an identity
for any $F$, in the limit $\bar\xi\to 0$ since, in this limit $ {\cal V}(z)
\approx z$. This shows that we are free to modify our pseudo-linear profile
at large scales into any other form [essentially determined by the input 
linear power
spectrum] without affecting any of our conclusions.

Finally, we will discuss a different way of thinking about
pseudolinear profiles.

In studying the evolution of the density contrast $\delta(a,{\bf x})$, it is 
conventional
to expand in in term of the plane wave modes as 
\begin{equation}
\label{name1}
\delta(a,{\bf x})=\sum_{\bf k} \delta(a,{\bf k}) \exp(i {\bf k}\cdot{\bf x})
\end{equation}
In that case,
the {\it exact} equation governing the evolution of $\delta(a,{\bf k})$  is 
given by \cite{Peeb80}
\begin{equation}
\label{deltakeq}
\frac{d^2 \delta_{\bf k}}{d a^2}+\frac{3}{2 a} \frac{d \delta_{\bf k}}{d 
a}-\frac{3}{2 a^2}\delta_{\bf k}={\cal A}
\end{equation}
where ${\cal A}$ denotes the terms responsible for the 
nonlinear coupling between different
modes. The expansion in equation (\ref{name1}) is, of course, motivated by the 
fact that
in the linear regime we can ignore ${\cal A}$  and each of the modes evolve
independently. For the same reason, this expansion is not of much value
in the highly nonlinear regime.

This prompts one to ask the question: Is it possible to choose some other
set of basis functions $Q(\alpha,{\bf x})$, instead of $\exp\;i{\bf k}\cdot{\bf 
x}$, and expand $\delta(a,{\bf x})$ in the form 
\begin{equation}
\delta(a,{\bf x})=\sum_{\alpha} \delta_{\alpha}(a)\; Q(\alpha,{\bf x})
\end{equation}
so that the
nonlinear effects are minimised ? Here $\alpha$ stands for a set of parameters 
describing the basis functions. This question is extremely difficult to answer, 
partly because it is ill-posed. To make any progress, we have to first give 
meaning to the concept of ``minimising the effects of nonlinearity''. One 
possible approach we would like to suggest is the following: We know that when 
$\delta(a,{\bf x}) \ll 1 $,then $\delta(a,{\bf x})\propto a\:F({\bf x})$ for 
{\it any} arbitrary $F({\bf x})$; that is all power spectra grow as $a^2$ in the 
linear regime. In the intermediate and nonlinear regimes, no such general statement can 
be made. But it is conceivable that there exists certain {\it special} power 
spectra for which $P({\bf k},a)$ grows (at least approximately) as $a^2$ even in 
the nonlinear regime. For such a spectrum, the left hand side of 
(\ref{deltakeq}) vanishes (approximately); hence the right hand side should
also vanish. {\it Clearly, such power spectra are 
affected least by nonlinear effects.} Instead of looking for such a special 
$P(k,a)$ we can, equivalently look for a
particular form of $\xb$ which evolves as closely to the linear theory
as possible. Such correlation functions and corresponding power spectra [which 
are the pseudo-linear
profiles] must be capable of capturing most of the essence of nonlinear 
dynamics. In this sense, we can think of our pseudo-linear profiles as
the basic building blocks of the nonlinear universe. The fact that the
correlation function  is closely related to isothermal spheres, indicates
a connection between local gravitational dynamics and large scale gravitational
clustering.

\section{Results and Summary}
It seems reasonable to hope that the late stage evolution of collisionless point particles, interacting via Newtonian
gravity in an expanding background, should be understandable in terms of a simple paradigm. This chapter  tries to realise this 
within some well defined framework. It should be viewed as a first step in a new direction.

There are three key points which emerge from this analysis. The first is 
the fact that we have been able to find approximate correlation functions
which evolve preserving their shapes. We achieved this by looking at the
structure of an exact equation which obeys certain nonlinear scaling relations. As we emphasised before, the existence of such special class of solutions
to the equations of gravitational dynamics is an important feature.

Secondly, we should take note of the role
played by the ``isothermal'' profile $(1/x^2)$ in our solution. Such a profile can lead to
correlation functions which go as $(1/x)$ at small scales and $(1/x^2)$ in the
intermediate scales. If this profile is indeed ``special'' then one expects it
to lead to a  pseudo-linear profile for the correlation function. Our analysis shows that there
is indeed good evidence for this feature. If one accepts this evidence, then
the next level of enquiry would be to ask why $(1/x^2)$ profiles are ``special''.
In the statistical mechanics of gravitating systems, one can show that these
profiles arise as end stages of violent relaxation which operates at dynamical
time scales. Whether a similar reasoning holds in an expanding background,
independent of the index for power spectrum, is open to question. We emphasise
that our equations, along with NSR, naturally lead to a pseudo-linear
profile, which can be interpreted and understood in terms of isothermal 
density profiles for halos; we did not have to assume anything a priori
regarding the halo profiles.

In a more pragmatic way, one can understand the pseudo-linear profile
from the dependence of the rate of growth of the correlation function
on the local slope. The NSR suggest that $\bar\xi$
grows (approximately) as $a^{6/(n_{eff}+4)}$ in the intermediate regime and as
$a^{6/(n_{eff}+5)}$ in the nonlinear regime. This scaling shows that $n_{eff}
=-1$ grows as $a^2$ in the intermediate regime and $n_{eff}=-2$ grows as $a^2$
in the nonlinear regime. This is precisely the form our pseudo-linear profile
has.  Also, in the intermediate regime, the correlation grows 
faster than $a^2$ 
if $n_{eff}<-1$ and slower than $a^2$ if $n_{eff}>-1$. The net effect is, of course, to straighten out a curved
correlation and drive it to $n=-1$. Similar effect drives the correlations
to $n=-2$ in the nonlinear regime.[see \cite{JsbTp} for a
more detailed discussion of this aspect in the intermediate regime]. Of course,
one still needs to understand the dependence of growth rate on the $n_{eff}$
from more physical considerations to get the complete picture. We have not addressed what is the timescale over which clustering can lead to the psuedo-linear profile assuming that it does.

The last aspect has to do with what one can achieve using the pseudo-linear
profiles. In principle, one would like to build the nonlinear density field
through a superposition of pseudo-linear profiles but this is a mathematically
complex problem. As a first step one should understand why the nonlinear term
in equation (\ref{deltakeq}) is subdominant for such a profile. This itself is complicated
since we have only fixed the power spectrum --- but not the phases of the
density modes --- while the nonlinear terms do depend on the phase.

\par
High resolution numerical simulations serve to test the framework presented in this chapter but resource limitations hamper such experiments. The following chapter discusses the general theory of two dimensional simulations which may be used to explore these issues in greater detail by bypassing the resource constraints plaguing 3-D simulations.

\chapter{A formal analysis of  two dimensional gravity}
\label{chap:twod}
\baselineskip=12pt
{\scriptsize Computers make it easy to do a lot of things but most of the things they make it \newline  easier to do don't need to be done -- Andy Rooney}
\par
\baselineskip=24pt


\section{Introduction}
\label{sec:intro}

The analytical explorations of the previous chapter has clearly shown  that he equations describing
the growth of density perturbations in the highly nonlinear stage are
analytically intractable and hence large scale numerical simulations
are resorted to for exploration of this regime. 
\par
These N--Body simulations require  large amount of computing resources
(CPU, memory and storage space) if one is to get the requisite amount of
dynamical range, {\it i.e.} good resolution in force and mass,  large range
in values of density  etc. Time and resource constraints usually
limit our ability to probe structure formation issues more deeply using computers, once the required resources are at the limits of technological feasibility. 
The key parameter which decides the feasibility level of numerical
simulations is the size of a simulation, which --- in turn --- is characterised
by:  (i)~The number of particles in the simulation volume, which is
generally specified as $N^D$, where  $D$ is the dimensionality of the
simulation (usually $2$ or $3$) and (ii)~The number
of mesh points ($M$) along any axis which determines the minimum length
scale at which the results can be treated as reliable indicators of
physical phenomena. In order to create a simulation volume that is a
fair sample of the universe one needs about  $10^7$ particles and in
order to have a high enough force resolution one needs to increase the
number of grid points adequately  \cite{nbodyrev}. Let us suppose we have
 $160^3$ particles in three dimensions and our grid is $160$ units on a side. Then, for the same amount of computational resources  one can
simulate a two dimensional  situation with $2048^2$ particles on a $2048^2$
grid ($160^3 \approx 2048^2$). So, if we can  extract useful
(i.e. generalisable to the three dimensional case) physical insights from results in two dimensions,
then simulations of two dimensional gravity will be helpful. This hope  has led to a large number of two dimensional simulations in the field of gravitational
clustering ( \cite{2dsc}, \cite{valiniaetal}, \cite{sathyaetal}, \cite{alimietal}, \cite{shandetal}).
\par
There are three ways in which two dimensional gravity can be operationally defined
and corresponding numerical simulations undertaken: (i)~Consider a
system of point particles in a three dimensional (expanding) background with the force of
interaction being given by Newton's law of gravitation ({\it i.e} $F\propto 1/r^2$). The initial positions and velocities of the
particles are such that they all lie in the same plane and all the
velocities are in the plane {\it i.e} there are no velocity components
orthogonal to the plane. This system  will evolve
with the particles being confined to the plane with clustering occurring
in the plane. Thus, we have a two dimensional clustering scenario.
(ii)~Another system we can consider consists of infinite, thin
`needles' located parallel to each other. The mass elements in the
`needles' still interact through the $1/r^2$ force, but the interaction
between `needles' (obtained by summing over the mass elements) is given by a $1/r$ force. In this case as well, the
background space expands uniformly in three dimensions. The two dimensional clustering that we study is the clustering of these `needles', examined by
taking a slice orthogonal to the `needles'. (iii)~The third possibility
involves writing down the Einstein's equations in two dimensions,
finding the homogeneous and isotropic cosmological solution, taking
the Newtonian limit (in which  the potentials due to density perturbations
and background metric can be  superposed), finding the
corresponding perfect fluid equations and solving them. In this case,
we will have a background spacetime expanding in {\it two} dimensions unlike
the other two cases.
(There is yet another, fourth possibility, which  can be defined only in an {\it ad hoc} manner discussed towards the end of the chapter.)
\par 
The first case is not of much interest for cosmological simulations
since the system is anisotropic, confined to a single plane and the
clustering takes place in a specific plane {\it only because the initial
conditions were specifically selected} to give this result. Hence we
will not discuss it and it is mentioned here only for completeness. The way  simulations in two dimensions are carried out usually is by simulating the second case and then defining
the `particles' as the intersection of the `needles' with any plane
orthogonal to them. In this case -- as in the first case -- the background
spacetime expands in three dimensions (for a flat dust dominated universe
the scale factor $a(t)$ goes as $t^{2/3}$). The clustering that we
observe and quantify in two dimensions, is basically the clustering of these
needles in three dimensions. But this is also an  anisotropic situation since
the background spacetime is expanding in three dimensions. 
As an alternative we may try to write down the equations derived from
Einstein's equations in two dimensions and examine how a system of
particles interacting in a two dimensional expanding background spacetime is to be simulated.
\par
In the rest of this chapter we shall examine  the third alternative.
We will take a very general approach by developing the formal theory
of $(D+1)$ gravity and considering $D=2$  as a special case \cite{desjhooft}.
\par
The basic layout of the chapter is as follows: 
In section~(\ref{sec:Dgravity}) we first define the analogue of Einstein's gravity in $(D+1)$ dimensions, discuss the Newtonian limit and its corresponding Poisson's equation and then go on to analyse the Friedmann metric in $(D+1)$ dimensions for a flat universe with dust. 
In section~(\ref{sec:structure}) we write down  the $D$ dimensional fluid equations and obtain the equation governing the density perturbations. This equation is then solved in the linear approximation and using the Spherical Top Hat model. 
Then in section~(\ref{sec:3D}) and section~(\ref{sec:2D}) we specialise to the cases $D=3$ and $D=2$ respectively. 
Finally, in section~(\ref{sec:conclusions}) we summarise and discuss the implications of the results obtained in the earlier sections.  

\section{Formal $(D + 1)$ dimensional gravity} \label{sec:Dgravity}
\noindent We start our  analysis of $(D + 1)$ dimensional ($1$ time dimension and $D$ space dimensions) gravity from the action principle which we assume has the same form as that used in $(3+1)$ dimensions.   Using this action we  construct the corresponding $(D+1)$ dimensional Einstein equations  which will be subsequently used to study structure formation and spherical collapse.  
Thus, we begin with the action principle,
\begin{equation}
{\cal S} = {\cal S}_{\rm g} + {\cal S}_{\rm m} =
-\frac{c^4}{2\kappa(D)}\int \! d^{(D+1)}x \; R \sqrt{|g|} \; + \; \int
\! d^{(D+1)}x \; {\cal L}_{\rm m} 
\label{eqn:action1}
\end{equation}
where ${\cal S}_{\rm g}$ is the action for the gravitational field, ${\cal S}_{\rm m}$ is the action for the matter fields, $g$ is the determinant of the metric tensor $g_{ik}$, $R$ is the Ricci scalar, $\kappa (D)$ is  a suitable constant which can be, in general, a function of $D$ (when $D=3$, $\kappa=8\pi G$, $G$ being the usual gravitational constant) and ${\cal L}_{\rm m}$ is the lagrangian density for the matter fields. The metric signature we adopt is $(+,-,-,-,\ldots,-)$.  We adopt the following convention regarding indices. Latin alphabets $i,j,k\ldots$ are used to represent $(D+1)$ dimensional indices which take on the values $(0,1, 2, \ldots, D)$ while greek letters are used to denote $D$ dimensional indices taking on the values $(1, 2, \ldots, D)$. Varying the total action ${\cal S}$ with respect to $g_{ik}$ we obtain Einstein's equations, 
\begin{equation}
G_{ik}\equiv R_{ik}-\frac{1}{2}\;g_{ik}\;R = {\kappa(D) \over c^4}
T_{ik} 
\label{eqn:einstein}
\end{equation}
where $T_{ik}$ is the energy momentum tensor of the matter fields and is defined by 
\begin{equation}
\frac{1}{2} \sqrt{|g|} T_{ik} = \frac{\partial (\sqrt{|g|}{\cal L}_{\rm m})} {\partial g^{ik}} \; - \; 
\frac{\partial}{\partial x^l} \left( \frac{\partial (\sqrt{|g|}{\cal
L}_{\rm m})}{\partial \left( \partial g^{ik}/\partial x^l \right) }
\right)
\label{eqn:emtensor}
\end{equation}
$G_{ik}$ is the Einstein tensor and $R_{ik}$ is the usual Ricci
tensor. Note that the $(1/2)$ that appears in Einstein's equations
arises due to the square root in the term $\sqrt{|g|}$ and has nothing to do with the dimension of the spacetime.  
\par
We will use the above equations in the subsections to follow.  
In subsection~(\ref{subsec:poisson}), we will study the Newtonian limit of the metric tensor and then construct the corresponding Poisson equation that relates the Newtonian gravitational field $\phi$ to the matter density $\rho$.  
Then, in subsection~(\ref{subsec:friedmann}), we analyse the Friedmann metric in $(D+1)$ dimensions and the corresponding Newtonian limit of this metric is derived.  
\subsection{Poisson equation in $D$ dimensions}
\label{subsec:poisson} 
\noindent In this section, we derive the Poisson equation relating the
gravitational potential $\phi$ to the matter density $\rho$. We keep
all factors of $c$ since the Newtonian limit involves the limit $c\to
\infty$. The analysis here follows closely the treatment in \cite{landau2}.
  Consider the metric.
\begin{equation}
ds^2 = \left( 1 + {2\phi \over c^2} \right)c^2dt^2 - dl^2
\label{eqn:Nmetric}
\end{equation}
where $\phi$ is a function of space and time with dimensions of velocity square. The
term  $dl^2$ is the $D$ dimensional spatial line element given by the formula
\begin{equation}
dl^2 = \sum_{\alpha=1}^D (dx^{\alpha})^2 
\label{eqn:Dline}
\end{equation}
We will now show that the metric written above is the Newtonian limit of Einstein's gravitational equations. We do this by showing that, in the Newtonian limit, the equation of motion of a particle follows Newton's force law with the force $-m\nabla \phi$...  
In relativistic mechanics, the motion of a particle of mass $m$ is determined by the action function $S$
\begin{equation}
S = -mc \int \! ds = -mc \int \! c\,dt \sqrt{\left( 1 + {2\phi
\over c^2}  -{v^2 \over c^2}\right)}
\label{eqn:Nmetric1}
\end{equation}
where $v^2$ is the square of the magnitude of the particle's velocity in $D$ dimensions. In arriving at the second equality we have used the form of the metric in equation~(\ref{eqn:Nmetric}). In the limit $c \to \infty$, the action $S$ can be approximated as
\begin{equation}
S \approx -mc^2 \int \! dt \left( 1 + {2\phi -v^2 \over 2c^2} \right)
= \int \! dt \left( -mc^2 + {1 \over 2}mv^2 - m\phi\right)
\label{eqn:Nmetric2}
\end{equation}
The equation of motion for the particle can be immediately written down and we obtain
\begin{equation}
m{d{\bf v} \over dt} = - m\nabla \phi 
\label{eqn:Nmetric3}
\end{equation}
where ${\bf v}$ is the velocity vector in $D$ space dimensions.  Thus, Newton's force law is recovered in the non-relativistic limit and from this we conclude that the metric given in equation~(\ref{eqn:Nmetric}) is the Newtonian limit of Einstein's gravitational equations with $\phi$ acting as the Newtonian gravitational potential. 

The relation between $\phi$ and the mass density $\rho$ is found by taking the $c\to\infty$ limit of
Einstein's equations. This procedure, in $(3+1)$ gravity, determines the constant $\kappa(D)$ since the Poisson equation is explicitly known. In other dimensions however, a definite criterion, like Gauss's law for example, must be imposed in order to determine $\kappa(D)$. We now consider the limit $c \to \infty$ of Einstein's equations in the following manner.
First, we use the line element given in equation~(\ref{eqn:Nmetric}) to calculate the Ricci tensor component $R_{00}$
\begin{equation}
R_{00} = \frac{1}{c^2}\frac{1}{c^2 + 2 \phi} (\partial_\mu \phi)(\partial^\mu \phi) - \frac{1}{c^2} \partial_\mu\partial^\mu \phi 
\label{eqn:poisson1}
\end{equation}
where the summation convention has been invoked in the above equation and the sum over $\mu$ is only over the {\it spatial} dimensions. 
Then, using equation~(\ref{eqn:einstein}), we obtain,
\begin{equation}
R = -{2\kappa(D) \over c^4(D-1)} T
\end{equation}
where we have used the fact that $g_{ik}g^{ik} = D + 1$ and assumed $D \neq 1$. Thus, Einstein's equations can be written in the equivalent form,  
\begin{equation}
R_{ik} = {\kappa(D) \over c^4} \left( T_{ik} - {1\over D-1}g_{ik}T \right). \label{eqn:poisson2}
\end{equation}
where $T$ is the trace of $T_{ik}$.  
The energy momentum tensor of point particles is $T_{ik} = \rho c^2 u_iu_k$ where $\rho$ is the mass density and $u_i$ is the four velocity.  Since, in the non-relativistic limit, the macroscopic motion is slow, the space components of $u_i$ can be neglected and only the time component should be retained.  Therefore, $u_0 = \sqrt{g_{00}}$ and $u_\mu \approx 0$ for all $\mu$. Consequently, only $T_{00} = g_{00}\rho c^2$ is non-zero.  Substituting 
for $T_{ik}$ into equation~(\ref{eqn:poisson2}) and using the expression in equation~(\ref{eqn:poisson1}) for $R_{00}$, we get,  
\begin{equation}
{1 \over c^2} \partial_\mu\partial^\mu \phi = -\left({D-2 \over
D-1}\right) \left(1 + {2\phi \over c^2}\right) {\kappa(D) \over c^4}
\rho c^2 \approx  -\left({D-2 \over D-1}\right){\kappa(D)\over c^2}
\rho 
\label{eqn:poisson3}
\end{equation}
That is,
\begin{equation}
\nabla^2 \phi = \left({D-2 \over D-1}\right)\kappa(D) \rho 
\label{eqn:poisson4}
\end{equation}
where $\nabla^2$ is the usual Laplacian operator in $D$ dimensions. This equation is the Poisson equation in $D$ dimensions. Note that when substituting for the value of $R_{00}$ from equation~(\ref{eqn:poisson1}), we neglected the first term in comparison with the second since the former is of order $c^{-4}$ while the latter is only of order $c^{-2}$.  

\subsection{Friedmann Universe in $(D+1)$ dimensions} 
\label{subsec:friedmann}
\noindent Let us next consider the maximally symmetric Robertson-Walker metric in $(D+1)$ dimensions, specialising to flat space with $k=0$ (we set $c=1$ in this and in subsequent sections),
\begin{equation}
ds^2 = dt^2 - a^2(t)dl^2 
\label{eqn:robertson}
\end{equation}
where $a(t)$ is the scale factor and $dl^2$ is the $D$ dimensional line element given in equation~(\ref{eqn:Dline}). Calculating the components of the Einstein tensor, we obtain,
\begin{eqnarray}
G_{00} &=& \frac{D(D-1)\dot{a}^2}{2a^2} \nonumber \\
G_{11} = G_{22} =\ldots = G_{DD} &= & (1-D)a \ddot{a} +\nonumber \\
 \left(1 - \frac{D}{2}\right) (D-1) \dot{a}^2 
\label{eqn:fried1}
\end{eqnarray}
where $\dot{a}$ stands for $da(t)/dt$ and similarly $\ddot{a}$ is the second derivative of $a(t)$ with respect to time. All the other components are zero.  For consistency, the energy
momentum tensor must have the form $T^i_{\; k} = {\rm diag}(\rho,
-p,-p,-p,\ldots)$ where $\rho$ is the matter density and $p$ is the
pressure. 
\par
Substituting in Einstein's equations, we obtain,
\begin{eqnarray}
\frac{D(D-1)\dot{a}^2}{2a^2}& =& \kappa(D) \rho  \label{eqn:friedmann1} \\
\frac{\ddot{a}}{a} +\frac{D-2}{2} \frac{\dot{a}^2}{a^2}&=& -\frac{\kappa(D) p}{D-1} \label{eqn:friedmann}
\end{eqnarray}
The above two equations, together with the equation of state in the form $p =p(\rho)$ completely specify the system.  Solving these three equations, we can determine $a(t)$, $\rho(t)$ and subsequently $p(t)$.
Combining equations~(\ref{eqn:friedmann1},\ref{eqn:friedmann}), we get the single equation,
\begin{equation}
\frac{\ddot{a}}{a} = -\frac{\kappa(D)}{D(D-1)}\left[ (D-2)\rho + Dp\right] .\label{eqn:friedmann2}
\end{equation}
\noindent We now specialise to the case of pressureless dust with the equation of state $p =0$.  Using the principle of conservation of energy and momentum expressed by the relation
\begin{equation}
T^k_{\;\; i;k}=0 \label{eqn:fried2}
\end{equation}
we derive the following relation,
\begin{eqnarray}
T^k_{\;\; i;k} &=& {1 \over \sqrt{|g|}} {\partial \over \partial x^k}\left( \sqrt{|g|} T^k_{\;\; i} \right) - {1 \over 2}{\partial g_{kl} \over \partial x^i} T^{kl} \nonumber \\
&=& {1 \over a^D} {\partial \over \partial x^k}\left(a^D T^k_{\;\;
i}\right) - {1 \over 2}{\partial g_{kl} \over \partial x^i} T^{kl} = 0
 \label{eqn:fried3}
\end{eqnarray}
Noting that the only non-zero component of $T^i_{\; k}$ is $T^0_{\; 0} = \rho$ we finally get
\begin{equation}
\rho a^D = {\rm constant} = C_1 \label{eqn:densityrel}
\end{equation}
Substituting the above relation into equation~(\ref{eqn:friedmann1}), and solving for $a(t)$ and subsequently for $\rho(t)$, we obtain the solutions,
\begin{equation}  
a(t) = \left({ D\kappa(D) C_1 \over 2(D-1) }\right)^{1/D} \, t^{2/D} \quad ; \quad \rho(t) = \left({2 (D-1) \over D \kappa(D)}\right) t^{-2} 
\label{eqn:friedsoln}
\end{equation}
Let us next consider the Newtonian limit of the Friedmann metric. This limit is important because the length scales of interest in structure formation are small compared to the Hubble radius and the velocities in the system are also much smaller than {\it c}. This permits us to study the formation of large scale structures in the universe in a Newtonian framework where the effective potential due to the expanding background universe, $\Phi_{\rm FRW}$, and the potential due to the density perturbations, $\varphi$, can be simply superposed.  In order to obtain $\Phi_{\rm FRW}$, we first recast the Friedmann metric in equation~(\ref{eqn:robertson}) into the more convenient form
\begin{equation}
ds^2 = dt^2 - a^2(t)\left( dX^2 + X^2 d\Omega^2 \right)
\end{equation}
where $X$ is the radial distance in $D$ dimensions and $\Omega$ is the
corresponding solid angle.  We then apply the transformations (see \cite{Probbook}, pg 80,346)
\begin{equation}
r = Xa(t) \quad , \quad  T = t - t_0 + {1\over 2} a\dot{a} X^2 + {\cal O}(X^4)
\end{equation}
where only terms up to quadratic in $X$ are retained.  Direct calculations, correct upto this order, transforms the Friedmann line element to the form,
\begin{equation}
ds^2 \approx \left(1 - {\ddot{a} \over a}r^2\right) dT^2 - dr^2 - r^2 d\Omega^2
\end{equation}
which upon comparison with the metric in equation~(\ref{eqn:Nmetric}) gives the equivalent Newtonian potential $\Phi_{\rm FRW}$ in $D$ dimensions as 
\begin{equation}
\Phi_{\rm FRW}=-\frac{1}{2} \frac{\ddot{a}}{a}\;r^2   
\label{eqn:bgpotential}
\end{equation}
\noindent We will now use the results developed in the last two subsections to study structure formation and spherical collapse using the STH model.

\section{Structure formation in $D$ dimensions} \label{sec:structure}
\noindent Having determined the form of the Poisson equation in the Newtonian limit and analysed the Friedmann equations in $(D+1)$ dimensions, we proceed to derive the equation for the growth of inhomogeneities in the expanding universe. After this we consider a specific model, the STH model, to study spherical collapse of matter.

\subsection{Equation for density perturbations in $D$ dimensions}  
\noindent Let us assume that matter in the universe is a perfect, pressureless fluid with density  $\rho_{\rm m}$ and  flow velocity ${\bf U}$.  We can formally write down the $D$ dimensional fluid equations describing a perfect fluid in an 
external potential field $\Phi_{\rm tot}$  in a proper coordinate system labelled by the $D$ dimensional vector ${\bf r}$. Therefore, we have,  
\begin{eqnarray}
\left(\frac{\partial\;\rho_{\rm m}}{\partial t}\right)_{\bf r}+\, \nabla_{\bf r}\cdot (\rho_{\rm m}\;{\bf U})&=& 0  \label{eqn:continuity} \\
\left(\frac{\partial \bf U}{\partial t}\right)_{\bf r}+ \, \left({\bf U} \cdot \nabla_{\bf r}\right) {\bf U} & = &-\nabla_{\bf r} \Phi_{\rm tot} \label{eqn:euler}
\end{eqnarray}
where equation~(\ref{eqn:continuity}) is the usual continuity equation while equation~(\ref{eqn:euler}) is the Euler equation for the fluid. The potential in equation~(\ref{eqn:euler}), $\Phi_{\rm tot}$, is the total external Newtonian potential 
\begin{equation}
\Phi_{\rm tot}= \Phi_{\rm FRW} + \varphi
\end{equation}
where $\Phi_{\rm FRW}$ is the background potential associated with the smooth background matter density $\rho_{\rm bm}$ and is given in  equation~(\ref{eqn:bgpotential}) 
while $\varphi$ is the potential caused by density perturbations $(\rho_{\rm m} - \rho_{\rm bm})$.  The potential $\varphi$ satisfies the Poisson equation given in  equation~(\ref{eqn:poisson4}).  Thus, 
\begin{equation}
\nabla_{\bf r}^2 \varphi = \left(\frac{D-2}{D-1}\right) \kappa(D)(\rho_{\rm m} - \rho_{\rm bm}) = \left(\frac{D-2}{D-1}\right)\kappa(D)\rho_{\rm bm}\delta \label{eqn:Poisson}
\end{equation}
where $\delta$ is the density contrast defined by 
\begin{equation}
\delta = \frac{\rho_{\rm m} - \rho_{\rm bm}}{\rho_{\rm bm}} \label{eqn:dcontrast}
\end{equation} 
We now transform to comoving coordinates defined by ${\bf x}={\bf r}/a(t)$ and define the peculiar velocity ${\bf v}$ by the relation
\begin{equation}
{\bf U} = H(t){\bf r} + {\bf v} = \dot{a}{\bf x} + {\bf v}
\end{equation}
where ${\bf v}= a\dot{{\bf x}}$ and $H(t)=(\dot{a}/{a})$.
Then, equation~(\ref{eqn:continuity}) and equation~(\ref{eqn:euler}) become  
\begin{eqnarray}
&&\left(\frac{\partial \rho_{\rm m}}{\partial t}\right)_{\bf x} + \, D H \rho_{\rm m}+\frac{1}{a}  \nabla_{\bf x}\cdot(\rho_{\rm m} {\bf v})= 0 \label{eqn:cocontinuity}\\
&&\left(\frac{\partial {\bf v}}{\partial t}\right)_{\bf x} + \, H {\bf v}+\frac{1}{a} \left( {\bf v} \cdot\nabla_{\bf x} \right) {\bf v}=-\frac{1}{a} \nabla_{\bf x} \varphi \label{eqn:coeuler}
\end{eqnarray}
where we have used equation~(\ref{eqn:bgpotential}) to substitute for $\Phi_{\rm FRW}$.
Similarly, in co-moving co-ordinates, equation~(\ref{eqn:Poisson}) reduces to 
\begin{equation}
\nabla_{\bf x}^2 \varphi =  \left(\frac{D-2}{D-1}\right)\kappa(D) a^2\rho_{\rm bm}\delta \label{eqn:coPoisson}
\end{equation}

\noindent Using $\rho_{\rm m}=\rho_{\rm bm} (1+\delta)$, transforming the time variable from $t$ to $a(t)$ and defining a new velocity variable ${\bf u}$ by 
\begin{equation}
{\bf u}= \frac{d{\bf x}}{da} = \frac{{\bf v}}{a\dot{a}}
\end{equation}
we can obtain equations for $\delta (a)$ and ${\bf u}(a)$.  Therefore, using equation~(\ref{eqn:densityrel}) and performing the transformations, equation~(\ref{eqn:cocontinuity}) and equation~(\ref{eqn:coeuler}) further reduce to, 
\begin{eqnarray}
&&\frac{\partial \delta}{\partial a}+\nabla_{\bf x}\cdot [{\bf u}(1+\delta)]=0 
\label{eqn:cocontinuity1} \\
&& \dot{a}^2 \frac{\partial {\bf u}}{\partial a} + \left(\ddot{a} + 2\frac{\dot{a}^2}{a} \right){\bf u} + \dot{a}^2 ({\bf u}\cdot \nabla_{\bf x}) {\bf u} = -\frac{1}{a^2} \nabla_{\bf x}\varphi \label{eqn:coeuler1}
\end{eqnarray}
Now, we use the Friedmann equations in equations~(\ref{eqn:friedmann1},\ref{eqn:friedmann}) with $\rho$ replaced by $\rho_{\rm bm}$ and with $p=0$ to substitute for $\ddot{a}$ in the above equation. Further, we define a new potential $\Psi$ by the relation
\begin{equation}
\Psi = \left(\frac{D(D-1)}{6-D}\right)  \frac{1}{\kappa(D) \rho_{\rm bm} a^3}\, \varphi
\end{equation} 
so that, upon using equation~(\ref{eqn:coPoisson}), one obtains, 
\begin{equation}
\nabla_{\bf x}^2 \Psi =  \left(\frac{D(D-2)}{6-D}\right) \frac{\delta}{a} \label{eqn:coPoisson1}
\end{equation}
where all reference to $\kappa(D)$ has disappeared.  Hence the final system of equations we need to tackle are, 
\begin{eqnarray}
&&\frac{\partial \delta}{\partial a}+\nabla_{\bf x}\cdot [{\bf u}(1+\delta)]=0 
\label{eqn:cocontinuity2} \\
&&\frac{\partial {\bf u}}{\partial a} + ({\bf u}\cdot \nabla_{\bf x}) {\bf u} = -\frac{6-D}{2a} A \left[ \nabla_{\bf x}\Psi + {\bf u}\right]
\label{eqn:coeuler2}
\end{eqnarray}
where $A$ is given by the relation
\begin{equation}
A = \left(\frac{2\kappa(D) }{D(D-1)}\right) \frac{a^2}{\dot{a}^2}\, \rho_{\rm bm} = \frac{\rho_{\rm bm}(t)}{\rho_c(t)} \qquad ; \qquad \rho_c \equiv \left({D(D-1) \over 2\kappa(D)}\right) {\dot{a}^2 \over a^2}
\end{equation}
For the $k=0$ universe, we will set $A=1$.  
\par
To proceed further and determine the equation satisfied by $\delta$, we decompose the term $\partial_{\alpha} u_{\beta}$ (where $u_{\beta}$ is the $\beta$th covariant component of the vector ${\bf u}$ and $\partial_\alpha$ is short for $\partial/\partial x^{\alpha}$) as
\begin{equation}
\partial_{\alpha} u_{\beta} = \sigma_{\alpha\beta} + \Omega_{\alpha\beta} + \frac{1}{D}\delta_{\alpha\beta}\theta \qquad \alpha,\beta = (1,2, \ldots, D)
\end{equation}
where $\sigma_{\alpha\beta}$ is the traceless, symmetric shear tensor,
$\Omega_{\alpha\beta}$ is the antisymmetric rotation tensor, $\theta$ is the
(trace) expansion and $\delta_{\alpha\beta}$ is the Kronecker delta symbol.
Then, equation~(\ref{eqn:cocontinuity2}) and equation~(\ref{eqn:coeuler2}) are combined by taking the divergence of equation~(\ref{eqn:coeuler2}) and using the above decomposition of $\partial_\alpha u_\beta$ to obtain a single equation for $\delta$.  Straightforward algebra gives,  
\begin{equation}
\frac{d^2 \delta}{d a^2}+ \left(\frac{6-D}{2 a}\right) \frac{d \delta}{d a}-
\left(\frac{D (D-2)}{2 a^2}\right) \delta(1+\delta)= \nonumber \\
 \left(\frac{D+1}{D}\right)\frac{1}{(1+\delta)} \left( \frac{d \delta}{d a}\right)^2 
 + \, (1+\delta) (\sigma^2-2\Omega^2) \label{eqn:ddelta}
\end{equation}
where $\sigma^2 = \sigma_{\alpha\beta}\sigma^{\alpha\beta}$ and $\Omega^2 = (1/2)\Omega_{\alpha\beta} \Omega^{\alpha\beta}$. This equation is the full non-linear equation for $\delta$. 
Apart from the obvious nonlinear terms
containing $\delta^2$ and $(d\delta/da)^2$, the term
$(1+\delta)(\sigma^2-2\Omega^2)$,  which is the contribution from the
shear and rotation, is also non-linear. The non-linear terms in $\delta$ in the above equation render the
equation unsolvable in general.  Ignoring these non-linear
terms to a first approximation, we can get a linear equation for $\delta (a)$,   \begin{equation}
\frac{d^2 \delta}{d a^2} + \left(\frac{6-D}{2 a}\right) \frac{d \delta}{d a} - \left(\frac{D (D-2)}{2 a^2}\right)\delta = 0.
\end{equation}
Assuming a power law solution for delta in the form $\delta \propto a^p$, we get, 
\begin{equation}
p = \frac{D-4}{4} \, \pm \, \frac{1}{4} \sqrt{ 9D^2 - 24D + 16}
\end{equation}
as the required values for $p$. Notice that $\delta$ has a growing
mode as well as a decaying mode in general.  The above solutions hold
for all values of $D>1$ in the linear regime.  
\par
Though the full non-linear equation is not solvable, by neglecting the contribution from the shear and rotation terms and  by using a suitable ansatz for $\delta$, the resulting non-linear equation {\it can} be solved.  We proceed to do this within the framework of the STH model in the next section.    

\subsection{The Spherical Top Hat (STH) model}
\noindent In the STH (spherical collapse) model, we assume spherical symmetry by neglecting the shear and rotation terms in the equation for $\delta$. With this assumption the $\delta$ equation can be exactly solved.
\par   
Transforming equation~(\ref{eqn:ddelta}) by changing the independent variable back to $t$, dropping the rotation and shear terms and using the Friedmann equations given in equations~(\ref{eqn:friedmann1},\ref{eqn:friedmann}), we get,
\begin{equation}
\frac{d^2 \delta}{d t^2}+ 2 \frac{\dot{a}}{a} \frac{d \delta}{d t} -\left(\frac{D+1}{D}\right)\frac{1}{(1+\delta)}\left(\frac{d \delta}{d t}\right)^2=\left(\frac{D-2}{D-1}\right) \kappa(D)\rho_{\rm bm} \delta(1+\delta) \label{eqn:deltatime}
\end{equation}  
We now define a function $R(t)$ by the relation
\begin{equation}
1+\delta=\frac{\rho}{\rho_{\rm bm}}=\frac{M}{C_D R^D (t) \rho_{\rm bm}} \label{eqn:dansatz}
\end{equation}
where  $C_D=2 \pi^{D/2}/(D \Gamma[D/2])$ is the volume of a unit sphere in $D$ dimensions introduced for later convenience and $M$ is a constant.  
The expression for $\delta$ above can be rewritten using the relation $\rho_{\rm bm}a^D = \rho_0 a_0^D$ from equation~(\ref{eqn:densityrel}):
\begin{equation}
1+\delta=\frac{M}{C_D \rho_0 a_0^D} \left[\frac{a}{R}\right]^D\\
=\lambda \frac{a^D}{R^D} \label{eqn:dansatz1}
\end{equation}
where $\rho_0$ and $a_0$ are the matter density and scale factor at
some (arbitrarily chosen) ``present" epoch $t_0$.
Substituting equation~(\ref{eqn:dansatz1}) in equation~(\ref{eqn:deltatime}), we get an equation for the growth of $R(t)$ as,
\begin{equation}
\frac{d^2R}{dt^2} = -\frac{D-2}{D(D-1)} \frac{\kappa(D)}{C_D}\frac{M}{R^{D-1}}
\end{equation}
[As an aside we may note that if the universe contains matter or fields
with equations of state other than $p=0$, the equation for $R(t)$ becomes
\begin{equation}
\frac{d^2R}{dt^2} = -\frac{D-2}{D(D-1)} \frac{\kappa(D)}{C_D}\frac{M}{R^{D-1}} \; - \; \frac{\kappa(D)}{D(D-1)}\left( (D-2)\rho + Dp\right)_{\rm rest} R
\end{equation}
where the term $((D-2) \rho+Dp)_{rest}$ comes from the smoothly distributed
component with $p\neq 0$.]
\par
From the form of the equation of motion of $R(t)$ we can give the following interpretation. Since the entire system considered above is spherically symmetric, we interpret $R$ as the radius of a $D$ dimensional spherical region containing a mass $M$. The equation of motion of $R$ determines the motion of the surface of this region. In general, a spherical overdense region will be expected to initially expand because of the expansion of the background universe till the excess gravitational force due to the overdensity of enclosed matter stops the expansion and causes the region to collapse back on itself. We will discuss the cases $D=3$ and $D=2$ in the subsequent sections and determine the differences in the behaviour of the growth of inhomogeneities.   

\section{Summary of standard results in three dimensions} \label{sec:3D}
\noindent When $D=3$, all the standard equations are recovered.  First the Poisson equation satisfied by the Newtonian gravitational potential given by equation(\ref{eqn:poisson4}) reduces to the standard form,
\begin{equation}
\nabla^2 \phi = 4\pi G \rho \label{eqn:3Dpoisson}
\end{equation}
where we have defined $G$ by relating it to $\kappa(3)$ by  $\kappa(3)
= 8\pi G$. Similarly, equations (\ref{eqn:cocontinuity2}) and
equation (\ref{eqn:coeuler2}) reduce to (with $A=1$), 
\begin{eqnarray}
&&\frac{\partial \delta}{\partial a}+\nabla_{\bf x}\cdot [{\bf u}(1+\delta)]=0 
\label{eqn:3Dcocontinuity2} \\
&&\frac{\partial {\bf u}}{\partial a} + ({\bf u}\cdot \nabla_{\bf x}){\bf u} = -\frac{3}{2a}  \left[ \nabla_{\bf x}\Psi + {\bf u}\right]
\label{eqn:3Dcoeuler2}
\end{eqnarray}
while the equation for $\Psi$ becomes,
\begin{equation}
\nabla_{\bf x}^2 \Psi = \frac{\delta}{a}. \label{eqn:3DcoPoisson1}
\end{equation}
In a similar manner, the $\delta$ equation reduces to,
\begin{equation}
\frac{d^2 \delta}{d a^2}+\frac{3}{2 a} \frac{d \delta}{d a}-
\frac{3}{2 a^2}\delta(1+\delta)=\frac{4}{3 (1+\delta)} \left( \frac{d \delta}{d a}\right)^2+(1+\delta) (\sigma^2-2\Omega^2) \label{eqn:3Dddelta}
\end{equation}
and the solutions to the linear perturbation equation which is
obtained by dropping the nonlinear terms and the
$(\sigma^2-2\Omega^2)$ term, are
\begin{equation}
\delta \propto a^p,  \quad p = 1, -\frac{3}{2}
\end{equation}
which are well known.
The STH model for $D=3$ also reduces to the standard form
\begin{equation}
\frac{d^2R}{dt^2} = -\frac{G M}{R^2} \; - \; \frac{4\pi G}{3}\left(\rho + 3p\right)_{\rm rest} R
\end{equation}
which is again well known \cite{Probbook}.
Therefore, it is seen that the full $D$ dimensional equations reduce to the correct equations in three dimensions.  Now we will go on to discuss the important case of $D=2$.

\section{Two dimensional gravity} \label{sec:2D}
\noindent If we naively consider the limit $D \to 2$ in the $D$ dimensional equations,
assuming that $\kappa(D)$ is finite in this limit, we obtain the following
results.  First, the Poisson equation (\ref{eqn:poisson4}) reduces to
\begin{equation}
\nabla^2 \phi = 0
\label{eqn:2Dpoisson1}
\end{equation}
The above result shows that in two dimensions, the
gravitational potential does {\it not} couple to the matter density $\rho$.  In structure formation, this means that inhomogeneities cannot grow since the perturbed potential $\varphi$ is not related to  $\delta$ at all. 
The second interesting  result is that the background Newtonian potential $\Phi_{\rm FRW}$ vanishes. This occurs because, referring back to equation~(\ref{eqn:friedmann2}), $\ddot{a} = 0$ for pressureless dust and hence the background potential is zero. 
Further, the $\delta$ equation reduces to 
 \begin{equation}
\frac{d^2 \delta}{d a^2}+\frac{2}{a} \frac{d \delta}{d a}=\frac{3}{2 (1+\delta)} \left( \frac{d \delta}{d a}\right)^2+(1+\delta) (\sigma^2-2\Omega^2). \label{eqn:2Dddeltalin}
\end{equation} 
Linearising the equation as before by dropping the
$(\sigma^2-2\Omega^2)$ and $(d\delta/da)^2$ terms we obtain
\begin{equation}
\frac{d^2 \delta}{d a^2}+\frac{2}{a} \frac{d \delta}{d a}=0. \label{eqn:2Dddelta}
\end{equation} 
 The solutions to the linearised equation are 
\begin{equation}
\delta \propto a^p, \quad p =0, -1.
\end{equation}
Thus only a constant or the decaying mode is present.  This is
consistent  with the result that the perturbed gravitational potential does not couple to $\delta$. If one considers the STH model, it is easy to see that the growth equation for $R(t)$ reduces to 
\begin{equation}
\frac{d^2R}{dt^2} = -\kappa(2) p_{\rm rest} R
\end{equation}
For, $p_{\rm rest} = 0$, the solution to the above equation is just
$R(t) = B_1 t + B_2$ where $B_1, B_2$ are constants. This is to be
expected since there is no gravitational force which can lead to clustering
and as a consequence the radius simply grows with time just like the
background universe.
Thus, if $\kappa(D)$ is {\it finite} in the limit $D\to 2$,
it is not possible to have gravitational clustering
that can grow with time. 
\par
We can, however, try some alternative approaches to examine whether it is possible to
have a consistent physical picture of growing structures for two
dimensional gravity. 
\par
One possibility is that instead of assuming $\kappa(D)$ to be finite, let us assume that the expression $(\kappa(D) (D-2))$ remains finite when $D \to 2$. This finite value can be fixed, for example, by invoking Gauss's theorem in $D$ dimensions. This gives
\begin{equation}
\kappa (D)=\left(\frac{D-1}{D-2}\right)\frac{2 \pi^{D/2} G}{\Gamma [D/2]} \label{eqn:poisson5}
\end{equation}
Thus, $\kappa(D) \to \infty$ when $D\to 2$, but the Poisson equation acquires the form
\begin{equation}
\nabla^2 \phi = 2 \pi G \rho \label{eqn:2Dpoisson2}
\end{equation}
This is, of course, the same form which is obtained by applying Gauss's law in two dimensions. Hence, as in three dimensions, the gravitational potential is determined by the matter density and thus inhomogeneities can in
principle grow.
There are, however, difficulties with this approach. To begin
with, the constant factor $c^4/(2 \kappa (D))$ in the action
$\cal S$ in equation(\ref{eqn:action1}) vanishes for $D=2$.
But this is not too serious a problem.  The gravitational part of the action certainly vanishes but because only the variations about the action are of significance this difficulty can be ignored. But a more serious problem arises when the solutions to the Friedmann equation are considered.  
The solutions for $a(t)$ and $\rho(t)$ in $D$ dimensions are given in
equation~(\ref{eqn:friedsoln}). Using equation~(\ref{eqn:poisson5}), these reduce to, 
\begin{equation}  
a(t) = \left({ D\pi^{D/2} G C_1 \over (D-2) \Gamma[D/2] }\right)^{1/D}  t^{2/D}
\quad ; \quad \rho(t) = C_1 a^{-D} = { (D-2) \Gamma[D/2] \over D
\pi^{D/2} G} \, t^{-2} 
\end{equation}
When $D\to 2$ then,  $a \to \infty$ and $\rho \to 0$ irrespective of the dependence on $t$.  This implies that
one cannot solve the equations describing the growth of structure in a
consistent and non--singular way. 
Hence, we conclude that it is not possible to have a theoretical
formulation of two dimensional gravity as the Newtonian limit to
Einstein's equations in two dimensions. 
\par
An alternative that remains is to use the Newtonian fluid equations
in $D$ dimensions directly and rewrite them for an expanding
background with an {\it arbitrary}
scale factor $a(t)$.  Note that $a(t)$ is not obtained from the Friedmann equations and is completely arbitrary.  We can superpose the potentials for the
background universe and the perturbations in this case as before. The further assumptions we need to make are 
(i)~the potential of the background
universe $\Phi_{\rm bg}$ is of the form 
\begin{equation}
\Phi_{\rm bg} = -\frac{1}{2} \frac{\ddot{a}}{a}\, r^2
\end{equation} 
and (ii)~the Poisson equation is given by 
\begin{equation}
\nabla^2 \phi=\kappa(D) \rho_{\rm bm} \delta
\end{equation}
where $\kappa(D)=2 \pi^{D/2} G/\Gamma [D/2]$. This form of $\kappa(D)$ is
obtained from the use of Gauss's law in $D$ dimensions.  We also need to
specify how the background density $\rho_{\rm bm}$ depends on  time. In analogy with the usual Friedmann equations, we will assume $\rho_{\rm bm} a^D=C_1$ where $C_1$ is a constant.
This gives an equation for $\delta$ with an arbitrary scale factor
$a(t)$ as
\begin{equation}
\frac{d^2 \delta}{d a^2}+\left(\frac{\ddot{a} a+2 \dot{a}^2}{a \dot{a}^2} \right) 
\frac{d \delta}{d a}-\kappa(D) C_1 \frac{1}{a^D \dot{a}^2} \delta
(1+\delta)= \nonumber \\
\left(\frac{D+1}{D}\right) \frac{1}{1+\delta} \left(\frac{d \delta}{d
a}\right)^2+(1+\delta) (\sigma^2-2\Omega^2) .\label{eqn:deltaD}
\end{equation}
The above equation can be solved in any dimension $D$ if the form of $a(t)$ is given. This gives us a non-singular way to analyse growth of structures in $D$ dimensions, including the case $D=2$. But
in three dimensions we observe that the above equation does not correctly
reduce to equation~(\ref{eqn:3Dddelta}). We may obtain the correct equation in three dimensions by making an additional ansatz, namely, that \begin{equation}
\left(\frac{\dot{a}}{a}\right)^2= \frac{2\kappa(D)}{D} \rho_{\rm bm}.
\end{equation} 
With this, equation~(\ref{eqn:deltaD}) reduces to
\begin{equation}
\frac{d^2 \delta}{da^2}+\left(\frac{6-D}{2a}\right)\frac{d \delta}{d a}-\frac{D}{2
a^2} \delta(1+\delta)=\nonumber \\
\left(\frac{D+1}{D}\right)\frac{1}{1+\delta} \left(\frac{d \delta}{da}\right)^2+(1+\delta) (\sigma^2-2\Omega^2) \label{eqn:ddeltaD}
\end{equation}
Notice that the above equation differs from the earlier equation for $\delta$ in $D$ dimensions, equation~(\ref{eqn:ddelta}), in that there is a factor of $(D-2)$ missing from the coefficient of the $\delta(1+\delta)$ term. Therefore, the above equation does correctly reduce to equation~(\ref{eqn:3Dddelta}) since $(D-2)$ equals unity when $D=3$. When $D=2$, equation~(\ref{eqn:ddeltaD}) gives
\begin{equation}
\frac{d^2 \delta}{da^2}+\frac{2}{a}\frac{d \delta}{d
a}-\frac{1}{a^2}\delta (1+\delta)=\frac{3}{2 (1+\delta)}\left( \frac{d
\delta}{d a}\right)^2+(1+\delta) (\sigma^2-2\Omega^2) \label{eqn:ddelta2}
\end{equation}
On linearising this equation by dropping the $(\sigma^2-2\Omega^2)$
term and the nonlinear terms and solving it we get both a growing mode
as well as a decaying mode for $\delta$.  
The solutions are
\begin{equation}
\delta \propto a^q, \qquad q=(-1\pm\sqrt{5})/2
\end{equation}
(It is interesting to note that one of the power law exponents is the golden ratio).
\par
The Spherical Top Hat (STH) equation in this case turns out to be
\begin{equation}
\ddot{R}=-\frac{GM}{R}+\frac{1}{2} \frac{R}{t^2} \label{eqn:sth2d}
\end{equation}
where $M$ is the constant mass inside a `spherical' shell of radius $R$. 
This equation, unfortunately, has no simple analytic solution.

While this procedure leads to nontrivial results, it has many {\it ad hoc} assumptions and cannot be obtained by taking appropriate limits of Einstein's theory in a systematic manner. Consequently it cannot be applied to numerical
investigations of $2$ dimensional gravity with the confidence that the results will have some implications for the three dimensional case.   

\section{Results and Summary} \label{sec:conclusions}
In this chapter we have analysed the case of two dimensional gravitational
clustering  starting from a formulation of the $D$ dimensional
Einstein's equations and taking the proper limits. The system of
equations thus arrived at for a $(D+1)$ dimensional universe has been
shown to reduce to the correct equations in three dimensions. But when the
$D\to 2$ limit of these equations is taken, we are forced to  conclude that irrespective of the value of $\kappa (D)$, a consistent two dimensional gravity theory in a cosmological context that supports growth of structures cannot
be constructed.
\par
If $\kappa(2)$ is assumed to be finite,  we observe that  the coefficient in Poisson's
equation goes to zero thus decoupling the potential from the
density.  This implies that perturbations do not
grow but decay in time due to the expansion of the background spacetime.
The alternative which is obtained by using the expression for $\kappa(D)$
given by equation~(\ref{eqn:poisson5}) gives rise to solutions for the
scale factor which are singular and therefore unacceptable.
\par
We have discussed all the ways in which two dimensional gravity may be simulated
including an {\it ad hoc} procedure without a strong foundation which
can give non--singular results as far as structure formation scenarios
in two dimensions are concerned. The results presented here leads us to conclude that the  only way to do a numerical simulation of two dimensional
gravity  is to simulate infinite `needles' in a background
spacetime expanding in three dimensions and consider the `particles'
in the  system to be intersections of the `needles' with any plane
orthogonal to them which is the approach taken in the next chapter.

\chapter{Scaling Relations in Two Dimensions}
\label{chap:twodsim}
\baselineskip=12pt
{\scriptsize When in doubt use brute force -- Ken Thompson}
\par
\baselineskip=24pt
\par
\section{Gravitational Clustering: Two vs Three Dimensions}
 A number of attempts have
been made in recent years to understand the evolution of constructs
like the two point correlation function using certain non-linear
scaling relations (NSR) as discussed in chapter 2 (\cite{Ham}),\cite{RajPad94},\cite{TPMNRAS}).  These studies have shown that the
relation between the non-linear and the linearly
extrapolated correlation functions is reasonably model independent.
This relation divides the evolution of correlation function into three
parts \cite{apmpap}: the linear regime, the intermediate regime and the 
non-linear regime.  The evolution in the intermediate regime can be
understood in terms of radial collapse around density peaks
\cite{TPMNRAS}, if it is assumed that the evolution of profiles of
density peaks follows the same pattern as an isolated peak.  It is
customary to invoke the hypothesis of stable clustering \cite{Peeb80}
to model the non-linear regime.  A large number of studies have
examined clustering in this regime and the general consensus is
that the stable clustering limit does not exist \cite{PadOst}.

However, the limited dynamic range of currently available
3-dimensional N-Body simulations poses serious difficulties in
investigating this problem in greater detail.  It was pointed out
\cite{tpdonald} that we can circumvent this problem by simulating a
two dimensional system, wherein a much higher dynamic range can be
achieved. For example, since $160^3\approx 2048^2$, the computational
requirements are the same for a 2D simulation with box size of 2048
and 3D simulation with box size of 160. Assuming that one can reliably
use, say, half of box size as {\it good} dynamic range we have a
dynamic range of factor 1000 in 2D against a factor of about 80 in
3D. This allows us to probe higher nonlinearities in 2D compared to
3D. As long as we stick to generic features (like the non-linear
scaling relations investigated here) which are independent of
dimension, 2D has a definite advantage over 3D.  Higher dynamic range
is the basic motivation for studying gravitational clustering in two
dimensions. 

When we go from three to two dimensions, we have, as discussed in chapter 3, two
different ways of modeling the system:
\par
(i) We can consider two dimensional perturbations in a three
dimensional expanding universe. Here we keep the force between
particles to be $1/r^2$ and assume that all the particles, and their
velocities,  are confined to a single plane at the initial
instant. 

(ii) We can study perturbations that do not depend on  one of the
three coordinates, i.e., we start with a set of infinitely
long straight ``needles'' all pointing along one axis. The
force of interaction falls as $1/r$. The evolution keeps the
``needles'' pointed in the same direction and we study the clustering
in an orthogonal plane.  Particles in the N-Body simulation represent
the intersection of these ``needles'' with this plane. 
In both these approaches the universe is three dimensional and the
background is expanding isotropically. 

The study of 2-D perturbations (like those due to pancakes, for example)
in a 3-D expanding universe faces an operational problem: To begin
with, we do not gain the dynamic range if we stick to 3D, even if we
consider perturbations in a plane; the force between particles still
has to computed by the solution of Poisson equation in three dimensions. 
Also, relevance of the interaction of matter outside the plane
with these perturbations makes it, essentially, a 3D problem.

Thus we are left with the second possibility. The two dimensional
system is the intersection of an orthogonal plane and the ``needles''
and the force between the  ``particles''  in this plane is given by
the solution of the Poisson equation in two dimensions.  Such 
a system is somewhat dichotomous with the background universe
expanding isotropically.  However, convenience is not the only reason
for studying this somewhat strange system --- relevant results for the
evolution of density profiles around peaks in 2-D have also been
computed for this type of a system \cite{fg_selfsim}.  
\par
Generalization of the NSR to the 2-D system was done using
relations for cylindrical collapse by Padmanabhan \cite{TPMNRAS} and we will
test these predictions here.
\par
Although the system of infinite needles is appropriate for testing the
predictions in the intermediate regime, the same cannot be said for
the asymptotic regime.  We are dealing with a system that occupies a
smaller number of dimensions in the phase space {\it and} the
interaction of the  constituents follows a different force law.
Therefore, it is difficult to interpret, or carry over, results
regarding stable clustering to the full 3-D system. 

\subsection{Non-linear Scaling Relations}

The non-linear and the linear correlation functions at two different
scales can be related by NSR.  The relation between these scales is
given by the characteristics of the pair conservation equation
\cite{RajPad94}.  For the two dimensional system of interest, this
equation can be written as \cite{tpdonald} 
\begin{equation}
{\partial D \over \partial A} -h (A, x) {\partial D \over \partial X}
= 2h (A, X), 
\end{equation}
Here $D=\log(1 + \bar\xi)$, $h = - v_p/Hr$ is the scaled pair
velocity, $\bar\xi(x)=2 x^{-2} \int^x r \xi(r) dr$ is the mean
correlation function ($\xi$ is the 
correlation function),  $H$ is the Hubble's constant, $X=\log(x)$ and
$A=\log(a)$.  The characteristics
of this equation are $x^2 (1 + \bar \xi(x,a)) = l^2 $, where $x$ and
$l$ are the two scales used in NSR.  The self similar models due to
Filmore and Goldreich \cite{fg_selfsim} imply that for collapse of cylindrical
perturbations the turn around radius and the initial density contrast
inside that shell are related as $x_{\rm ta} \propto
l/\bar\delta_i \propto l/\bar\xi_L (l)$.  (Here $\bar\xi_L$ is the
linearly extrapolated mean correlation function).  Noting that in
two dimensions $M \propto x^2$, we find $\bar\xi (x) \propto
\left[ \bar\xi_L (l) \right]^2$ in the regime dominated by infall. 
Stable clustering limit implies $\bar\xi_{NL}(a, x)
\propto \bar\xi_L(a, l)$ \cite{tpdonald}.  Thus in 2-D the scaling
relations are
\begin{equation}
\bar \xi (a,x) \propto \left\{ \hbox{  }
\begin{array}{ll} 
\bar \xi_L (a,l) & \hbox{ (Linear)} \\
\bar \xi_L(a,l)^2  & \hbox{ (Radial Infall)} \\
\bar \xi_L(a,l) &  \hbox{ (Stable Clustering)} \\
\end{array} \right.
\label{hamilton}
\end{equation}
A more general assumption compared to stable clustering involves
taking $h=$~constant asymptotically. In a system reaching steady state with
both virialisation and mergers contributing to the evolution, one may reach
a constant value for $h$, though it will not be unity if mergers are a
dominant phenomenon. (This assumption has been discussed in, for example, 
\cite{TPMNRAS}.)
It also allows a larger parameter space to compare simulation results.
If $h$=constant asymptotically, then $\bar\xi(x) \propto {\bar\xi_L}^h(l)$ 
in this limit.  Note that in 3D, the indices for three regimes are 
$1$, $3$ and $3h/2$ respectively.

All features of clustering in three dimensions are present
here as well. In particular,
\par
(i) If the asymptotic value of $h$ scales with $n$ such that $h (n +
2 )= {\rm constant}$ then the final slope of the non-linear correlation
function will be independent of the initial slope.
\par
(ii) If NSR exists then it will predict a {\it specific} index in the
intermediate and asymptotic regimes which will depend on the initial
power spectrum.  In other words, existence of NSR implies that
gravitational clustering does not erase memory of initial conditions.
\par
(iii) It is, however, possible that spectra which are not scale free
acquire universal critical indices at which the correlation
functions grow in a `shape invariant' manner. This comes about because
the growth rate of correlation function varies with the local index
and for an index that is not globally constant the correlation functions
may `straighten out' by this process.
\par
(iv) In 3-D clustering, $n=-1$ in the intermediate regime and $n=-2$
in the asymptotic regime \cite{critindex} are the critical indices.
These are the same for clustering in two dimensions. 

\section{Simulations and Results}

We carried out a series of numerical experiments to test the ideas
outlined above.  We used a particle mesh code \cite{nbodyrev} to
simulate power law models.  The simulations were done with $1024^2$ or
$2048^2$ particles in order to ensure that we had sufficient dynamic
range to study all the three regimes in evolution of non-linear
clustering.  In particular, it is necessary to use larger simulations
for power law spectra with a negative index.  Here, we will present
results for three models: $n=1$, $n=0$ and $n=-0.4$. 

All the models are normalized by requiring the linearly extrapolated
root mean square fluctuations in density, computed using a Gaussian
filter, to be unity at a scale of $10$ grid points at $a=1.0$.
The results we present are for $a=1$, $2$ and $5$ for $n=0$ and $n=1$,
and $a=1$, $2$ and $3$ for $n=-0.4$.

A significant source of errors in large simulations is the addition of
a small displacement in each step (fraction of a grid length) to a
large position (up to 2048 grid lengths).  We avoid this problem by
using net displacement for internal storage.  

We will show the correlation function and the pair velocity only for
length scales larger than four grid lengths.  We do this to avoid
error due to shot noise and other artifacts introduced by various
effects at smaller scales.  This ensures that errors in our results
are acceptably small.  (Variations between different realizations give
a dispersion of less than $10\%$ in the correlation function.)

\begin{figure}[htbp]
\label{abcd}
\epsfxsize=5truein
\epsfbox[40 28 502 544]{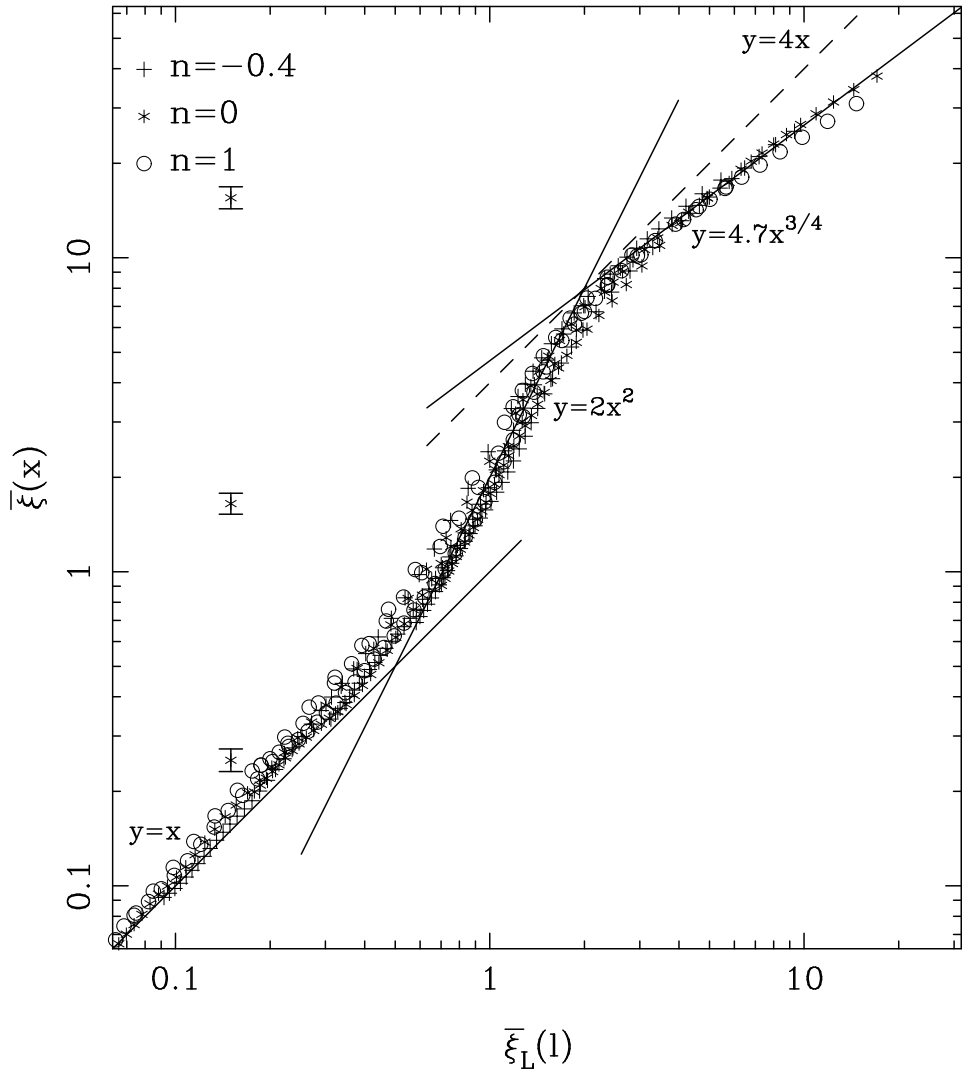}
\caption[{\sf Two dimensional scaling relation for power law spectra}]{\sf This figure shows the non-linear correlation function
$\bar\xi(x)$ as a function of the linearly extrapolated correlation
function $\bar\xi_L(l)$.  Here the scales $x$ and $l$ are related by
$x^2 (1 + \bar \xi) = l^2 $.  Data for $n=1$ is represented by
circles, that for 
$n=0$ by stars and $+$ marks the points for $n=-0.4$.  For each of
these models we have plotted data for the three epochs mentioned in
the text.  The estimated 2 $\sigma$ error bars are shown as vertical lines at  three representative values of $\bar\xi$ {\it viz.} at $\bar\xi$=15.582, 1.65 and  0.25, covering the nonlinear, intermediate and linear regimes. The error bars are shown away from the NSR plot for the sake of visibility. It is clear
from this figure that there are no systematic differences between
the three models and they trace out a simple curve with three distinct
slopes.  The slope of the curve in the intermediate regime is same as
that predicted by the radial infall model.  The stable clustering
limit is shown as the dashed line and it is clear that the data points
deviate from this curve.}
\end{figure}

In fig.\ref{abcd} we have plotted the non-linear correlation function
$\bar\xi(x)$ as a function of the linearly extrapolated correlation
function $\bar\xi_L(l)$.  Here the scales $x$ and $l$ are related by
$x^2 (1 + \bar \xi) = l^2 $.  Data for $n=1$ is represented by
circles, that for $n=0$ by stars and `$+$' marks the points for
$n=-0.4$.  Clearly, there are no systematic differences between
the three models and the data points trace out a simple curve with
three distinct slopes (We have also marked the 2 $\sigma$ errors
calculated by averaging over several data sets. The error bars  are plotted 
away from the NSR plot, for visibility and clarity.). The NSR, shown as thick lines, is
\begin{equation}
\bar \xi (a,x) = \left\{ \hbox{  }
\begin{array}{ll} 
\bar \xi_L (a,l) & \hbox{$\bar\xi_L(l) \leq 0.5$; $\bar\xi(x)\leq 0.5$}\\
 2 {\bar\xi_L(a,l)}^2  & \hbox{$0.5 \leq \bar\xi_L(l) \leq 2$; $0.5 \leq
\bar\xi(x)\leq 8 $} \\
 4.7 {\bar\xi_L(a,l)}^{3/4} &  \hbox{$2 \leq \bar\xi_L(l)$; $8 \leq 
\bar\xi(x)$ } \\
\end{array} \right.
\label{ourfit}
\end{equation}
The slope in the intermediate regime is as expected.  The asymptotic
regime has a different slope than that predicted by stable clustering,
which is shown as a dashed line.
Unlike the observed relations for clustering in three dimensions, the
coefficient for the intermediate regime is large.  This has
important implications for the critical index.

\begin{figure}[hp]
\label{data}
\epsfxsize=5.5truein
\epsfbox{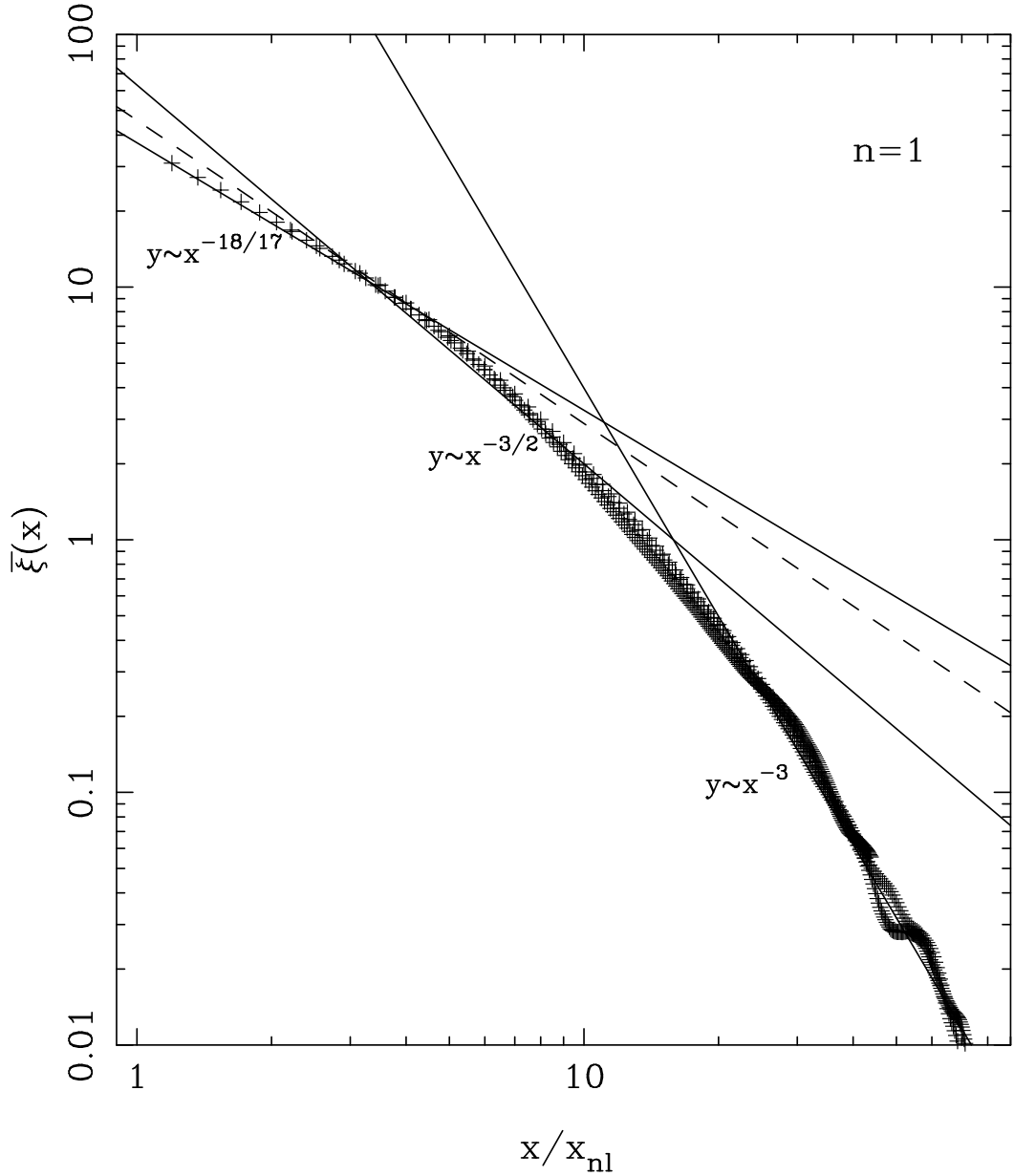}
\caption{\sf The correlation function $\bar\xi(x)$ as a
function of $x/x_{nl}$ for $n=1$ model.  Here $x_{nl} \propto
a^{-2/(n+2)}$.  Thick lines mark slopes
expected from the non-linear scaling relations shown in fig. 4.1.  The
dashed line marks the expected slope of the correlation function in
the stable clustering limit.  The mismatch between the expected slope
and the true slope in the intermediate regime may arise from the fact
that the assumption of $\bar\xi \gg 1$ used in computing the slope is
not valid at the lower end of the regime.}
\end{figure}

Panels of fig. \ref{data} show $\bar\xi(x)$ as a function of $x/x_{nl}$ for
the three models.  These confirm that the slope of $\bar\xi(x)$
is consistent with the NSR shown in fig. 4.1.  In each of these
panels, the slope expected in the stable clustering limit is shown as
a dashed line.

As mentioned above, the existence of the NSR (eqn.(\ref{ourfit}))
implies that the slope of the correlation function will depend on the
initial  spectral index. To this extent, gravitational clustering does
not erase memory of initial conditions.  However, the differences of
slope are significantly reduced by non-linear evolution.

\setcounter{figure}{1}

\begin{figure}[htbp]
\epsfxsize=5.5truein\epsfbox[40 24 502 562]{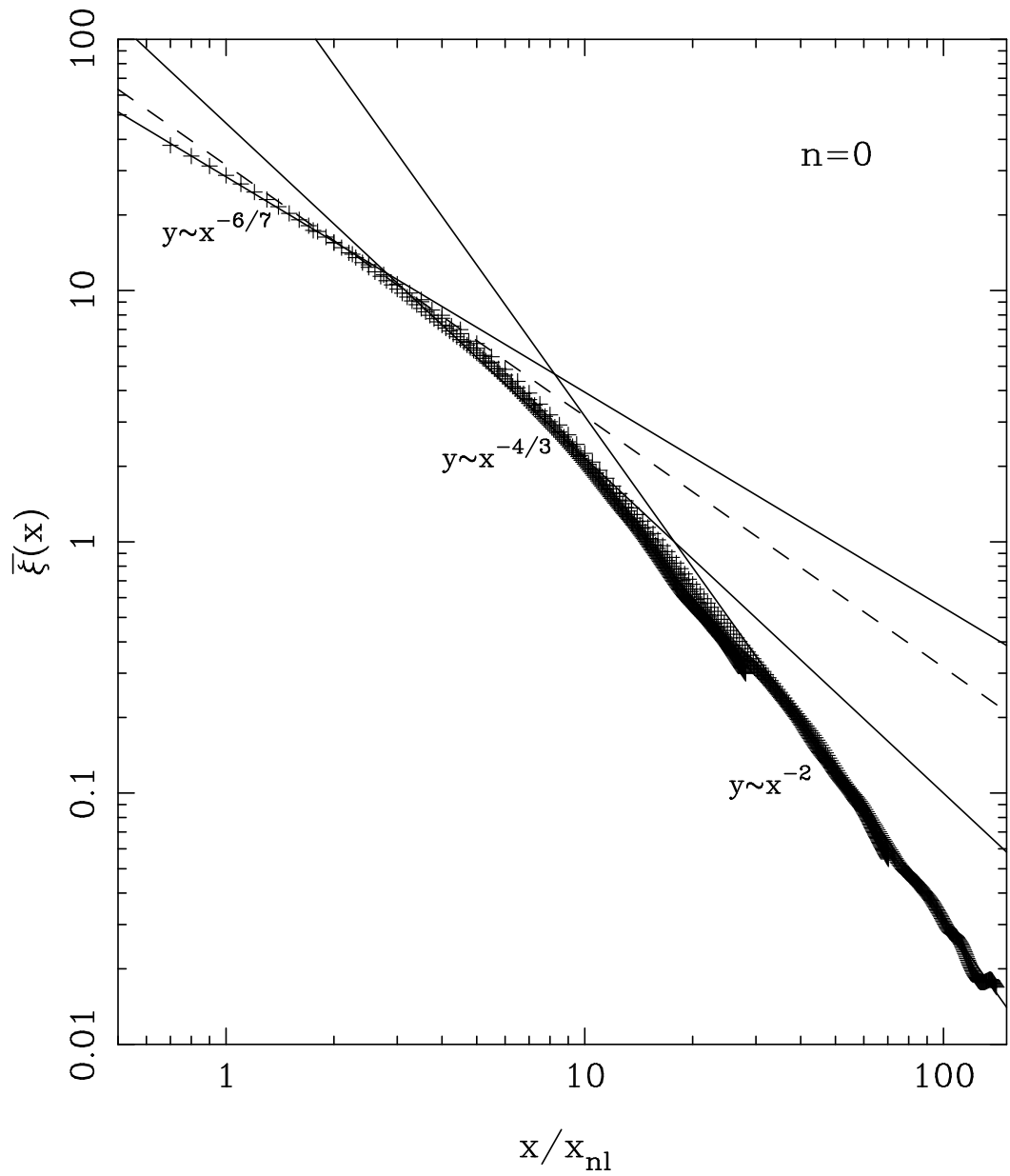}
\caption{\sf Similar plot for $n=0$.}
\end{figure}
\setcounter{figure}{1}

\begin{figure}[htbp]
\epsfxsize=5.5truein\epsfbox[40 24 502 548]{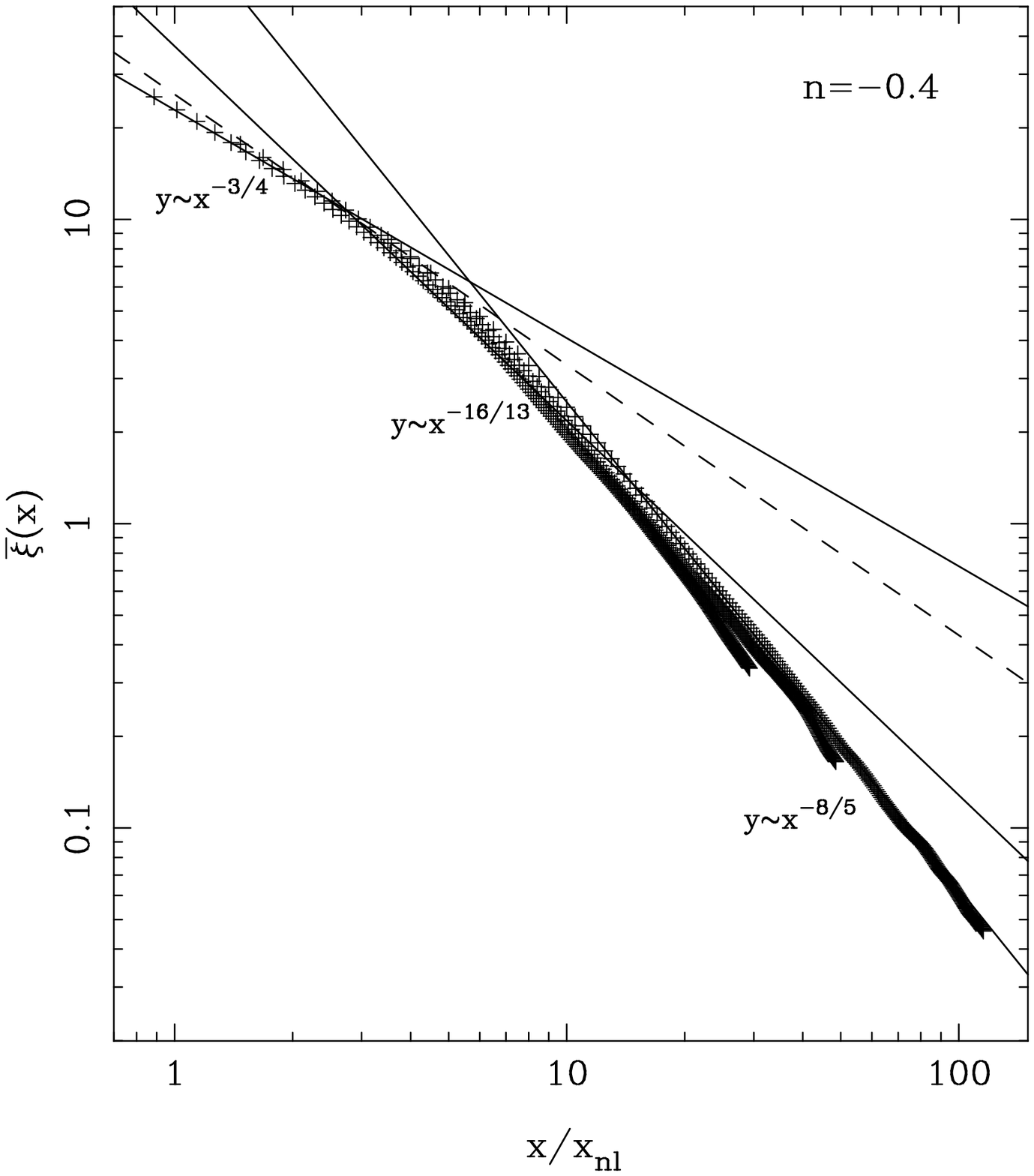}
\caption{\sf  $n=-0.4$.}
\end{figure}
\section{Summary}

Results obtained in this chapter can be summarized as follows:
\par
(i) We have verified  that NSR for the correlation function exist
for clustering in 2D in all the three regimes, just like in 3D. This
NSR  is independent of the power law index -- at least for the three
indices studied here.
\par
(ii) In the intermediate regime, the NSR in the form of
eqn.(\ref{ourfit}), can be understood in terms of radial infall
around peaks. Our simulations verify the predictions \cite{tpdonald}
for this regime.  
\par
(iii) In the asymptotic regime, our results do {\it not} agree with
the stable clustering hypothesis.  The slope of the NSR in the
asymptotic regime in fig. 4.1 implies $h=$ constant.  We find that, in
this regime, $h \simeq 3/4$ for all the models studied here.
\par
(iv) The existence of NSR implies that the asymptotic slope of the
correlation function depends on the initial slope.  However, this is
strictly true only for pure power law models; for other models it is
possible for the spectra to be driven to a universal form.  
The NSR in the asymptotic regime seems to be linked to the logarithmic
nature of the potential. 
\par
A theoretical model for existence of such scaling relations, as mentioned before, was derived on the basis of Spherical Collapse Model. In the next chapter we analyse the physical features of spherical collapse model in detail and derive a physically motivated approach to halting the collapse as opposed to the stardard `virialization' argument.

\chapter{An improved spherical collapse
model}
\label{chap:spherical}
\baselineskip=12pt
{\scriptsize Turbulence is life force. It is opportunity. Let's love turbulence \newline and use it for change -- Ramsay Clark }
\par
\baselineskip=24pt





\section{Introduction}
\label{sec:Intro}
Analytic modelling of the non--linear phase of gravitational clustering
has been a challenging but interesting problem upon which a considerable
amount of attention has been bestowed in recent years. The simplest, yet
remarkably successful, model for non-linear evolution is the 
 Spherical Collapse Model (SCM, hereafter), which has been  applied in
the study  of  various empirical results in the gravitational instability 
paradigm.  Unfortunately, this approach has  serious flaws --- both 
mathematically and conceptually. Mathematically, the SCM has a singular 
behaviour at finite time and predicts infinite density contrasts for all 
collapsed objects. Conceptually, it is not advisable to model the real 
universe as a sphere, in spite of the standard temptations to which 
theoreticians often succumb. The two issues are, of course, quite 
related, since, in any realistic situation, it is the deviations from spherical 
symmetry which lead to virialised stable structures getting formed. 
In conventional approaches, this is achieved by an {\it ad hoc\/} method
which involves halting the collapse at the virial radius by hand and mapping the
resulting non-linear and linear overdensities to each other. This leads to
the well known rule-of-thumb  that, when the linear overdensity is about 
$1.68$, bound structures with non-linear overdensities of about $178$ would 
have formed. The singular behaviour, however, makes the actual trajectory
of a spherical system quite useless after the turnaround phase ---
a price we pay for the arbitrary procedure used in stabilizing the system. 
But the truly surprising feature is that, despite its inherent arbitrariness, 
the SCM, when properly interpreted, seems to give useful insights into the behaviour 
of real systems. The Press--Schechter formalism \cite{Press}, for 
the abundance of bound structures, uses SCM implicitly; more recently, it was
shown that the basic physics behind the non-linear scaling relations (NSR)
obeyed by the two point correlation function can be obtained from a judicious
application of SCM \cite{TPMNRAS}. These successes, as well as the 
inherent simplicity of the underlying concepts, make the SCM an attractive 
paradigm for studying non-linear evolution in gravitational clustering and 
motivate one to ask: Can we improve the basic model in some manner so that 
the behaviour of the system after turnaround is `more reasonable' ?
\par
It is clear from very general considerations that such an approach has
to address  fairly non-trivial technical issues. To begin with, 
{\it exact } modelling of deviations from spherical symmetry is quite
impossible since it essentially requires solving the full BBGKY hierarchy.
Secondly, the concept of a radius $R(t)$ for a shell, evolving
only due to the gravitational force of the matter inside, becomes ill-defined
when deviations from spherical symmetry are introduced. Finally, our real 
interest is in modelling the statistical features of the density growth; 
whatever modifications we make to SCM should eventually tie up with known 
results for the evolution of, for instance, the two point correlation function. 
That is, we have to face the question of how best to obtain the {\it statistical\/} 
properties of the density field from the behaviour of a {\it single\/} system.
\par
In this chapter, these problems are addressed in a limited but focussed manner. 
The deviations from spherical symmetry are tackled by  the retention of a 
term (which is usually neglected) in the equation describing the growth of the 
density contrast. Working in the fluid limit, we show that 
this term is physically motivated and  present some arguments  to derive 
an acceptable form for the same. The key new idea is to introduce a Taylor 
series expansion in ($1/\delta$) (where $\delta$ is the density contrast) to 
model the non-linear evolution. We circumvent the question of defining the
`radius' of the non-spherical regions by working directly with density
contrasts. Finally, we attempt to make the connection with statistical descriptors
of non-linear growth, by using the non-linear scaling relations known from 
previous work. More precisely, we show that the modified equations predict a 
behaviour for the relative pair velocity (when interpreted statistically) which 
agrees with the results of N-body simulations.
\par
The chapter is divided into the following sections. The relevant equations
 describing the SCM are set out in Section 5.2; we also summarise the physical 
and {\it ad hoc\/} aspects of the SCM here. Next, we recast the equations in a
different form and introduce two functions (i) a ``virialization term''  and (ii) 
a function $h_{\rm SC}(\delta)$, whose asymptotic forms are easy to determine. 
The behaviour of $h_{\rm SC}(\delta)$ 
in the presence and absence of the ``virialization term'' is also detailed here. 
In Section 5.4,  we present the  arguments that give the functional  
forms for the above term over a large range of $\delta$; we then go on to present 
the results in terms of a single collapsing body and show how 
this term stabilizes a collapse which would have otherwise 
ended up in a singularity in terms of the growth of the density contrast 
with time. When this term is carried through into the equation for $R(t)$ for 
a single system, it can be seen the radius reaches a maximum and gracefully 
decreases to a constant, remaining so thereafter. In the standard SCM, the radius 
decreases from the maximum all the way down to zero, thereby causing the 
density to diverge. Section 5.5 summarises the results and discusses their 
implications. 
\section{The Spherical Collapse Model}
The scales of interest in the current work are much smaller than 
the Hubble length and the velocities in question are non-relativistic; Newtonian 
gravity can hence be used for the following analysis. We will consider the case 
of a dust-dominated, $\Omega = 1$ universe and treat the system in the fluid limit as 
being made up of pressureless dust of dark matter, with a smoothed density, 
$\rho_m(t,{\bf x})$, and a mean velocity, ${\bf v}(t,{\bf x})$. (This approach, 
of course, ignores effects arising from shell crossing and multi-streaming; these 
will be commented on later.)
The density contrast, $\delta({\bf x},t)$ is defined by 
\begin{equation}
\rho_m(t,{\bf x}) = \rho_b(t) [1 + \delta({\bf x},t)]
\end{equation}
where $\rho_b$ denotes the smooth background density of matter. We 
define a velocity field $u^i = v^i/(a \dot{a})$, where $v^i$ is the peculiar 
velocity (obtained after subtracting out the Hubble expansion) and $a(t)$ 
denotes the scale factor. Taking the divergence of the field $u^i$ and
writing it as
\begin{equation}
\partial_{i} u_{j}=\sigma_{ij}+\epsilon_{ijk} \Omega^{k}+\frac{1}{3}
\delta_{ij} \theta
\end{equation}
where $\sigma_{ij}$ is the shear tensor, $\Omega^k$ is the rotation 
vector and $\theta$ is the expansion, we can manipulate the fluid equations \cite{Probbook} to obtain the following equation for $\delta$\\

\begin{eqnarray}
\label{deltaequation}
\frac{d^2\delta}{d a^2}+\frac{3}{2 a}\frac{d \delta}{d a}-\frac{3}{2 a^2} 
\delta (1+\delta)\quad = \quad\qquad\qquad\qquad\qquad\nonumber \\
\qquad\qquad\qquad\frac{4}{3} \frac{1}{(1+\delta)} 
\left( \frac{d \delta}{d a}\right)^2+(1+\delta) (\sigma^2-2\Omega^2)
\end{eqnarray}
The same equation can be written in terms of time $t$ as 
\begin{eqnarray}
\label{deltatime}
\ddot{\delta}-\frac{4}{3} \frac{\dot{\delta}^2}{(1+\delta)}+\frac{2
\dot{a}}{a} \dot{\delta} \quad=\quad\qquad\qquad\qquad\qquad\qquad\qquad\nonumber \\
\qquad\qquad\qquad\qquad 4 \pi G \rho_{b} \delta (1+\delta)+\dot{a}^2
(1+\delta) (\sigma^2-2 \Omega^2)
\end{eqnarray}
This equation turns out to be the same as the one for density contrast
in the SCM, except for the additional term in $(1+\delta)(\sigma^2-2\Omega^2)$, 
arising from the angular momentum  and shear of the system. To see this explicitly,
 we introduce a function $R(t)$ by the definition
\begin{equation}
\label{deltadefn}
1+\delta= {{9GM{t^2}}\over {2R^3}} \equiv \lambda \frac{a^3}{R^3}
\end{equation}
where $M$ and $\lambda$ are constants. Using this relation 
between $\delta$ and $R(t)$, equation (\ref{deltatime}) can be converted 
into the following equation for $R(t)$ 
\begin{equation}
\label{reqn}
\ddot{R}=-\frac{GM}{R^2}-\frac{1}{3} \dot{a}^2 \left(
\sigma^2-2\Omega^2\right) R 
\end{equation}
\noindent where the first term represents the gravitational attraction
due to the mass inside a sphere of radius $R$  
and the second gives the effect of the shear and angular momentum. \\
In the case of spherically symmetric evolution, the shear and 
angular momentum terms can be set to zero; this gives
\begin{equation}
\label{standardSCM}
\frac{d^2 R}{d t^2}=-\frac{GM}{R^2}
\end{equation}
which governs the evolution of a spherical shell 
of radius $R$,  collapsing under its own gravity; $M$ can now be identified 
with the mass contained in the shell; this is standard SCM.
\par
At this point, it is important to note a somewhat subtle aspect of these
equations. The original fluid equations are clearly Eulerian in nature: 
{\it i.e.} the time derivatives give the temporal variation of the quantities  
at a fixed point in space. However, the time derivatives in equation 
(\ref{deltatime}), for the density contrast $\delta$, are of a different kind. 
Here, the observer is moving with the fluid element and 
hence, in this, Lagrangian case, the variation in density contrast seen 
by the observer has, along with the intrinsic time variation, a component 
which arises as a consequence of his being 
at different locations in space at different instants of time. When the 
$\delta$ equation is converted into an equation for the function $R(t)$, 
the Lagrangian picture is retained; in SCM, we can  interpret $R(t)$ as 
the radius of a spherical shell, co--moving with the observer. The mass 
$M$ within each shell remains constant in the absence of shell crossing 
(which does not occur in the standard SCM for reasonable initial
conditions) and the entire formalism is well defined. The physical 
identification of $R$ is, however, not so clear in the case where the 
shear and rotation terms are retained, as these terms break the spherical 
symmetry of the system. We will nevertheless continue to think of $R$ 
as the ``effective shell radius`` in this situation, {\it defined  by\/} 
equation (\ref{deltadefn}) governing its evolution. Of course, there is 
no such ambiguity in the {\it mathematical} definition of $R$ in  this formalism.
\par 
Before proceeding further, let us briefly summarize the results of 
standard SCM. Equation (\ref{standardSCM}) can be integrated to obtain $R(t)$ in the 
parametric form 
\begin{eqnarray}
\label{stheqn1}
R&=&\frac{R_i}{2 \delta_i} (1- \mbox{cos}\;\theta)\\
\label{stheqn2}
t&=& \frac{3 t_i}{4 \delta_i^{3/2}} (\theta-\mbox{sin}\;\theta)
\end{eqnarray}
where $R_i$, $\delta_i$ and $t_i$ are the initial radius, initial 
density contrast and initial time, respectively, with 
$R_i^3=(9GMt_i^2/2)(1+\delta_i)^{-1} \simeq (9GMt_i^2/2) $ for $\delta_i \ll 1$. 
Given $M$, there are only two independent constants, {\it viz\/} $t_i$ and $\delta_i$.
All the physical features of the SCM can be easily derived from the above
 solution. Each spherical shell expands at a progressively slower rate against the 
self-gravity of the system, reaches a maximum radius and then collapses under its 
own gravity, with a steadily increasing density contrast. The maximum radius, 
$R_{max}=R_i/\delta_i$, achieved by the shell,  occurs at a density 
contrast $\delta =(9\pi^2/16)-1 \approx 4.6$, which is in the ``quasi-linear'' 
regime. In the case of a perfectly spherical system, there exists no 
mechanism to halt the infall, which proceeds inexorably towards a 
singularity, with all the mass of the system collapsing to a single point. 
Thus, the fate of the shell (as described by equations (\ref{stheqn1}) and 
(\ref{stheqn2})) is to collapse to zero radius at $\theta = 2\pi$ with an infinite 
density contrast; this is, of course, physically unacceptable.
\par
In real systems, however, the implicit assumptions 
that (i) matter is distributed in spherical shells and (ii) the non-radial 
components of the  velocities of the particles are small, will 
break down  long before infinite densities are reached.
Instead, we expect the collisionless dark matter to reach virial equilibrium. 
After virialization, $|U|=2 K$, where $U$ and $K$ are, respectively, the potential 
and kinetic energies; the virial 
radius can be easily computed to be half the maximum radius reached by the system. 
\par
The virialization argument is clearly physically well-motivated for real systems. 
However, as mentioned earlier, there exists no mechanism in the standard SCM 
to bring about this virialization; hence, one has to
introduce  by hand the assumption  that, as the 
shell collapses and  reaches a particular radius, 
say $R_{max}/2$, the collapse 
is halted and the shell remains at this radius thereafter. This arbitrary 
introduction of virialization is clearly one of the major drawbacks of the standard
SCM and takes away its predictive power in the later stages of evolution.  We 
shall now see how the retention of the angular momentum  
term in equation (\ref{reqn}) can serve to stabilize the collapse of the system, 
thereby allowing us to model the evolution towards $r_{vir}=R_{max}/2$ smoothly.
\section{The $h_{\rm SC}(\delta)$ function.}
\label{sec:hfunction}
 As detailed in the previous section, the primary defect of  
the standard SCM is the {\it ad hoc\/} nature of the stabilization of the shell 
against its collapse under gravity, which arises on account of 
the assumption of perfect spherical symmetry, implicit in the neglect of the shear 
and angular momentum terms. We hence return to equation (\ref{deltaequation}), 
retain the above terms, and recast the equation into a form more 
suitable for analysis. Using logarithmic variables, $D_{\rm SC} \equiv {\rm ln}
\hskip 0.03 in (1 + \delta)$ and $\alpha \equiv {\rm ln}\hskip 0.03 in a$, equation 
(\ref{deltaequation}) can be written in the form (the subscript `SC'
stands for `Spherical Collapse')
\begin{eqnarray}
\label{deltalog}
\frac{d^2 D_{\rm SC}}{d \alpha^2}-\frac{1}{3} \left(\frac{d D_{\rm SC}}{d
\alpha }\right) ^2 + \frac{1}{2} \frac{d D_{\rm SC}}{d \alpha} \quad = 
\qquad\qquad\qquad \quad\nonumber \\
\qquad\qquad\qquad\qquad \frac{3}{2} \left[\exp (D_{\rm SC})-1 \right] + a^2 (\sigma^2-2 \Omega^2)
\end{eqnarray} 
It is convenient to  introduce the quantity, $S$, defined by
\begin{equation}
S \equiv a^2 (\sigma^2-2 \Omega^2)
\end{equation}
which we shall hereafter call the ``virialization term''.  The
 consequences of the retention of the virialization term are easy to
describe qualitatively. We expect  the 
evolution of an initially spherical shell to proceed along the lines of the standard SCM 
in the initial stages, when any deviations from spherical symmetry, present in the 
initial conditions, are small. However, once the maximum radius is reached and the 
shell recollapses, these small deviations are amplified by a positive feedback 
mechanism. To understand this, we note that all particles in a given spherical 
shell are equivalent due to the spherical symmetry of the system. This implies 
that the motion of any particle, in a specific shell, can be considered 
representative of the motion of the shell as a whole. Hence, the behaviour of the 
shell radius can be understood by an analysis of the motion of a single particle. 
The equation of motion of a particle in an expanding universe can be written as 
\begin{equation}
\ddot{{\bf x}_i}+2\frac{\dot{a}}{a} \dot{{\bf x}_i}=-\frac{\nabla \phi}{a^2}
\end{equation}
\par
where $a(t)$ is the expansion factor of the locally overdense ``universe".
The $\dot{{\bf x}_i}$ term  acts as a damping force when it is positive; 
{\it i.e.} while the background is expanding. However, when the
overdense region reaches the point of maximum expansion and turns around, this 
term becomes negative, acting like a {\it negative\/} damping
term, thereby amplifying any deviations from spherical symmetry 
which might have been initially present. Non-radial components of velocities 
build up, leading to a randomization of velocities which finally results 
in a virialised structure, with the mean relative velocity between any 
two particles balanced by the Hubble flow. It must be kept in mind, 
however, that the introduction of the virialization term  changes the 
behaviour of the solution in a global sense and it is  not strictly 
correct to say that this term starts to play a role {\it only after}
  recollapse, with the evolution proceeding along the lines of the 
standard SCM until then. It is  nevertheless reasonable to expect that, 
at early times when the term is small, the system will evolve as standard SCM  
 to reach a maximum radius, but will fall back smoothly to a constant size  later on. 
\par
The virialization term, $S$, is, in general, a function of $a$ and ${\bf x}$, especially 
since the derivatives in equation (\ref{deltatime}) are total time derivatives, 
which, for an expanding Universe, contain partial derivatives with respect 
to both ${\bf x}$ and $t$ separately.  
Handling  this equation exactly will take us back to the full non-linear equations for the
fluid and, of course, no progress can be made. Instead, we will make the
 {\it ansatz\/}   that the virialization term depends on $t$ and ${\bf x}$
only through $\delta(t,{\bf x})$.
\begin{equation}
S(a,{\bf x}) \equiv S(\delta(a,{\bf x})) \equiv S(D_{\rm SC})
\end{equation}
In other words, $S$ is a function of the density contrast alone. 
This {\it ansatz\/}  seems well  motivated because  the density contrast, $\delta$,
 can be used to characterize the SCM at any point in its evolution and one might 
 expect the virialization  term to be a function only of the system's state, at
least to the lowest order. Further, the results obtained with this assumption 
appear to be sensible and may be treated as a test of the {\it ansatz\/} in its 
own framework.\\ 
To proceed further systematically, we {\it define} a function $h_{\rm SC}$
by the relation
\begin{equation}
\label{defh}
{{dD_{\rm SC}}\over {d\alpha}} = 3h_{\rm SC}
\end{equation}
For consistency, we  shall assume the {\it ansatz\/}  $h_{\rm SC}(a,{\bf x}) \equiv
 h_{\rm SC}\left[\delta(a,{\bf x})\right]$.
The definition of $h_{\rm SC}$ allows us to write equation (\ref{deltalog}) as 
\begin{equation}
\label{hequation}
\frac{d h_{\rm SC}}{d \alpha}=h_{\rm SC}^2-\frac{h_{\rm SC}}{2}+\frac{1}{2} 
\left[\exp (D_{\rm SC}) -1\right] + \frac{S(D_{\rm SC})}{3}
\end{equation}
Dividing (\ref{hequation}) by (\ref{defh}), we obtain the following 
equation for the function $h_{\rm SC}(D_{\rm SC})$
\begin{eqnarray}
\label{dhdDeqn}
\frac{dh_{\rm SC}}{dD_{\rm SC}} = \frac{h_{\rm SC}}{3}-\frac{1}{6}+ \frac{1}{6 h_{\rm SC}}
\left[\exp(D_{\rm SC})-1\right]+\frac{S(D_{\rm SC})}{9 h_{\rm SC}}
\end{eqnarray}
\par
If we know the form  of either $h_{\rm SC}(D_{\rm SC})$ or $S(D_{\rm SC})$,
this equation allows us to determine the other. Then, using equation (\ref{defh}),
one can determine $D_{\rm SC}$. Thus, our modification of the standard SCM 
essentially involves providing the form of $S(D_{\rm SC})$ or 
$h_{\rm SC}(D_{\rm SC})$. 
We shall now discuss several features of such a modeling in order to arrive 
at a suitable form.
\par 
The behaviour of $h_{\rm SC}(D_{\rm SC})$ can be qualitatively understood from 
our knowledge of the behaviour of $\delta$ with time. In the linear regime 
($\delta \ll 1$), we know that $\delta$ grows linearly with $a$; hence 
$h_{\rm SC}$ increases with $D_{\rm SC}$. At the extreme non-linear end ($\delta \gg 1$), 
 the system ``virializes'', {\it i.e.\/}  the proper radius and the density of the system become
constant. On the other hand, the density $\rho_b$, of the background, falls like $t^{-2}$ 
(or $a^{-3}$) in a flat, dust-dominated Universe. The density contrast  
is defined by $\delta = (\rho/\rho_b) - 1 \sim \rho/\rho_b$ (for $\delta \gg 1$) 
and hence
\begin{equation}
\delta \propto t^2 \propto a^3
\end{equation}
in the non-linear limit. Equation (\ref{defh}) then implies that 
$h_{\rm SC}(\delta)$ tends to unity for $\delta \gg 1$. Thus, we expect that 
$h_{\rm SC}(D_{\rm SC})$ will start with a value far less than unity, grow, reach a 
maximum a little greater than one and then smoothly fall back to unity. 
[A more general situation  discussed in the literature corresponds to $h 
\rightarrow {\rm constant}$ as $\delta \rightarrow \infty$, though the 
asymptotic value of $h$ is not necessarily unity. Our discussion can be 
generalised to this case.]
\par
This behaviour of the $h_{\rm SC}$ function can be given another useful 
interpretation whenever the density contrast  has a monotonically 
decreasing relationship with the scale, $x$, with small $x$ implying large 
$\delta$ and vice-versa. Then, if we use a local power law approximation 
$\delta \propto x^{-n}$ for $\delta \gg 1$ with some $n >0$, $D_{\rm SC} 
\propto \ln (x^{-1})$ and 
\begin{equation}
h_{\rm SC} \propto {{dD_{\rm SC}} \over 
{d\alpha}} \propto - {{{d \ln} ({1\over x})}\over {d \ln a}} \propto 
\frac{\dot{x} a}{\dot{a} x} \propto - {v \over {{\dot a}x}}
\end{equation}
where $v \equiv a{\dot x}$ denotes the mean relative velocity.
Thus, $h_{\rm SC}$ is proportional  to the ratio of the peculiar velocity 
to the Hubble velocity. We know that this ratio is small 
in the linear regime (where the Hubble flow is dominant) and later 
increases, reaches a maximum and finally falls back to unity with the 
formation of a stable structure; this is another argument leading to
the same  qualitative behaviour 
of the $h_{\rm SC}$ function.
\par
Note that, in standard SCM (for which $S = 0$), equation 
(\ref{dhdDeqn}) reduces to
\begin{equation}
\label{dhdDscm}
3h_{\rm SC}\frac{dh_{\rm SC}}{dD_{\rm SC}}=h_{\rm SC}^2-{h_{\rm SC}\over 2}+{\delta \over 2}
\end{equation}
The presence of the linear term in $\delta$ on the RHS of the 
above equation causes $h_{\rm SC}$ to increase with $\delta$, with $h_{\rm SC} \propto 
\delta^{1/2}$ for $\delta \gg 1$. If virialization is imposed as  
an {\it ad hoc\/}  condition, 
then  $h_{\rm SC}$ should fall back to unity discontinuously --- which is 
clearly unphysical; the form of $S(\delta)$ must hence be chosen so as to ensure 
a smooth transition in $h_{\rm SC}(\delta)$ from one regime to another.
\par
As an aside, we would like to make some remarks on the nature of the
``virialization term'', $S(\delta)$, in a somewhat wider context. As is well-known, 
gravitational clustering can be described at three different levels of 
approximation, by different mathematical techniques. The first
approach tracks the clustering by following the true particle trajectories;
this is what is done, for example, in N-body simultations. This method does
not involve any approximation (other than the validity of the 
Newtonian description at the scales of interest); it is, however, clearly analytically 
intractable.
At the next level, one may describe the system by an one-particle distribution funtion
and attempt to solve the collisionless Boltzmann equation for
the distribution function $f(t,{\bf x},{\bf v})$; the approximation here lies in the
neglect of gravitational collisions, which seems quite reasonable as the time scale
for such collisions is very large for standard dark matter particles. Finally, 
one can treat the system in the fluid limit described by five functions:
the density $\rho(t,{\bf x})$, mean velocity ${\bf v}(t,{\bf x})$,
and gravitational potential $\phi(t,{\bf x})$, thus neglecting multi-streaming
effects. Our analysis was based on this level of approximation. The key difference 
between the last two levels of description lies in the fact that the distribution function
allows for the possibility of different particle velocities at any point in space
({\it i.e.} the existence of {\it velocity dispersions}), while the fluid
picture assumes a mean velocity at each point. It is also known
that the gradients in velocity dispersion can provide a kinetic pressure
which will also provide support against gravitational collapse.
While a detailed analysis of these terms is again exceedingly difficult, one can
incorporate the lowest order effects of the gradient in the velocity dispersion by
modifying equation (\ref{deltalog}) to the form
\begin{eqnarray}
\label{new_deltalog}
\frac{d^2 D_{\rm SC}}{d \alpha^2}-\frac{1}{3} \left(\frac{d D_{\rm SC}}{d
\alpha }\right) ^2+\frac{1}{2} \frac{d D_{\rm SC}}{d \alpha} \quad = \qquad
\qquad\qquad\nonumber \\
\qquad\qquad\frac{3}{2} \left[\exp (D_{\rm SC})-1 \right] + 
a^2 (\sigma^2-2 \Omega^2) + f(a,x)
\end{eqnarray}
where $f(a,x)$ contains the lowest order contributions from the dispersion terms. We
can then define
\begin{equation}
S(a,x) = a^2 (\sigma^2-2 \Omega^2) + f(a,x)
\end{equation}
and again invoke the {\it ansatz} $S(a,x) \equiv S(\delta) $. Note that
$S(\delta)$ now contains the lowest order contributions arising from shell crossing,
multi-streaming, etc., besides the shear and angular momentum terms, {\it
i.e.} it contains all effects leading to virialization of the system.
It is explicitly demonstrated that velocity dispersion
terms arise naturally in the ``force'' equation (for the function $h
\equiv - v/{\dot a}x$), derived from the BBGKY hierarchy, and play the
same role as the function $S(\delta)$ in the fluid picture. This clearly
justifies the above procedure and shows that our approach could have a
somewhat larger domain of validity than might be expected from an analysis
based on the fluid picture.
\section{Detailed derivation of equation for $h$ function}
The zeroth and first moments of the 2$^{\rm nd}$ BBGKY equation 
\cite{ruam}  can be combined to obtain the 
following equation for the dimensionless function $h = -v/{\dot a}x$ 
\begin{eqnarray}
\label{hevolution}
3h(1 + \overline\xi) {{dh}\over {d\overline\xi}} + {h \over 2} - h^2 -
{{3\overline\xi} \over F} - {{9MQe^{-X}}\over {4\pi F}} \quad =  \qquad \quad \nonumber \\
 \qquad\qquad \qquad \qquad \qquad h^2_{\parallel} {\Big (} 4 + 
{{{\partial \mbox{ln}\;\;  F}\over{\partial{X}}}} 
 {\Big )} + {{{\partial h^2_{\parallel}}\over{\partial{X}}}} -  2 h^2_{\perp} 
\end{eqnarray}
\noindent where we have used the {\it ansatz}, $h \equiv h({\overline\xi})$ (\cite{Ham}, 
\cite{RajPad94}). In the above, we have defined
\begin{equation}
\label{barxi}
\overline{\xi}(x,a) = {3 \over {x^3}}\int_0^x dx \xi(x,a) x^2 
\end{equation}
\begin{equation}
F = {{{\partial {\overline\xi}}\over{\partial{X}}}} + 3(1 + \overline\xi)\:\; 
, \:\; X = {\rm ln}\; x
\end{equation}
\begin{equation}
h^2_{\parallel} = {{\Pi} \over {{\dot a}^2x^2}} \:\;,\:\; 
h^2_{\perp} = {{\Sigma} \over {{\dot a}^2x^2}}
\end{equation}
where $\Pi$ and $\Sigma$ are parallel and perpendicular peculiar 
velocity dispersions (\cite{ruam},\cite{kanekar}). Finally, we have assumed 
that the 3-point correlation function has the hierarchical form (\cite{DP},
 \cite{ruam})
\begin{equation}
\zeta_{123} = Q(\xi_{12}\xi_{13} + \xi_{13}\xi_{23} + \xi_{12}\xi_{23})
\end{equation}
and defined
\begin{equation}
M = \int d^3z{\Big [} \xi(x) + \xi(z){\Big ]}\xi({\bf z}-{\bf x})
{{\mbox{cos}\;\theta} \over {z^2}}
\end{equation}
In the non-linear regime, $\overline\xi \gg 1$, the stable clustering ansatz 
yields a scale-invariant power-law behaviour for $\overline\xi$ \cite{DP}, 
with $\overline\xi \propto a^{(3-\gamma)}x^{-\gamma}$, if $h \rightarrow 1$ as 
$\overline\xi \rightarrow \infty$. In this limit, we have \\
\begin{equation}
F = (3-\gamma)\overline\xi + 3
\end{equation}
\noindent and
\begin{equation}
{{{\partial \mbox{ln} \; F}\over{\partial{X}}}} = -\gamma
{\Big[} 1 + ({3 \over {3 - \gamma}})({1 \over \overline\xi}) {\Big]}^{-1}
\end{equation}
Further, we can write \cite{YG} $M = M'x\overline\xi^2$, where $M'$ 
is a constant. Thus, equation (\ref{hevolution}) reduces, in the non-linear regime,
to
\begin{eqnarray}
3h\overline\xi \frac{dh}{d\overline\xi} - h^2 + \frac{h}{2} - 
\frac{3}{3-\gamma} \left[ \frac{3M'}{4\pi} \left\{ \overline\xi - \frac{3}{3-\gamma}\right\}
+ 1 \right]   = \nonumber \\
\left[ \left(4 - \gamma\right) + \frac{3\gamma}{(3-\gamma)\overline\xi}\right]h^2_\parallel
+ {{{\partial h^2_{\parallel}}\over{\partial{X}}}} -  2 h^2_{\perp} + 
{\cal O}\left(\frac{1}{\overline\xi}\right)
\end{eqnarray}
where we have retained terms upto order constant in $\overline\xi$. We now 
assume that $h^2_\parallel$ and $h^2_\perp$ are functions of $\overline\xi$ alone, to first 
order. This yields
\begin{eqnarray}
\label{geqn}
3h\overline\xi \frac{dh}{d\overline\xi} -h^2 + \frac{h}{2} - 
\frac{3}{3-\gamma} \left[ \frac{3M'}{4\pi} \left\{ \overline\xi - \frac{3}{3-\gamma}\right\}
+ 1 \right] = \nonumber \\
G(\overline\xi) + {\cal O}\left(\frac{1}{\overline\xi}\right) 
\end{eqnarray}
with
\begin{equation}
G(\overline\xi) = \left[ \left(4 - \gamma\right) + \frac{3\gamma}{(3-\gamma)
\overline\xi}\right] h^2_\parallel(\overline\xi)
- \gamma\overline\xi \frac{d h^2_\parallel}{d \overline\xi}-  2 h^2_{\perp}(\overline\xi)
\end{equation}
Clearly, if $h \rightarrow 1$ as $\overline\xi \rightarrow \infty$, we must have
\begin{eqnarray}
G(\overline\xi) = - \frac{3}{3-\gamma} \left[ \frac{3M'}{4\pi} \left\{ \overline\xi - 
\frac{3}{3-\gamma}\right\} + 1 \right]  - \frac{1}{2} + {\cal O}\left(\frac{1}{\overline\xi}
\right)
\end{eqnarray}
{\it i.e.} $G(\overline\xi) \approx - {9M'\overline\xi}/{4\pi(3-\gamma)}$ 
for $\overline\xi \gg 1 $. \\
 Since $G(\overline\xi)$ tends to the above asymptote at late times, the residual 
part can be expanded in a Taylor series in $ 1/\overline\xi$. Retaining the first two terms 
of the expansion in equation (\ref{geqn}), we obtain
\begin{equation}
3h\overline\xi \frac{dh}{d\overline\xi} -h^2 + \frac{h}{2} + \frac{1}{2} = 
\frac{A}{\overline\xi}+\frac{B}{\overline\xi^2} +{\cal O}(\overline\xi^{-3})
\end{equation}
 This is exactly the same as equation (\ref{new_dhdDeqn}), with $\overline\xi$ 
replacing $\delta$. $G(\overline\xi)$ thus plays the same role as $S(\delta)$ in the 
stabilising of the system against collapse. This clearly implies that the 
velocity dispersion terms, $h^2_\parallel$ and $h^2_\perp$, will contribute to the 
support term; we are hence 
justified in writing the virialization  term in the more general form
\begin{equation}
S = a^2\left(\sigma^2 - 2\Omega^2\right) + f(a,x)
\end{equation}
where $f(a,x)$ contains contributions from effects arising from shell crossing, multi-streaming, 
etc.
\section{The virialization term}
We will now derive an approximate functional form for the virialization 
function from physically well-motivated arguments. 
If the virialization term is retained in equation (\ref{reqn}), we have
\begin{equation}
\label{theRequation}
{{d^2 R}\over {d t^2}}=-{{GM}\over {R^2}} - {{H^2 R} \over 3} S  
\end{equation} 
where $H=\dot a/a$. 
Let us first consider the late time behaviour of the system. When virialization 
occurs, it seems reasonable to
assume that  $R\rightarrow  {\rm constant} $  and $\dot{R} \rightarrow 0$. 
This implies that, for large density contrasts, 
\begin{equation}
S \approx  -{{3GM} \over {R^3 H^2}} \;\; \qquad(\delta \gg 1) 
\end{equation}
Using $H=\dot{a}/a=(2/3t)$, and equation  (\ref{deltadefn})
\begin{equation}
 S  \approx  -{{27GM t^2} \over {4 R^3}} = -{3 \over 2} (1 + \delta
)\approx -{3\over 2}\delta \;\; \qquad(\delta \gg 1)
\end{equation}
Thus, the ``virialization'' term tends to a value of ($ -3 \delta/2$) in the non-linear 
regime, when stable structures have formed. This asymptotic form for 
$S(\delta)$ is, however, insufficient to model its behaviour 
  over the  larger range of density contrast (especially the 
quasi-linear regime) which is of interest to us. Since $S(\delta)$ 
tends to the above asymptotic form at late times, the residual part, {\it i.e.} 
the part that remains after the asymptotic value has been subtracted away, 
can be expanded in a Taylor series in $(1 / \delta)$ without any loss of generality.
Retaining the first two terms of expansion, we write the complete virialization term as 
\begin{equation}
\label{netrotation2}
S(\delta)=-\frac{3}{2} (1+\delta) -\frac{A}{\delta}+\frac{B}{\delta^2}
+{\cal O}(\delta^{-3})
\end{equation}
Replacing for $S(\delta)$ in equation ({\ref{deltalog}), we obtain, 
for $\delta \gg 1$
\begin{equation}
\label{new_dhdDeqn}
3h\delta \frac{dh_{\rm SC}}{d\delta} - h^2_{\rm SC} + \frac{h_{\rm SC}}{2} + 
\frac{1}{2}  = -\frac{A}{\delta}+\frac{B}{\delta^2}
\end{equation}
[It can be easily demonstrated that  the first order term in the 
Taylor series is alone insufficient  to model the turnaround behaviour of the 
$h$ function. We will hence include the next higher order term and use the 
form in equation (\ref{netrotation2}) for the virialization  term. The signs are 
chosen for future convenience, since it will turn out that both $A$ and $B$ are 
greater than zero.] In fact, for sufficiently large $\delta$, the 
evolution depends only on the combination $q\equiv(B/A^2)$. This is most
easily seen by rewriting equation (\ref{deltaequation}), replacing $S(\delta)$ 
with the above form. Taking the limit of 
large $\delta$, {\it i.e.} $\delta \gg 1$, and rescaling  $\delta$
to $\delta/A$, we  obtain
\begin{eqnarray}
\label{deltaequationscaled}
\frac{d^2\delta}{d b^2}+\frac{3}{2 b}\frac{d \delta}{d b}-\frac{4}{3 \; \delta} \left(
\frac{d \delta}{d b}\right)^2 \:&=&\:  -\frac{1}{a^2}+\frac{B}{A^2}\frac{1}{a^2\,\delta} 
\\
&=&-\frac{1}{a^2}+\frac{q}{a^2\delta}
\end{eqnarray}
From the form of the equation it is clear that the constants 
$A$ and $B$ occur in the combination $q=B/A^2$ and hence the non-linear
regime is modelled by a one parameter family for
 the virialization  term. 
\par
Equation (\ref{theRequation}) can be written as
\begin{equation}
\label{theRequation2}
\ddot{R}=-\frac{GM}{R^2}-\frac{4R}{27t^2} \left[ -\frac{27 GMt^2}{4 R^3}- \frac{A}{\delta}+ \frac{B}{\delta^2} \right]
\end{equation}
Using $\delta=9GMt^2/2R^3$ and $B=qA^2$ we may express equation (\ref{theRequation2}) 
completely in terms of $R$ and $t$. We now rescale $R$ and $t$ in the form 
$R=r_{vir}y(x)$ and $t=\beta x$, where $r_{vir}$ is the final virialised 
radius [{\it i.e.} $R \rightarrow r_{vir}$ for $t \rightarrow \infty$], where 
$\beta^2=(8/3^5) (A/GM) r_{vir}^3$, to obtain the following equation for $y(x)$
\begin{equation}
\label{thescaledeqn}
y''=\frac{y^4}{x^4} -\frac{27}{4} q \frac{y^7}{x^6}
\end{equation}
We can integrate this equation to find a form for $y_q(x)$ (where $y_q(x)$ is the function $y(x)$ for a specific value of $q$) using the physically motivated boundary conditions $y=1$  and $y'=0$ as $x \rightarrow \infty$, which is simply an expression of the fact that the system reaches the virial radius $r_{vir}$ and remains here thereafter.
\begin{figure}[htbp]
\leavevmode\centering
\psfig{file=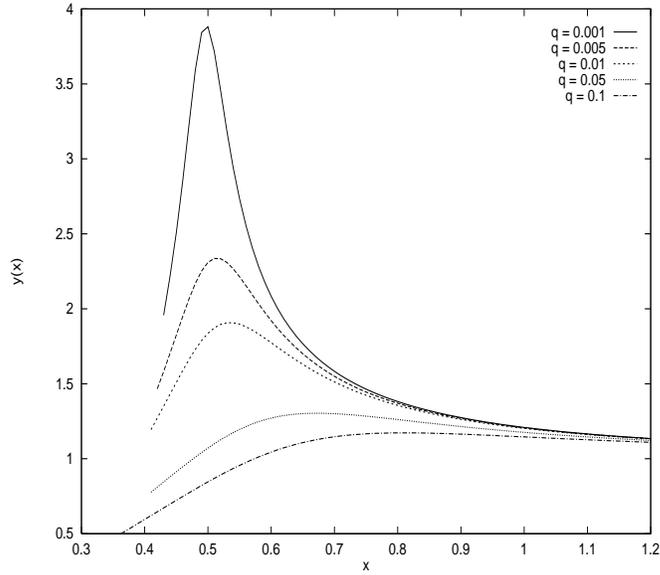,width=3.5truein,height=3.0truein,angle=-90}
\caption{\sf  $y_q(x)$ for some values of $q$. The x axis has scaled time, $x$ and the y axis is the scaled radius $y$.}
\label{figure1}
\end{figure}
The results of numerical integration of this equation for a range of $q$ 
values are shown in fig. \ref{figure1}. 
As  expected on physical grounds,  the function has a maximum and gracefully decreases 
to unity for large values of $x$ [the behaviour of $y(x)$ near $x=0$ is irrelevant since the 
original equation is valid only for $\delta \geq 1$, at least]. For a given value of 
$q$, it is possible to find the value $x_c$ at which the function reaches its maximum, 
as well as the ratio $y_{max}=R_{max}/r_{vir}$. The time, $t_{max}$,  at which the 
system will reach the maximum radius is related to $x_c$ by the relation $t_{max}=
\beta x_c = t_0 (1+z_{max})^{-3/2}$, where $t_0=2/(3 H_0)$ is the present age of 
the universe and $z_{max}$ is the redshift at which the system turns around. 
Figure (\ref{figure2a}) shows the variation of $x_c$ and $y_{max}\equiv 
(R_{max}/r_{vir})$ for different values of $q$. The entire evolution of the 
system in the modified spherical collapse model (MSCM) can be expressed in terms of 
\begin{equation}
\label{MSCMsoln}
R(t)=r_{vir}\; y_q(t/\beta)
\end{equation} 
where $\beta=(t_0/x_c) (1+z_{max})^{-3/2}$.
In SCM, the conventional value used for ($r_{vir}/R_{max}$) is ($1/2$), 
which is obtained by enforcing the virial condition that $\vert U \vert=2K$, where 
$U$ is the gravitational potential energy and $K$ is the kinetic energy. It must 
be kept in mind, however, that the ratio ($r_{vir}/R_{max}$) is not really 
constrained to be {\it precisely} ($1/2$) since the 
actual value will depend on the final density profile and the precise definitions
used for these radii. While we expect it to be around $0.5$, some
 amount of variation, say between 0.25 and 0.75, cannot be ruled out theoretically.
\begin{figure}[htbp]
\leavevmode\centering
\psfig{file=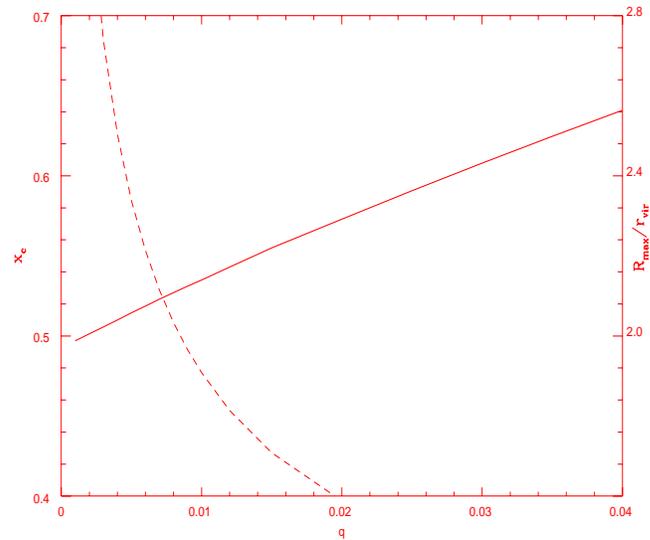,width=3.5truein,height=3.0truein,angle=-90}
\caption{\sf The parameters ($R_{max}/r_{vir}$) (broken line)  and $x_c$ 
(solid line) as a function of $q=B/A^2$. This clearly demonstrates that the single 
parameter description of the virialization term  is constrained by the value that is 
chosen for the ratio $r_{vir}/R_{max}$.}
\label{figure2a}
\end{figure}
Figure (\ref{figure2a}) shows the parameter ($R_{max}/r_{vir}$),   
plotted as a function of $q=B/A^2$ (dashed line),
obtained by numerical integration of  
equation (\ref{theRequation}) with the {\it ansatz\/}  (\ref{netrotation2}).
The solid line  gives the dependence of $x_c$ (or equivalently $t_{max}$) 
on the value of $q$. It can be seen that one can obtain a suitable value for 
the ($r_{vir}/R_{max}$) ratio by choosing a suitable value for $q$ and vice versa.   
Using equation (\ref{defh}) and the definition $\delta \propto t^2/R^3$, we obtain 
\begin{equation}
h_{\rm SC}(x)=1-\frac{3}{2} \frac{x}{y} \frac{d y}{d x}
\end{equation}
which gives the form of $h_{\rm SC} (x)$ for a given value of $q$; this, in turn, 
determines the function $y_q(x)$.
Since $\delta$ can be expressed in terms of $x$, $y$ and $x_c$ as $\delta=
(9 \pi^2/2 x_c^2) x^2/y^3$, this allows us to implicitly obtain a form for 
$h_{SC}(\delta)$, determined only by the value of $q$.
\begin{figure}[htbp]
\leavevmode\centering
\psfig{file=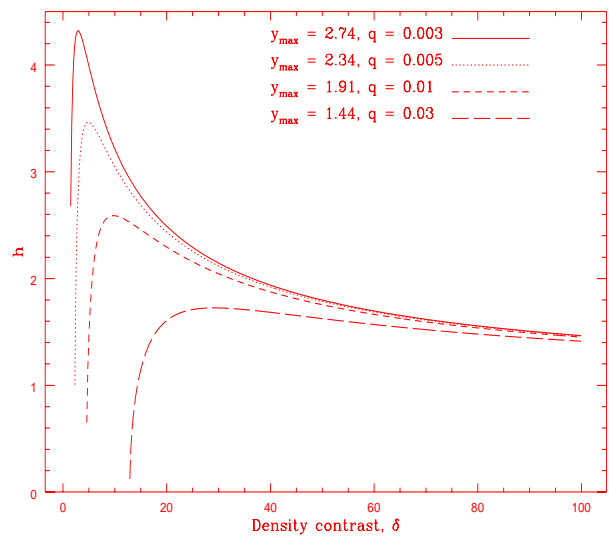,width=3.5truein,height=3.0truein,angle=-90}
\caption{\sf The $h_{\rm SC}$ function, obtained for  various 
values of $q$. The values of $q$ and $y_{max} \equiv R_{max}/r_{vir}$ for 
the curves are indicated at the top right hand corner. (Further discussion in text) }
\label{figure2}
\end{figure}
Figure (\ref{figure2}) shows the  behaviour of $h_{SC}$
 functions obtained by integrating equation (\ref{dhdDeqn}) backwards,
 assuming that $h_{SC} \rightarrow 1$ as $\delta \rightarrow \infty$.
 It is seen that all the curves  have the same turnaround behaviour  
expected on the basis of the physical arguments presented in the earlier 
section. 
\par
If the functional form for  $h_{SC}$ -- determined, say, from N-body simulations -- is
used as a further constraint, we should be able to obtain the values of $q$. 
The major hurdle in attempting to do this is the fact that the available 
simulation results are given in terms of the averaged two point correlation 
function, $\bar\xi$, and the averaged pair velocity, $h(a,x)$, defined by 
\begin{equation}
\bar\xi=\frac{3}{r^3}\int_0^r \xi(x,a) x^2 dx \; ;\qquad h(a,x) =
-\frac{\left< v(a,x) \right>}{\dot{a} x}
\end{equation}
where the two-point correlation function $\xi$ is defined as 
the Fourier transform of the power spectrum, $P(k)$, of the distribution.
 The results published in the literature  assume that $h(a,x)$ 
depends on $a$ and $x$ only through $\bar\xi(a,x)$, that is, $h(a,x) 
\equiv h[\bar\xi(a,x)]$. This assumption has been invoked in several works 
in the past (\cite{Ham}, \cite{RajPad94}, \cite{MoJain}, \cite{TPMNRAS}, \cite{Dreams})
and seems to be validated by numerical simulations. The
fitting formula for $h(\bar\xi)$ can be obtained from related fitting formulas 
available in the literature \cite{Ham}. These are, however, statistical 
quantities and are not well 
defined  for an isolated overdense region. Hence we have to first make the 
correspondence between $h_{SC}(\delta)$ and $h(\bar\xi)$, which we do as follows. 
\par
It is possible to show by standard arguments 
\cite{RajPad94} that
\begin{equation}
\label{paddyh}
{{d \bar\xi}\over {d\alpha}} = 3h(1 + \bar\xi)
\end{equation}
\noindent that is,
\begin{equation}
\label{hxieqn}
{{d D}\over {d\alpha}} = 3h
\end{equation}
where $D = {\rm ln} (1+\bar\xi)$ and $\alpha={\rm ln}\;a$. Equation (\ref{hxieqn}) 
is very similar to equation (\ref{defh}), which defines the function 
$h_{\rm SC}(\delta)$, except for the different definitions of $D$ and 
$D_{\rm SC}$ 
in terms of $\bar\xi$ and $\delta$ respectively. This suggests  that one 
can obtain a relation between $h_{\rm SC}(\delta)$ and $h(\bar\xi)$ 
by  relating  the density contrast $\delta$ of an isolated 
spherical region to the two-point correlation function $\bar\xi$ 
averaged over the distribution at the same scale. We essentially need to find 
a mapping between $\bar\xi$ and $\delta$ which is valid in a statistical sense.
\par
Gravitational clustering is known to have three regimes in its
growing phase, usually called ``linear'', ``quasi-linear'' and ``non-linear'' 
respectively. The three regimes may be characterized by values of 
density contrast as  
$\delta \ll 1$ in the linear regime, $1<\delta < 100$ in the quasi-linear regime 
and  $100<\delta$ in the non-linear regime. The three regimes have different rates 
of growth for various quantities of interest such as $\delta$, $\xi$ and so on. 
In the linear regime, it is well known that
the density contrast grows proportional to the scale factor, $a$.
 This implies that the power spectrum, $P(k) \equiv |\delta_k|^2$ (where
$\delta_k$ is the Fourier mode corresponding to $\delta(x)$), grows as
$a^2$. Consequently, $\bar\xi$, which is related to $P(k)$ via a Fourier
transform, also grows as $a^2$, {\it i.e.} as the square of the density
contrast. In the quasi-linear and non-linear regimes, the density contrast 
does not grow linearly with the scale factor and the relation between 
$\delta$ and $\bar\xi$ is not so clearly defined. The quasi-linear regime 
may be loosely construed as the interval of time during which the high peaks
of the initial Gaussian random field have collapsed, although mergers
of structures have not yet begun to play an important role.
(This idea was used in \cite{TPMNRAS} to model the non-linear scaling relations
successfully). If we consider
a length scale smaller than the size of the collapsed objects, the dominant 
contribution to $\bar\xi$ (at this scale)
arises from the density profiles centered on the collapsed
peaks. Using the relation
\begin{equation}
\rho \simeq  \rho_b\left( 1 + \bar\xi \right)
\end{equation}
for density profiles around high peaks, one can see that $\bar\xi \propto \delta$ in this regime.
 In the non-linear regime, $\delta$ and $\bar\xi$ 
have the forms $\delta(a,x)=a^3F(ax)$, $\bar\xi=a^3 G(ax)$, where $x$ is a 
co-moving and $r=ax$ is a proper coordinate. When the system is described by Lagrangian 
coordinates (which correspond to 
proper coordinates $r=ax$, {\it i.e.\/}  at constant $r$), $\bar\xi$ is proportional 
to $\delta$. Thus, the relation $\bar\xi \propto \delta$
appears to be satisfied in all regimes, except at the very linear 
end. Since we are only interested in the $\delta >1$ range, we use
$\bar\xi \approx \delta$
and compare equations (\ref{paddyh}) and (\ref{defh}) to identify
\begin{equation}
h_{\rm SC}(\delta) \approx h(\bar\xi) 
\end{equation}
It is now straightforward to choose the value of $q$ such that the
known fitting function for the $h$ function is reproduced as closely as possible.
We use the original function given by Hamilton \cite{Ham} to obtain the following 
expression for $h(\bar\xi)$:
\begin{equation}
\label{hamiltonfit}
h(\bar\xi)=\frac{2}{3} \left(\frac{d \ln {\cal V}(\bar\xi)}{d \ln (1+\bar\xi)}\right)^{-1}
\end{equation} 
where ${\cal V}(\bar\xi)$ is given by the fitting function 
\begin{equation}
\label{Vhamilton}
{\cal V}(\bar\xi)=\bar\xi \left( \frac{1+0.0158\;{\bar\xi}^2+0.000115\;{\bar\xi}^3}{1+0.926\;{\bar\xi}^2-0.0743\;{\bar\xi}^3+0.0156\;{\bar\xi}^4}\right)^{1/3} \end{equation} 
\begin{figure}[htbp]
\leavevmode\centering
\psfig{file=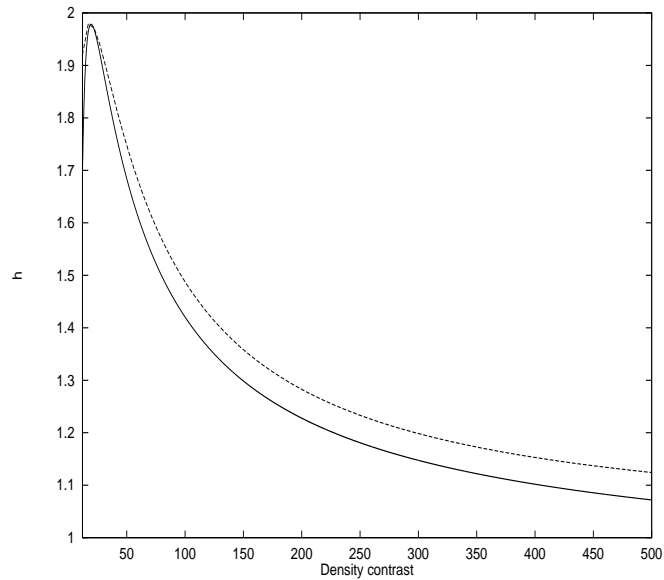,width=3.5truein,height=3.0truein,angle=-90}
\caption{\sf The best fit curve for  the $h$ function (dashed line) to the simulation data  (solid line). The simulation results are obtained from Hamilton \cite{Ham} and the fit is obtained  by adjusting the value of $q$ parameter until the curves coincide. }
\label{figure3}
\end{figure}
Figure (\ref{figure3}) shows the simulation data 
represented by the fit (solid line) \cite{Ham} and the best fit (dashed line), 
obtained in our model, for $q\simeq 0.02$. We note that the fit is better 
than 5\% for all values of density contrast $\delta  \ge 15$. The change in the fit is very marginal if one imposes the boundary condition $h(\delta) 
\rightarrow 1$ for $\delta \gg 1$, instead of constraining the curves to match 
at their peaks (for example, the change in the peak height is $\sim 1 \%$, 
if we impose the above condition at $\delta = 10000$). \\ 
\begin{figure}[htbp]
\leavevmode\centering
\psfig{file=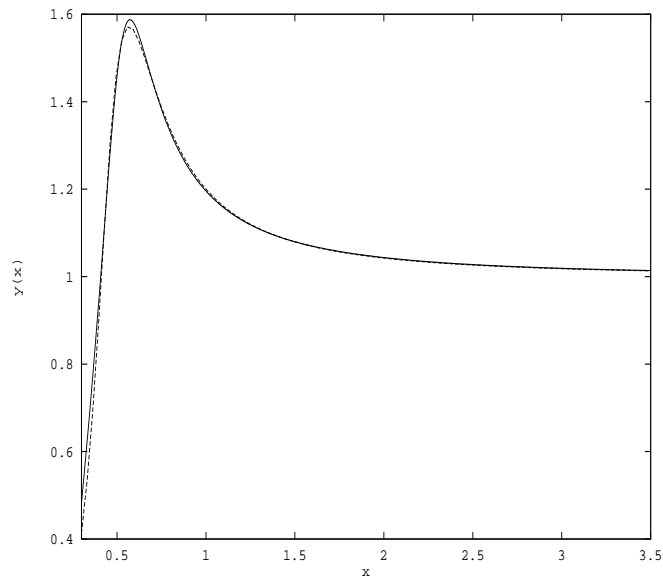,width=3.5truein,height=3.0truein,angle=-90}
\caption{\sf  Plot of the scaled radius of the shell $y_q$ as a function of scaled time $x$ (solid line) and the fitting formula $y_q=(x+ax^3+bx^5)/(1+cx^3+bx^5)$, with $a=-3.6$, $b=53$ and $c=-12$ (dashed line) (See text for discussion) }
\label{figure5}
\end{figure}
Figure (\ref{figure5}) shows the plot of scaled radius $y_q(x)$ vs $x$, 
obtained by integrating equation(\ref{thescaledeqn}), with $q=0.02$.
The figure also shows an accurate fit (dashed line) to this solution of the form
\begin{equation}
\label{yqfit}
y_q(x)=\frac{x+a x^3+b x^5}{1+c x^3+b x^5}
\end{equation} 
with $a=-3.6$, $b=53$ and $c=-12$. This fit, along with values for $r_{vir}$ 
and  $z_{max}$, completely specifies our model through equation (\ref{MSCMsoln}).
It can be observed  that ($r_{vir}/R_{max}$) is approximately $0.65$.
It is  interesting to note that the value  obtained for the 
($r_{vir}/R_{max}$) ratio is not very widely off the usual value of $0.5$ used 
in the standard spherical collapse model, {\it  in spite of the fact that no 
constraint was imposed on this value, {\it ab initio},  in arriving at 
this result.\/} Part of this deviation {\it may} also originate in the fit 
which has been used for $h(\bar\xi)$; Hamilton et al. \cite{Ham} noticed that 
objects virialised at $ R_{max}/r_{vir} \sim 1.8$, instead of 2, in their 
simulations.
\par 
\begin{figure}[htbp]
\leavevmode\centering
\psfig{file=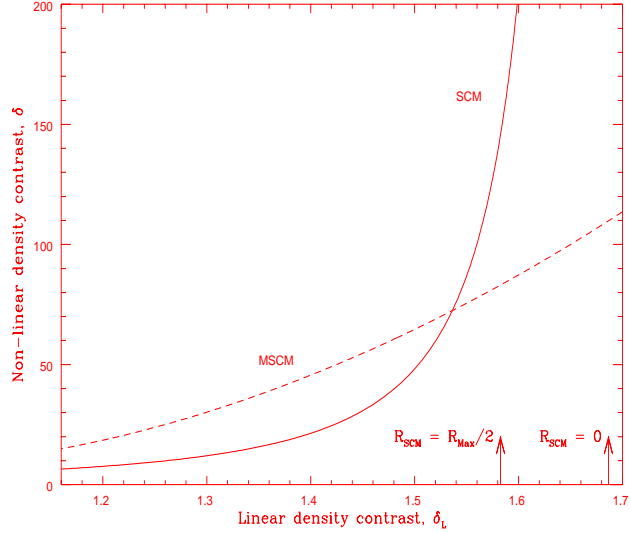,width=3.5truein,height=3.0truein,angle=-90}
\caption{\sf The non-linear density contrast in the SCM 
(solid line) and in the modified SCM (dashed line), plotted against 
the linearly extrapolated density contrast $\delta_L$. }
\label{figure6}
\end{figure}
Finally, figure (\ref{figure6}) compares the non-linear  density
contrast  in the modified SCM (dashed line) with that in the standard SCM 
(solid line), by plotting both against the linearly extrapolated density contrast,
$\delta_L$. 
It can be seen (for a given system with the same $z_{max}$ and 
$r_{vir}$) that, at the epoch where the standard SCM model has a
singular behaviour ($\delta_L \sim 1.686$), our model has a smooth 
behaviour with $\delta \approx 110$ (the value is not very sensitive 
to the exact value of $q$). This is not widely off from the value 
usually obtained from the {\it ad hoc}  procedure applied in the standard 
spherical collapse model. In a way, this explains the unreasonable
effectiveness of standard SCM in the study of non-linear clustering. 
\par
As mentioned earlier, deviations from spherical symmetry
are expected to be small at early epochs and to grow as the
system evolves. One would thus expect the two curves of figure (\ref{figure6})
to approach each other as $ \delta \rightarrow 1$ (from above).
Further, the curves should overlay in the linear regime ($\delta_L
\ll 1$). It can be seen from the figure that the 2 curves {\it do}
approach each other as $\delta_L$ reduces towards unity. However, 
the MSCM has been obtained using a Taylor expansion in $(1/\delta)$; 
it is clearly {\it not} applicable for $\delta \ll 1$. Further,
the region $\delta > 15$ has been used to fit the function $h(\delta)$ 
to the data of Hamilton \cite{Ham}. Hence, one cannot compare the 
curves in the linear regime.
\par
Figure (\ref{figure6}) also shows a comparison between the 
standard SCM and the MSCM in terms of $\delta$ values in the MSCM at 
two important epochs, indicated by vertical arrows. (i) 
When $R = R_{max}/2$ in the SCM, {\it i.e.} the epoch at which 
the SCM {\it virializes}, $\delta({\rm MSCM}) \sim 83$.
(ii) When the SCM hits the singularity, ($\delta_L \sim 1.6865$), 
$\delta({\rm MSCM}) \sim 110$.
\par
We note, finally, that figure (\ref{figure6}), which shows the effects  
of evolution as a mapping from linear to non-linear density contrasts, 
contains a subtle implicit assumption regarding the definition of the 
non-linear density contrast. The radius $R$ of a system is not a 
rigorously defined quantity in the absence of spherical symmetry, 
and obviously, any argument involving `virialization' precludes 
strict spherical 
symmetry. It is, however, a conventional 
practice to define the `radius', $R$, 
even for a virialized system without strict spherical symmetry. 
For example, this approach is used to define the density contrast at 
`virialization' (which has the value, $\delta_{vir} \approx \; 200$) in 
the standard SCM. In our model we have an explicit equation for $R$; once $R$ 
and $M$ are given, the non-linear density contrast is a well-defined quantity.
\section{Results and Summary}
It has been shown how the Taylor expansion of a 
term in the equation for the evolution of the  
density contrast, $\delta$, in inverse powers of $\delta$, allows us to have a 
more realistic picture of spherical collapse, which is free from arbitrary 
abrupt ``virialization'' arguments. Beginning from a well motivated {\it ansatz\/}  
for the dependence of the ``virialization'' term on the density contrast we have shown 
that a spherical collapse model will gracefully turn around and collapse to a 
constant radius with  $\delta \sim 110$ at the same epoch when the standard 
model  reaches  a singularity. Figure (\ref{figure5}) shows clearly that the singularity  
 is avoided in our model due to the enhancement of deviations from spherical symmetry, 
and consequent generation of strong non-radial motions.
\par
We derive an approximate functional form for the virialization  term 
starting from 
the physically reasonable assumption that the system reaches a constant radius. This 
assumption allows us to derive an asymptotic form for the virialization term, with the 
residual part  adequately expressed by keeping only the first and second order terms 
in a Taylor series in ($1/\delta$).  It is shown that there exists a scaling relation 
between the coefficients of the first and second order terms, essentially reducing 
the virialization term to a one parameter family of models. 
\par
The form of the $h$ function published in the literature, along with 
a tentative  mapping from $\delta$ to $\bar\xi$, in the 
non-linear and quasi-linear regimes, allow us to further constrain our model, 
bringing it in concordance with the available simulation results. Further, it is 
shown that this form for the virialization  term  is sufficient to model 
the turnaround behaviour of the spherical shell and leads to a reasonable 
numerical value for density contrast at collapse. 
\par
There are several new avenues suggested by this work.

\noindent (i) The assumption $h_{\rm SC} \rightarrow 1$, $R \rightarrow r_{vir}$ is 
equivalent to ``stable clustering'', in terms of the statistical behaviour. 
Since stable clustering has so far not been proved conclusively in simulations and 
is often questioned, it would be interesting to see the effect of changing this 
constraint to $h \rightarrow {\rm constant}$ for $t \rightarrow \infty.$

\noindent (ii) The technique of Taylor series expansion in ($1/\delta$) seems 
to hold promise. It would be interesting to try such an attempt with the 
original fluid equations and (possibly) with more general descriptions.

\noindent (iii) It must be stressed that we used the $\delta-\bar\xi$ mapping
--- possibly the weakest part of our analysis, conceptually --- only to fix 
a value of $q$. We could have used some high resolution simulations to actually 
study the evolution of a realistic overdense region. We conjecture that such 
an analysis will give results in conformity with those obtained here. 

\noindent (iv) Finally, the curves of figure (5.6) can be used to describe the spatial
distribution of virialised haloes (\cite{MoJain},\cite{Sheth}). It would be interesting to investigate
how things change when the MSCM is used in place of the standard
spherical collapse model.
\par
In the next chapter we shall approach the problem of structure formation from a completely different point of
view, namely that of Quasi Steady State Cosmology (QSSC).

\chapter{Structure formation in  QSSC}
\label{chap:QSSC}
\baselineskip=12pt
{\scriptsize Just because everything is different doesnt mean anything\newline has changed -- Paradoxial quotations}
\par
\baselineskip=24pt
 


\section{Introduction}

The Quasi-Steady State Cosmology (QSSC) was first proposed in 1993 and explored 
further
by Fred Hoyle, Geoffrey Burbidge and Jayant Narlikar in a series of papers 
(\cite{hbn93},\cite{hbn94a},\cite{hbn94b},\cite{hbn95a},\cite{hbn95}). The QSSC offers an alternative to 
the commonly accepted big bang cosmology, and the above work claims to provide 
a singularity-free  cosmological model, which is consistent with the data on 
discrete source populations, can explain the production of light nuclei as well 
as the spectrum and anisotropy of the microwave background.  Because the 
dynamical and physical conditions in this cosmology are considerably different 
from those in the standard cosmology, the theoretical reasoning required to 
understand what is observed may differ too.  In short, one may not simply lift 
a theoretical line of reasoning from standard cosmology and expect to apply it to the same problem in the QSSC.

One of the outstanding problems in modern big bang cosmology is the 
problem of formation of large scale structure in the universe.  The standard approach consists in starting with prescribed primordial fluctuations of spacetime geometry and matter density, evolving them through an inflationary era, 
having them interact with nonbaryonic dark matter, then carrying out N-body simulations of interacting masses which may eventually form into groups to be identified with large scale structures like galaxies, clusters, superclusters and voids, etc.  Although a lot of this work has gone into cosmology textbooks (\cite{Peeb80},\cite{structurebook}), it is a fair comment to say that no unique and generally acceptable structure formation scenario has yet emerged in standard cosmology.  

The problem of structure formation poses a challenge in the QSSC also and 
it should be viewed against the background of the above standard approach.  As 
we shall see in Section 6.2, the QSSC does not have an era when the baryonic 
matter density in the universe was $\sim 10^{81}$ times its present value, as 
it was in the big bang cosmology in the immediate post-inflation era. Thus the 
growth of fluctuations in the form of gravitational instabilities will not be 
similar in this cosmology to that in the big bang cosmology.

Recently the gravitational stability of the QSSC models 
against small perturbations was examined in detail in a paper by Banerjee and Narlikar \cite{banj97}. They found the cosmological solution to be stable and thus 
there was no net growth in density fluctuations.  The model is basically oscillatory and perturbations of density and metric grow only to a finite amount during the contraction phase and then decay during  the expansion phase. These authors concluded that gravitational instability alone cannot lead to formation of structures in the QSSC. Instead, explosive matter creation in the 
so-called {\it minibangs} is expected to be the principal cause of 
forming  structures.  In this chapter we try to understand the pattern of formation and growth of structures in the QSSC   though numerical simulations by using a simplified toy model.

The organization of this chapter is as follows: In section  6.2 we briefly
review the basic theory of QSSC. The numerical toy model will be introduced
in section 6.3. Section 6.4 is devoted to computing the two point correlation 
function for the distributions arising in the toy model and its comparison with observations. In section 6.5 we summarise the results by highlighting the success of this approach and indicating how it can be further improved.

\section{The Basic Theory of the QSSC}

The  basic formulation of the QSSC is via the Machian theory  of  gravity 
first  proposed  by Hoyle and Narlikar (\cite{hn64},\cite{hn66}) in  which  the 
origin of inertia is linked to a long range scalar  interaction 
between matter and matter.  Specifically, the theory is derivable 
from an action principle with the simple action:
\begin{equation}
{\cal A} = -\sum_{a}\int m_a ds_a,
\end{equation}
\noindent where  the  summation  is over all the particles  in  the  universe, 
labeled by the index $a$, the mass of the $a$th particle being $m_a$.  The 
integral  is  over  the  world  line  of  the  particle,   $ds_a$ 
representing the element of proper time of the $a$th particle.

The  mass  itself arises from interaction with  other  particles.  
Thus the mass of particle $a$ at point $A$ on its
worldline arises from all other particles $b$ 
in the universe:
\begin{equation}
m_a = \sum_{b\ne a} m_{(b)}(A),
\end{equation}
\noindent where  $m_{(b)}(X)$  is  the  contribution of inertial mass from 
particle $b$  to  any particle  situated  at a general spacetime 
point $X$.   The  long range effect is Machian in nature and is 
communicated by the scalar mass function $m_{(b)}(X)$ which satisfies 
the conformally invariant wave equation
\begin{equation}
\Box m_{(b)}  + \frac{1}{6}Rm_{(b)}  
+ m_{(b)}^3  =  N_{(b)}.
\end{equation} 
Here  the wave operator is with respect to the general  spacetime 
point $X$. $R$ is the scalar curvature of spacetime and the right 
hand side gives the number density of particle $b$. 
The field equations are obtained by varying the action 
with  respect  to the spacetime metric $g_{ik}$.   The  important 
point  to  note  is  that  the  above  formalism  is  conformally 
invariant.   In particular, one can choose a conformal  frame  in 
which the particle masses are constant.  If the constant mass  is 
denoted by $m_p$, the field equations reduce to
\begin{equation}
R^{ik}- {1\over2}g^{ik}R+ \Lambda g^{ik} = -{{8\pi 
G}\over{c^4}}[T^{ik}- f (C^iC^k- {1\over 
4}g^{ik}C^lC_l)],
\end{equation}
\noindent where $c$ is the speed of light and $C$ is a scalar field which 
arises explicitly from the ends of  broken  world  lines, that is when  there  is  creation  (or, annihilation) of particles in the universe.  The constant $f$ denotes
the coupling of the $C$-field to spacetime.  Thus the  divergence of  the  matter tensor $T^{ik}$ need not always be zero,  as  the creation or annihilation of particles is compensated by the  non-zero  divergence  of  the  $C$-field  tensor  in  Eq.(6.4).    The quantities  $G$ (the gravitational constant) and  $\Lambda$  
(the cosmological   constant)   are  related  to   the   large   scale 
distribution of particles in the universe.  Thus,
\begin{equation}
G = {3\hbar c\over 4\pi m_p^2},
~~~~\Lambda = -{3\over N^2m_p^2},
\end{equation}
\noindent $N$ being the number of particles within the cosmic horizon.

Note  that the signs of the various constants are  determined  by 
the theory and not put in by hand.  For example, the constant  of 
gravitation  is positive, the cosmological constant negative  and 
the  coupling  of  the $C$-field energy tensor  to  spacetime  is 
negative.

\subsection{Matter Creation}

The action principle tells us
that  matter creation is possible at  a  given  spacetime 
point provided the ambient $C$-field satisfies the equality $C_iC^i$ = 
$m_p^2$  at  that point. In normal  circumstances,  the  background 
level of the $C$-field will be {\it below} this level.   However, 
in  the strong gravity obtaining in the neighbourhood of  compact 
massive  objects,  the value of the field can be  locally  raised.  
This  leads  to  creation of matter along with  the  creation  of 
negative $C$-field energy.  The latter also has negative stresses 
which have the effect of blowing the spacetime outwards (as in an 
inflationary  model) with the result that the created  matter  is 
thrown out in an explosion. Qualitatively, the creation and ejection proceeds along the following lines.

The process normally begins by the creation of the $C$-field along with matter in the neighbourhood of a compact massive object.  The former, being propagated by the wave equation, tends to travel outwards with the speed of light, leaving the created mass behind.  However, as the created mass grows, its gravitational redshift begins to assert itself, and the $C$-field gets trapped in the vicinity of the object.  As its strength grows, its repulsive effect begins to manifest itself, thus making the object less and less bound and unstable. Finally, a stage may come when a part of the object is ejected from it with tremendous energy. It is thus possible for a parent compact mass to eject a bound unit outwards.  This unit may act as a center of creation in its own right.

We shall refer to such pockets of creation as {\it minibangs}  or 
{\it mini-creation events} (MCEs). A spherical (Schwarzschild type) 
compact matter distribution will lead  to  a  spherically  symmetric  
explosion  whereas  an  axi-symmetric  (Kerr type) distribution 
would  lead  to  jet-like ejection   along  the  symmetric  axis. 
Because  of the conservation of angular momentum of a  collapsing 
object, it is expected that the latter situation will in  general 
be more likely.

In either case, however, the minibang is {\it nonsingular}. There 
is no state of infinite curvature and terminating worldlines, as 
in the standard big bang, nor is there a black hole type horizon. 
The  latter  because  the presence of the  $C$-field  causes  the 
collapsing object to bounce outside the event horizon.
                                                     
\subsection{The Cosmological Solution}

The feedback of such minibangs on the spacetime as a whole is  to 
make it expand.  In a completely steady situation, the  spacetime 
will be that given by the de Sitter metric.  However, the creation 
activity  passes through epochs of ups and downs with the  result 
that the spacetime also shows an oscillation about the long  term 
steady state. Sachs \cite{sachc96} has computed the general solutions
of this kind and the simplest  such solution with the line element given by
\begin{equation}
ds^2 = c^2dt^2-S^2(t)[dr^2+r^2(d\theta^2+sin^2\theta 
d\phi^2)], \label{metric}
\end{equation}
\noindent where  $c$ stands for the speed of light has the scale factor   
given by
\begin{equation}
S(t)= e^{t/P}\bigg[1+\eta 
\cos {2\pi\tau(t)\over Q}\bigg].
\end{equation}
The  constants  $P$ and $Q$ are related to the  constants  in  the 
field equations, while $\tau(t)$ is a function $\sim t$ which  is 
also determined by the field equations.We shall, however, use the approximation $\tau(t) = t$
which is adequate for the approach used in this chapter.
The parameter $\eta$ may be  taken  positive 
and  is  less than unity.  Thus the scale  factor  never  becomes 
zero: the cosmological solution is without a spacetime singularity.
The form of the scale factor, $S(t)$, in the metric (\ref{metric}) is shown
in Figure \ref{fig-scale}.
\begin{figure}[htbp]
\leavevmode\centering
\psfig{file=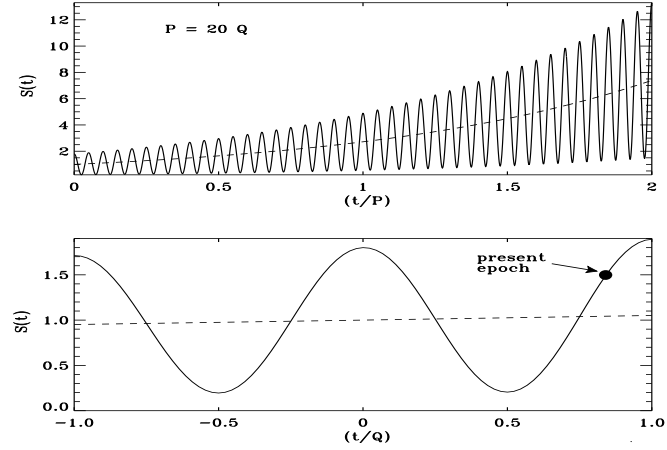,width=0.6\textwidth,height=0.4\textwidth}
\caption{\sf The scale factor $S(t)$ of the QSSC in the upper panel
against $t$ to show how several oscillatory cycles of short period $Q$
are accommodated in the longer $e$-folding time $P$ of the exponential
expansion. In the lower panel are sketched a few oscillations on an
expanded timescale with our present epoch marked.}
\label{fig-scale}
\end{figure}
\subsection{Observational Checks}
Hoyle (\cite{hbn94a},\cite{hbn94b}) have shown that the above cosmology  gives 
a  reasonably  good fit to the observations  of  discrete  source 
populations,  such  as  the  redshift-magnitude  relation,  radio 
source count, angular diameter-redshift relation and the  maximum 
redshifts so far observed, with the choice of the following set of 
parameters:

$$P{\bf \approx} 20Q,\; Q {\bf \approx} 
4.4\times10^{10}yrs,\;\eta=0.8,\;\Lambda=-0.3\times10^{-56}cm^{-2}, 
\;t_0=0.7Q.$$

Of  these, the last is the present epoch of observation.   It  is 
not essential that the model should have only these parameteric values.   Indeed, the  parameter  space is wide enough to make  the  model  robust.  
Moreover, the fitting of observations to theory does not  require 
postulating  ad hoc evolution which is commonly necessary in  the 
case of standard cosmology.

The above framework thus outlines a cosmological model without a beginning and 
without an end, in which a de Sitter type exponential expansion, characterized 
by a very long time scale $P$, is superposed with finite size oscillations of a 
shorter time scale $Q$.  Each cycles are statistically identical in their 
physical properties.  In this sense  the universe is `quasi-steady'.  We next 
see how structures might grow and proliferate in such a universe. 

\section{A Toy Model for Formation of Structures}
In an attempt to understand how  structures may possibly grow and distribute in
space we have carried out the following numerical experiment in two as well as three dimensional space.
We describe the  2$D$ case first and detail the 3$D$ versions subsequently.
\subsection{2$D$-Simulations}
A large number of points ($N \sim 10^5-10^6$), each one representing a 
mini-creation event, is distributed randomly over a unit square area .  The 
average nearest-neighbour distance for such a distribution will then be 
$(1/\sqrt{N})$.  Now suppose that in a typical mini-creation event, each particle generates another neighbour particle at random within a distance, $d = x /\sqrt{N}$ in 2$D$ . Here, the number $x$ is a fraction between $0$ and $1$ .  We shall call $x$ the separation parameter.  As explained in section 6.2.1, the above denotes an ejected piece lying
at a distance $\leq d$ from the original compact object. 
 
The sample area is then uniformly
stretched by a linear factor $\sqrt{2}$ to represent expansion of space.  We now have the same density of points as 
before, i.e., $2N$ points over area  of 2 units.  From this enlarged square 
remove the periphery so as to retain only the inner unit square.This process 
thus brings us back to the original state but with a different distribution of 
an average $N$ points over a unit square.
This process is repeated $n$ times . Here the number of iterations, $n$,
plays the role of ``time'' as in the standard models of structure formation. 
The number distribution of points evolves as the `creation process' generates 
new points near the existing ones.  We will refer to each point as a `particle' or `unit'.

Not surprisingly, soon after, i.e., after $n = 3 - 4$ iterations of the above procedure, clusters and voids begin to emerge in the sample area and create a {\it Persian Carpet} type of patterns. As the
experiment is repeated, voids grow in size while clusters become denser.  
Figure \ref{fig-2di} illustrates a typical numerical simulation.  It shows that 
expansion coupled with creation of matter is a natural means of generating 
voids and clusters.  But what of the filaments ?
\begin{figure}[htbp]
\leavevmode\centering
\psfig{file=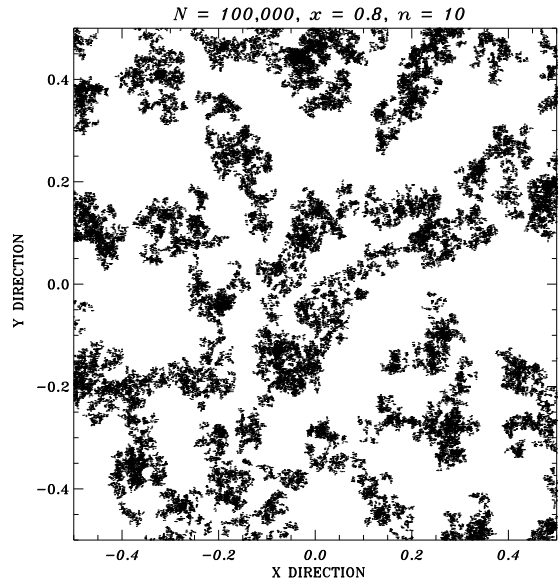,width=0.5\textwidth,height=0.5\textwidth}
\caption{\sf A cluster-void distribution generated in the 2$D$ toy model
for $N = 100, 000$ initially randomly distributed particles, with typical separation parameter $x = 0.8$, and the number of iterations $n = 10$. 
Each particle resembles a galaxy. For further discussion see the text.}
\label{fig-2di}
\end{figure}
Here we recall that the creation process near a typical compact massive object 
will not be isotropic if the mass is spinning. Matter will be preferentially 
ejected along the axis of spin.  To build this effect into the above simulation 
we adopt the following algorithm.

We assume that in a typical $n > 2$ iteration, 
 the creation of the new neighbour unit $C$ around a typical unit
$B$ is not entirely random, but, instead, related to the previous history of
creation of $B$ from an earlier generation unit $A$. So  the direction $BC$ is broadly aligned with the direction $AB$
in which $B$ itself was  ejected. Typically this is ensured by assuming that 
the ejection is at a random angle in the forward semicircle as explained in 
Figure \ref{fig-fwd}. We will refer to this as {\it aligned ejection}, as opposed to the {\it isotropic ejection} of Figure 6.2.
\begin{figure}[htbp]
\leavevmode\centering
\psfig{file=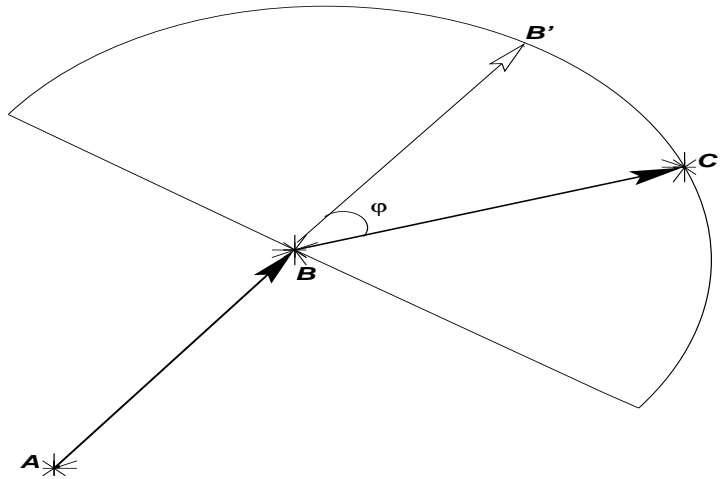,width=0.6\textwidth,height=0.4\textwidth}
\caption{\sf Schematic procedure for creating units
in aligned direction for $n>2$. Particle $A$ is a representative of first
generation units which are distributed randomly. $B$ is a representative
of the second generation units, being created in a random direction.
Point $C$ represents a third generation unit which  has been 
created in the half plane lying away from $A$ off the line perpendicular to $AB$.  $BC$ therefore makes an acute angle, $\varphi$, with the line $ABB'$.}
\label{fig-fwd}
\end{figure}
Physically this means that the unit $B$ ejected by $A$ retains `memory'
of its origin through its spin which is more or less aligned with the spin
of $A$.  Which is why when it ejects a unit $C$, it is more or less aligned with
the earlier ejection direction $AB$.

Although this algorithm does not demand strict alignment, it is interesting to 
note that the filamentary structure grows along with voids as $n$ increases. 
Features generated in this way have very suggestive similarities with the observed large scale structure as shown in a typical simulation of Figure \ref{fig-2df}.  We have also investigated the result of restricting  the secondary ejection to a narrower angle, e.g, by keeping the angle $\varphi$ of Figure \ref{fig-fwd} in the range $(-\pi/4,\pi/4)$.  Not surprisingly we find the filamentary structure more pronounced in such a case.  In general, we may
argue that the higher the angular momentum per unit mass of the compact object
causing ejection, the narrower is the angle of ejection, the greater is the alignment and hence more pronounced the filamentary structure.
\begin{figure}[htbp]
\leavevmode\centering
\psfig{file=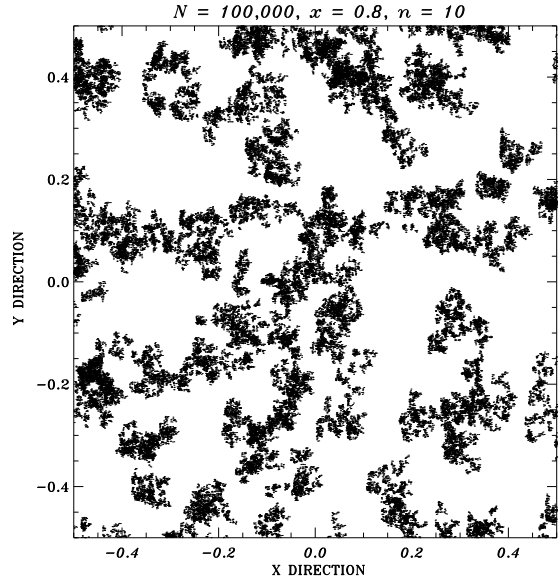,width=0.5\textwidth,height=0.5\textwidth}
\caption{\sf A computer simulated filament-void distribution with  $n (>2)$ iterations of aligned ejections having  new points following the rule of Figure 6.3, for the same 
parameters of Figure 6.2.}
\label{fig-2df}
\end{figure}
Figure \ref{fig-2ds} shows a typical two dimensional gravitational clustering
simulation data in standard big bang cosmology. 

It can be seen that both compact and extended structures are present in both
the approaches to structure formation. Since there is no observational data with which comparisons can be made in two dimensions we shall henceforth deal with the 3$D$ simulations only.
\begin{figure}[htbp]
\leavevmode\centering
\psfig{file=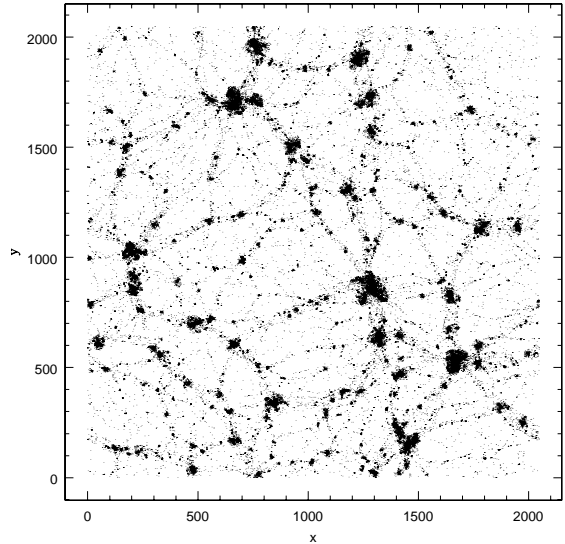,width=0.5\textwidth,height=0.5\textwidth}
\caption{\sf A power law 2$D$ simulation in the standard big bang cosmology for power index, $n = -0.4$ of density fluctuations.}
\label{fig-2ds}
\end{figure}
\subsection{3$D$-Simulations}
The 3$D$ simulation is similar, with the necessary modifications for the higher 
dimensionality.  Thus we start with a unit cube with $N$ points distributed at 
random within it,  the typical interpoint distance being $(1/\sqrt[3]{N})$.
Creating a new near neighbour for each particle by the same rule as in the 2$D$ 
case, we need to expand each edge of the cube by the factor $\sqrt[3]{2}$.  We 
next apply the same algorithm favouring aligned ejection, suitably modified for 3$D$.  
To compare the three dimensional distributions with the observed distributions 
made up from redshift surveys, we need to take a thin inside slice of the cube 
perpendicular to one of its edges and examine the distribution of points therein.  Before making such a comparison, however, we will first apply
$3D$-simulations within the framework of the QSSC. 
\subsection{Simulations of QSSC Cycles}
To bring the toy model closer to the reality of the QSSC, we proceed as follows. We 
expect the creation activity to be confined largely to a narrow era around a 
typical oscillatory minimum, when the $C$-field is at its strongest.
By considering the number density of collapsed massive objects at one
oscillatory minimum of QSSC to be $f$, the number density at the next 
oscillatory minimum would fall to  $f \exp{(-3Q/P)}$, if no new massive objects 
were added. Thus to restore a steady state from one cycle to the next, 
\begin{equation}
\alpha f \equiv [1 - \exp{(-3Q/P)}] f \sim (3Q/P) f, \label{frac}
\end{equation}
\noindent masses must be created anew. In other words, a fraction $(3Q/P)$ of the total number of massive objects must duplicate themselves in the above fashion.

Notice that, unlike the old steady state theory which had new matter appearing continuously, we have here discrete creation, confined to epochs of minimum of
scale factor. The `steady-state' is maintained from one cycle to next.  Which is why the above addition $\alpha f$ is required at the beginning of each cycle.
\begin{figure}[htbp]
\leavevmode
\centerline{
\psfig{file=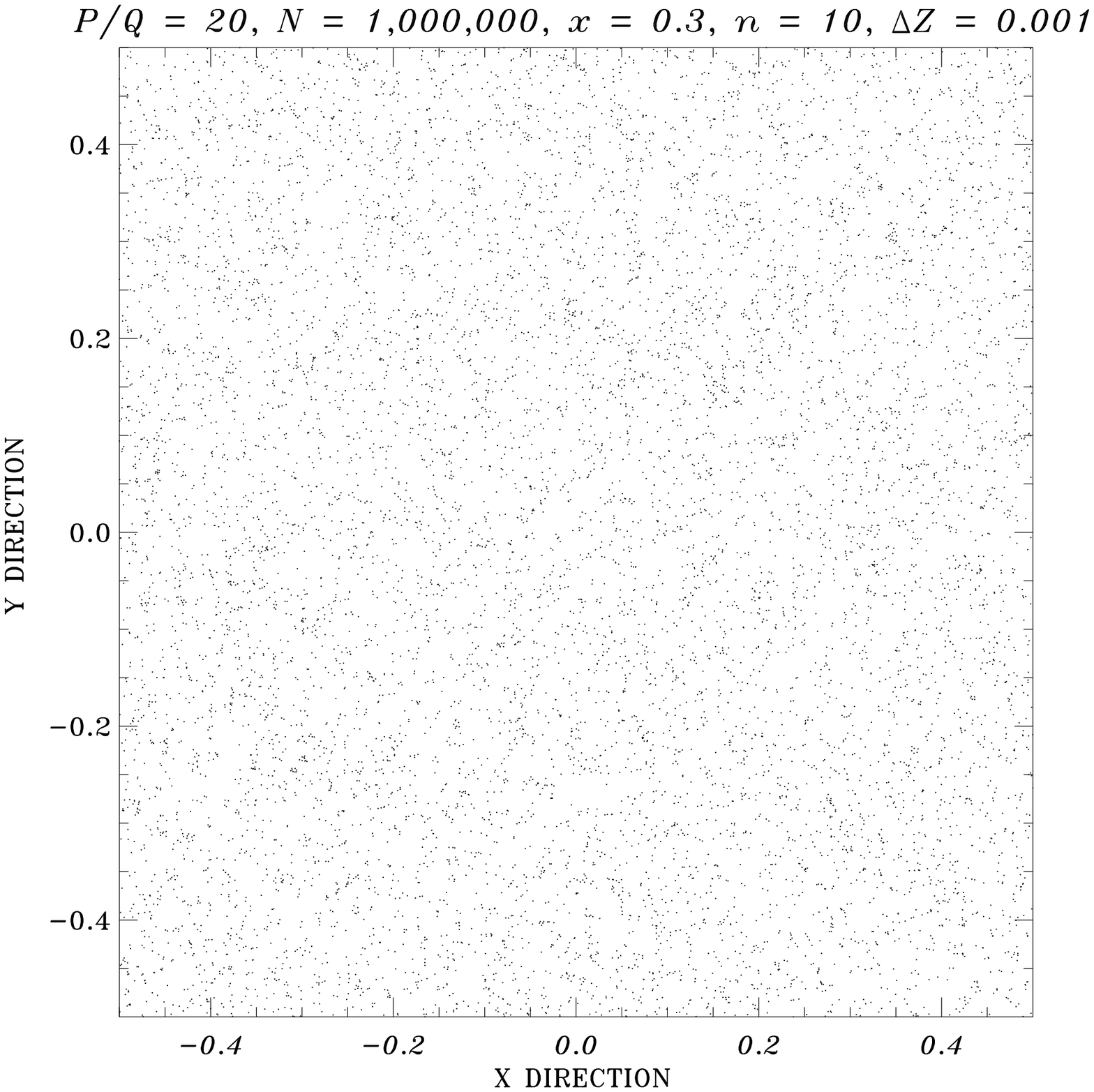,width=0.5\textwidth,height=0.5\textwidth}
\psfig{file=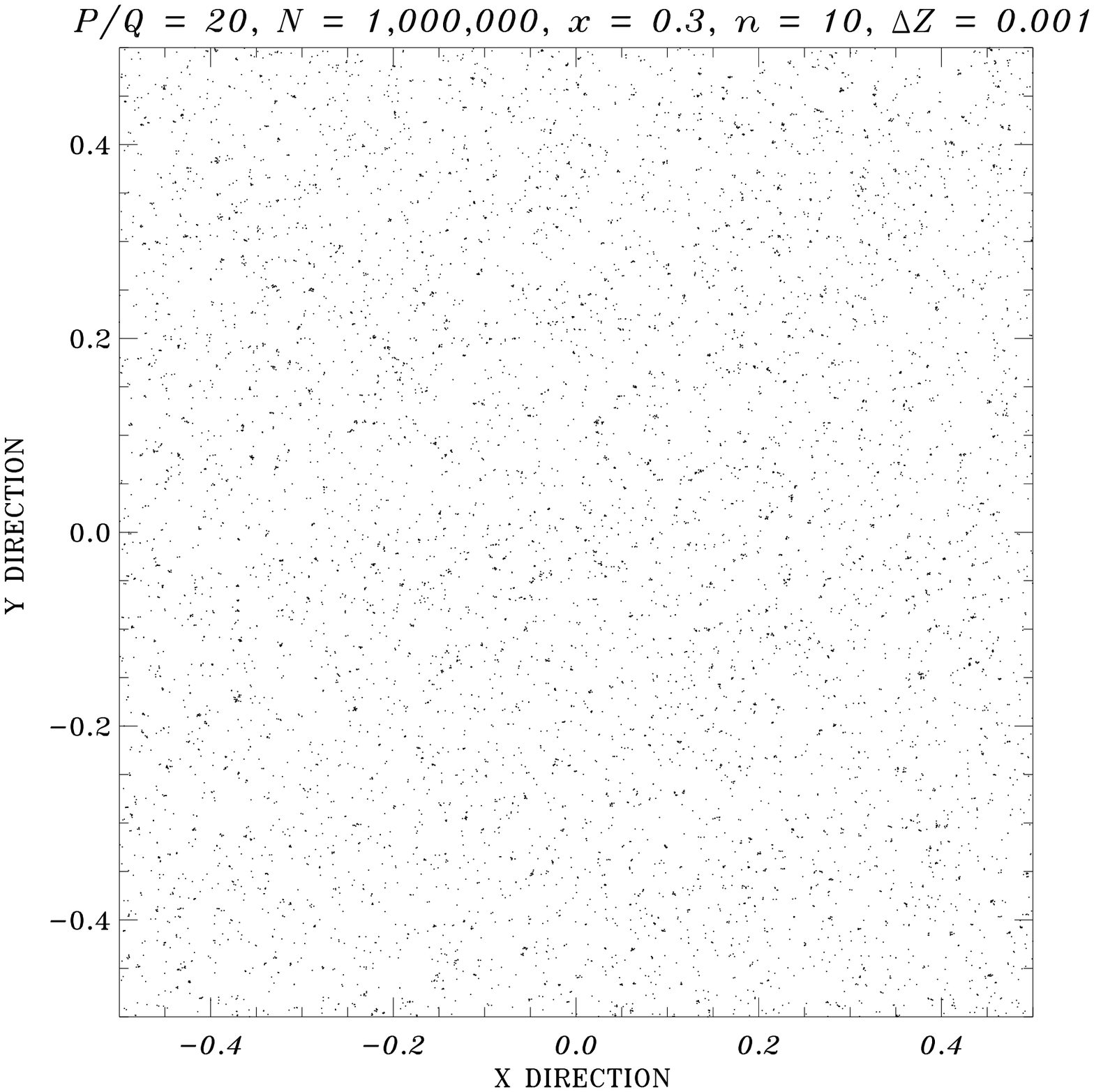,width=0.5\textwidth,height=0.5\textwidth}
}
\caption{\sf A 3-dimensional version, adapted for the 
QSSC with $N = 1, 000, 000$, $x = 0.3$, $n = 10$, and $P/Q = 20$. Slice thickness in the $Z$ direction is $\Delta Z = 0.001$. Evidently, voids are seen separated  by filamentary structures. The left panel is for the case of isotropic distribution of particles, whereas the right panel shows the case of aligned ejections.}
\label{fig-qssc}
\end{figure}
Therefore, instead of creating a new neighbour particle around each and every 
one of the original set of $N$ particles, we  do so only around $\alpha N$
of these points chosen randomly, where the fraction $\alpha$ is as defined in (\ref{frac}). 
Likewise, the sample volume is homologously expanded by the factor $\exp{(3Q/P)}$ only instead of by factor 2.  We choose the inner cube as before. Figure \ref{fig-qssc} shows the simulated distributions in cubical slices for 
isotropic as well as  aligned ejections. After a few iterations clusters and voids begin to appear, with the case for aligned ejections showing filaments. For a comparison, see an actually observed distribution of galaxies from a
redshift survey in Figure \ref{fig-survey}.
\begin{figure}[htbp]
\leavevmode\centering
\psfig{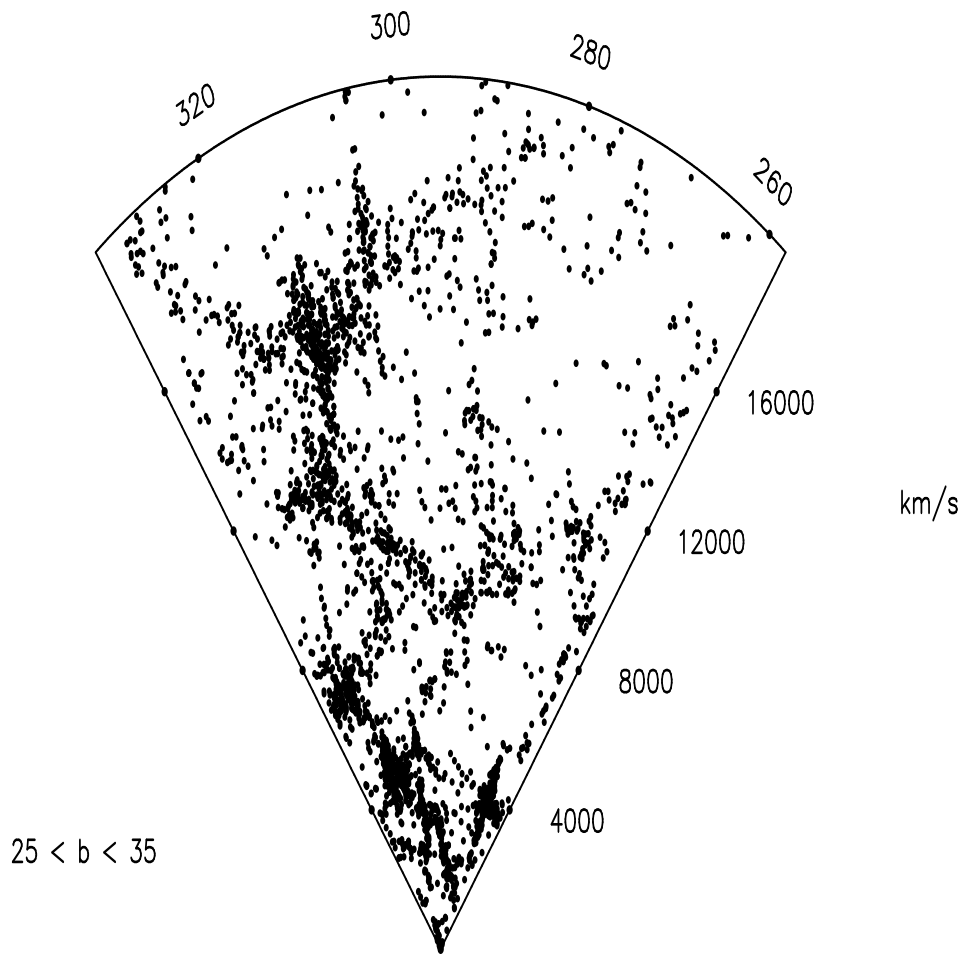}
\caption{\sf A FLAIR redshift survey in the direction of Hydra-Centaurus\cite{somak}}
\label{fig-survey}
\end{figure}
In the above approximation we have assumed that the creation activity is concentrated at the oscillatory minima.  It could be extended over an appreciable part of the oscillatory period, in which case one would see large scale structure in the radial direction as seen from an observer.  We have not
modified an algorithm to cover such cases, but feel that this should be investigated, especially since the recent analysis of the redshift-magnitude relation for supernovae has generated interest in the QSSC models of
this kind \cite{banj99}.
\section{The Two Point Correlation Functions}
Although visual inspection of Figures \ref{fig-qssc} and \ref{fig-survey} suggests that the simulation is proceeding along the right lines, a {\it quantitative} measure of the cluster-void distribution will help in comparing simulations with reality.

The dimensionless autocorrelation function
\begin{equation}
\xi (r) = <[\rho ({\bf r})-<\rho>][\rho ({\bf r}_1 + {\bf r})-<\rho>]>/<\rho>^2,
\end{equation}
\noindent where $<\rho>$ is the average density in the volume, is one convenient measure of  such irregularities in the space distribution.  Typically, different
classes of objects cluster at different characteristic lengths.  To fix ideas in
the present model, we will look at distribution of clusters of galaxies.
Observationally, it is believed that the two point correlation function for 
cluster distribution obeys the following scaling law:
\begin{equation}
\xi _{cc}(r) = \left(\frac{r}{r_0}\right)^{-\gamma}, \label{scale}
\end{equation}
with $\gamma \simeq 1.8$ and $r_0 = 25h^{-1}$ Mpc, where the Hubble constant
at the present epoch is taken to be $100h$ kms$^{-1}$Mpc$^{-1}$. In order to quantify the issues of formation of structures in this scenario we have taken the following measures. 
\par
It is known that instead of having a smooth
distribution of matter on large scales, the observed universe has structures of typical sizes of a few tens of megaparsecs .
These ``structures'' are regions of density considerably higher than the 
background density, with the maximum density contrast $\delta = (\delta \rho 
(r) /< \rho >)$ going from order unity (in the case of clusters) to a few thousand (in the case of the galaxies).
\par
Any process which generates structures must be able to produce to  the zeroth 
order , entities whose density contrast is of such magnitude  and with the 
property that on larger and larger distance scales, the density contrast becomes less significant. This is to ensure that on a large enough scale the universe is homogeneous.
\par
Given this prescription for generating structures without gravitational dynamics, we first
ensured that the visual impression created by the cluster simulations did 
imply that as the number of iterations were increased the number of high 
density regions also increased. In the initial configuration one expects to find
regions of high density arising only because of the Poisson noise. In the later ``epochs'' after a few iterations, however,
we expected and did find that the variation of the one point distribution function for density $(\rho/< \rho >)$ with $<\rho>$ the average density in the volume, showed a steady and significant increase in the the number of 
high and intermediate density regions, as is expected in a 
clustering scenario. We further observed that the value of maximum density also 
increased as a function of the number of iterations, which in this experiment 
corresponds to ``time''. The density field has been generated on a grid placed into the simulation volume using the algorithm of cloud in cell.
\par
Our simulations show the growth of structures through rise in the density maximum  as a function of number of iterations. The aligned ejection mode leads to faster clustering than the isotropic one.
\par
One must also examine the dependence of this ``growth'' on another important 
parameter in this prescription, namely the typical maximum separation between a
 creation site and the unit which is created. This was indicated by the parameter $x$ in our earlier discussion.
\par
Again our studies investigate results of the structure formation 
algorithm when the parameter $x$ is changed. We find that higher densities are achieved when this distance (in units of boxsize) is made smaller. This is intuitively expected.
\par
In the QSSC case,  clustering is stronger in the early epochs for the isotropic ejection model, although at a later stage the density function for the aligned ejection model catches up and ultimately exceeds the rate for the isotropic case. 
\par
The next quantitative measure that we computed from these data 
set was the two point correlation function. The following figures summarise
the results of these computations. It can be seen that the observationally 
obtained power law dependence of the two point correlation function 
$\xi(r)=(r_0/r)^{1.8}$ can be obtained provided a sufficient number of 
iterations has been performed, {\it i.e}, a sufficient length of ``time'' has 
elapsed.
\par
Figure \ref{fig-corqssc} shows the two point correlation function for
the case of the QSSC based model. As ``time'' goes on, the slope of the correlation function gets closer and closer to $-1.8$.  From the value of the X-axis intercept of the two point correlation function 
it is possible to get a rough estimate of the size of the structures in units 
of the size of our simulation box. From our results we have estimated that the size of the structures formed is approximately $\beta =0.15-0.3$ times the boxsize. If one sets these values equal to the observationally accepted value of $r_0$, one can get a  better physical sense of the results. If we set, $\beta= 0.3$, say, and  $r_0 = 25 h^{-1}$ Mpc, then the linear size of the simulation box would be $ \sim 84 h^{-1}$ Mpc.

The above exercise is an attempt to relate our toy model to a realistic cosmological scenario.  The model per se talks of a `dimensionless' box
containing $N$ points.  With the above identification, we have $10^5$
points in a volume of $(84h^{-1})^3$ (Mpc)$^3$.  Let us assign a mass of
$10^n M_{\odot}$ to each point.  We then get a cosmological smoothed out
density of
\begin{equation}
\rho = \frac{10^{n+5} M_{\odot}}{(84h^{-1})^3(Mpc)^3} = 0.6 h \times 10^{n-12} \rho_c
\end{equation}
where $\rho_c$ is the critical density of the universe.  Thus we get the
density parameter $\Omega = 0.6 h \times 10^{n-12}$.  Setting this equal to
unity (the QSSC does not have any limit on baryonic matter either from
deuterium abundance or  from CMBR anisotropy) we get  for $h = 0.6$, a
typical mass as $1.5 \times 10^{13} M_{\odot}$, suitable for a cluster.

Of course, as the above exercise shows, the results can be scaled up/down
by rescaling the simulation parameters and thus are independent of the `absolute' size of the box. A more detailed dynamical theory of the
creation process will tell us how to relate the absolute size of
clustering to the theoretical parameters.
\begin{figure}[htbp]
\leavevmode\centering
\psfig{file=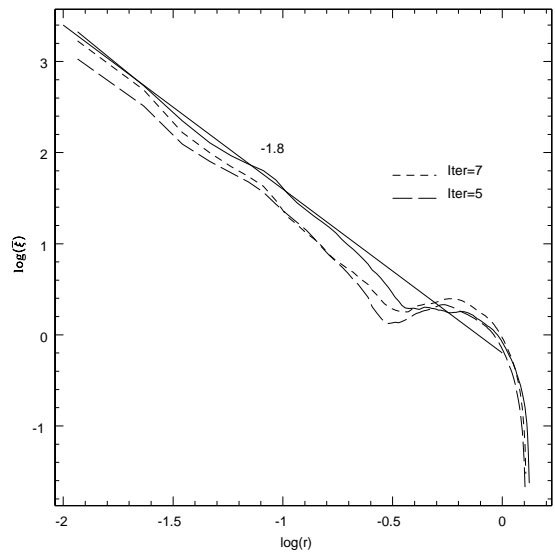,width=0.5\textwidth,height=0.5\textwidth}
\caption{\sf The two point correlation function for the QSSC based model. Here,
$N = 100, 000$ and $x = 0.3$. As ``time'' goes on, the curve approaches more and more closely the slope of $-1.8$. The solid curve shows the result after
10 iterations. }
\label{fig-corqssc}
\end{figure}

\section{Results and Summary}
It must be stressed that these results are to be viewed as a preliminary report
on a new scheme for generating structures in the Quasi Steady State Cosmology and quite a lot of follow up work has to be put into refining the model, so as to arrive at the values of the various parameters (chosen so far in an empirical way) from a deeper theoretical standpoint.  Whatever the details of the creation
process, the QSSC has repeated oscillations.  We are trying to understand with
the help of probability theory and stochastic processes, how clustering develops through such an iterative programme.
\par
The primary statistical indicator we have used in our analysis is the two
point correlation function $\bar\xi(r)$. In order to further examine the
statistical properties of the particle distribution, one must investigate
the behaviour of higher moments and other quantities such as
 the ``shape statistics''. Work analyzing the higher moments, scaling
relations , shape statistics etc using algorithms such as counts in
cells, is in progress.
\par
However, the problem of formation of large scale structure being a complex one, it is desirable to keep the theoretical options open in the underlying cosmology. At the risk of stating the obvious, we should contrast the present approach from the standard approach to structure formation in the big bang cosmology. In the standard approach primordial fluctuations are postulated to begin with and their growth is studied under the effect of the gravitational field.  Here, the main process which generates
structures in the universe is the creation of matter around MCEs rather than
gravitational instability. Our computer simulations show  very
clearly that the filament-cluster-void pattern observed in large scale 
structure can be generated simply from a creation algorithm. To what extent gravitational effects will further influence this picture remains to be seen.
Although we have given preliminary ideas in subsection 6.2.1, a more detailed cosmogonic theory is also needed to tell us how coherent objects are ejected by 
mini-creation events.
 
The success of the present toy model, however, holds out hope for a better understanding of structure formation via this alternative route.

\chapter{Conclusions}
\label{chap:concl}
\baselineskip=12pt
{\scriptsize I may not have gone where I intended to go, but I think I have ended up where I intended to be -- Douglas Adams}
\par
\baselineskip=24pt
In the previous chapters we have established some new and important results in the field of nonlinear  gravitational clustering and have shown that it is possible to gain insights into the role of gravity in clustering and consequent formation of large scale structure. The coherent theme presented in this work dealt with existence and models  for  the universal aspects of gravitational clustering, via various mechanisms. We also critically examined the role of gravity in structure formation theories by studying alternate models for structure formation, thus taking an open minded approach to the problem.
\par
We are led to the conclusion that there are universal phenomena  in gravitational clustering and we have analysed the reason behind their existence to some extent, analytically and semi analytically. 
In chapter 2 we established the existence of  {\it approximately} invariant profiles for two point correlation function which evolved in a manner similar to linear growth, at all scales. It was also shown that this  universal form is  related to the fixed point phenomena in power transfer, which is already well known in literature. The connection between these forms ``units of nonlinear universe'' and the  universal nonlinear scaling relation proposed by Hamilton and others  is also clearly established. Numerical verification of these results required high resolution N body simulations, which in turn led to analysis of two dimensional gravity in the next chapter.
\par
The next chapter tried to understand this result better by trying to understand the analytic framework underlying 2D simulations. In this work we have concluded that two dimensional structure formation simulations in an expanding background requires that we model it as parallel ``infinite needles'' with particles defined by the intersection of a plane orthogonal to these needles. We developed a (D+1) dimensional model of structure formation and have discovered that all the other approaches to 2D gravity violate either physical or consistency constraints.
\par
In the next chapter we have probed  this connection further by addressing the exitence of such Nonlinear Scaling Relations in two dimensions. Two dimensional simulations provided us with much higher resolution so that this question could be addressed in detail. We have conclusively shown  that although such a relation does exist in two dimensions as well, the behaviour at nonlinear end indicated that the clusters do not display the stable clustering behaviour as expected from 3D simulations. Thus the slope at the nonlinear end for the scaling relation is not unity as would have been expected from a model based on standard stable clustering behaviour, contrary to some of the results claimed in literature. On the other hand, the results obtained supported the existence of a generalised 'stable clustering' that had been predicted earlier.
\par
We followed the trail of results and attempted to  understand stable clustering and late time behaviour of clustering systems in the universe in more detail. This led to a detailed analysis of the behaviour of single cluster without making the usual additional assumptions about strict spherical symmetry and so on. A careful modelling of the asymmetries that are generated during the collapse phase of a density perturbation allowed us to derive a completely  natural (as opposed to the {\it ad hoc})  procedure for stabilisation of a spherically collapsing system which was consistent with the usual reults. However our results differed quantitatively from the values obtained in the standard spherical collapse model. This allowed us to establish the connection between the shear and vorticity terms in the equations and the averaged pairwise velocity ratio function which is a measure of stable clustering.  
\par
A critical examination of the role of gravity led us to investigate structure formation in Quasi Steady State Cosmology. The results of this investigation into a toy model without gravity (just expansion and creation of matter) indicate that it is possible to model some quantitative aspects of structure formation, such as the index of the two point correlation function by this process which is consistent with the accepted values. The visual images also reveal a clear cluster and void network which develops from  a uniform distribution of particles. Statistical indicators such as minimal spanning trees which examine essentially all moments of the distribution reveal the growth of clustering in a manner comparable to growth of clustering in standard scenarios. Thus we have established that the observed two point correlation function with a slope of $-1.8$ can be obtained even in a model without gravitationally induced clustering. 
\par
\par
The analysis of dark matter gravitational clustering process has led to the following results established in this thesis. There exist generic aspects to gravitational clustering  the most notable being the existence of fixed points in the way power is transferred from large scales to small scales. This transfer of power leads to scaling laws for statistical quantities such as two point correlation functions which are to a great degree independent of background cosmology and the matter power spectra. These scaling laws whose existence enable us to study late time behavior of the system easily have been vindicated in numerous simulations both in three and two dimensions. These two dimensional simulations which were intended to explore generic behavior of gravitational clustering led to a complete theoretical analysis of gravity in two dimensions which leads to greater coherence in the theoretical framework. 
\par
It remains an open problem to explain and understand how the actual particle motion in the simulation volume is connected with power transfer and with the existence of fixed points in evolution of these quantities. Some work  which addresses the issue has indicated that the growth of power in small scales are caused by matter flows from large distances, essentially the central peaks gathering in the matter particles which flow along the regions of shocks or `caustics' in the initial field. Thus the final configurations are connected to the initial Gaussian random field which are generated in inflationary scenarios for example. \par
It can be observed that the final structures that are formed at late times tend to be spherical (when they are not tidally disrupted or have significant amounts of angular momentum). The existing spherical collapse models suffer from the {\it ad hoc} nature of the  mechanism used to halt the spiraling collapse. A more rigorous analysis of spherical collapse model, including the amplification of asymmetries in the system in the collapsing phase leads to different thresholds for formation of collapsed structures with respect to the standard spherical model upon which theories such as the Press Schechter formalism is based. A study of this `virialisation' process is important in modeling the formation of clusters of galaxies and in analyzing quantities such as mass functions and their evolution.
\par
These mass functions provide a quantitative measure of the dark matter wells in which the baryons can settle, cool and fragment and form the myriad structures that are observed. Connecting the observations with the semi empirical models  of galaxy formation available currently as well as simulating the complete formation scenario from dark matter to clusters and galaxies is the fundamental program of structure formation theories in the standard model. 
\par
This work is expected  to lead us to  a comprehensive model for baryonic structures. It is necessary to understand further the transfer of power in detail to explore generic features of gravity. This involves understanding second order effects which may be modelled by approximation schemes such as Zeldovich approximation. It is possible to separate the particles into categories based on their net displacement and compute the contributions to the power spectrum from each category. This avenue must be explored to understand the details of connections between particle motion and power transfer. 
\par
Subsequent to the formation of these clusters at high peaks at early times, the baryonic structures form a population of super massive stars, which serve as nuclei for galaxies at late times. A complete model for structure formation should be able to compute the initial star formation as well as the formation and propagation of radiation and ionization fronts in the neutral medium. This initial population of stars will lead to formation of super massive black holes and quasars and trigger of further star formation which may reionize the universe at a later epoch, leaving a signature on the cosmic microwave background. 
\par
In parallel with this, an alternative cosmological structure formation scenario  explored in this thesis is the quasi steady state cosmology. In studying this model and attempting to understand the observational data in the light of this model, we were led to a greater understanding of the standard model itself and a set of stringent tests that may be designed to distinguish between the many cosmologies. A large number of statistical measures, such as minimal spanning trees, fractal dimension, cluster mass functions, measures of shapes and orientations etc are used as discriminators between models which leads to a larger array of tests of the standard model {\it per se} which may cast further light on the formation processes. This model also brought to light an interesting question regarding power law profiles for two point correlation function which was generated even in a model devoid of gravitational clustering.
\par
Existence of other universal aspects of gravity, such as density and potential profiles and their functional forms is another avenue which requires further exploration and is connected to the conclusions arrived at in this thesis.
\par
As the observational data drives us toward the era of precision cosmology, it will be possible to analyse these universal aspects in greater detail by comparing simulation models convolved with observational and instrument effects with maps of the sky which is expected to expose much of the weaknesses as well as reveal the strengths of the current paradigms of cosmology and structure formation.

\end{document}